\documentclass[a4paper,11pt]{article}
\usepackage{jheppub} 
\usepackage{amsmath,mathrsfs,braket,hyperref,bm,amssymb,graphicx}
\usepackage {slashed}

\usepackage{tikz}
\usepackage{tikz-feynman}
\usepackage{subcaption}  

\numberwithin{equation}{section}
\allowdisplaybreaks[4]
\newcommand{\bmu}{\bar\mu}
\newcommand{\gammaE}{{\gamma_\rmii{E}}}

\newcommand{\be}{\begin{equation}}
\newcommand{\ee}{\end{equation}}
\newcommand{\lp}{\left(}
\newcommand{\rp}{\right)}
\newcommand\diff{\,\mathrm{d}}
\newcommand{\MSbar}{\overline{\rm{MS}}}

\newcommand{\Tint}[1]{{\hbox{$\sum$}\!\!\!\!\!\!\int}_{\!\!\!\!#1}}  

\newcommand{\nn}{\nonumber}
\newcommand{\rmi}[1]{{\mbox{\scriptsize #1}}}
\newcommand{\rmii}[1]{{\mbox{\tiny\rm{#1}}}}

\newcommand{\gs}{g_\rmi{s}}

\newcommand{\mD}{m_\rmii{D}}
\newcommand{\mh}{\mu_{h}}
\newcommand{\g}{g}
\newcommand{\gp}{g'}
\newcommand{\gY}{g_{Y}}
\newcommand{\Nf}{N_{\rm f}}

\title{\boldmath Towards Accurate Gravitational Wave Predictions: Gauge-Invariant Nucleation in the Electroweak Phase Transition }

\author[a]{Jie Liu,}
\emailAdd{jieliu@stu.cqu.edu.cn}

\author[b]{Renhui Qin\footnote{Corresponding Author},}
\emailAdd{20222701021@stu.cqu.edu.cn}
\affiliation[b]{College of Science, Hunan Institute of Science and Engineering, Yongzhou 425199, P. R. China}

\author[a]{Ligong Bian\footnote{Corresponding Author},}
\emailAdd{lgbycl@cqu.edu.cn}
\affiliation[a]{Department of Physics and Chongqing Key Laboratory for Strongly Coupled Physics, Chongqing University, Chongqing 401331, P. R. China}

\begin{document}

\abstract{The vacuum decay in the early Universe should be gauge-invariant. In this work, we study the gauge dependence of the vacuum decay occurring through a first-order phase transition and the associated gravitational wave production. We investigate the gauge dependence of the bubble nucleation and phase transition parameters within the framework of the Standard model effective field theory in three dimension. By considering the power-counting and utilizing the Nielsen identity at finite temperature, we show that, depending on the power-counting scheme favored by the new physics scale, the perturbative computation methodology allow we get the gauge-independent nucleation rates and phase transition, this enables more accurate predictions of gravitational wave signatures.}

\maketitle
\flushbottom
\section{Introduction}
\label{sec:intro}

The first-order phase transition(PT) provides the thermal environment for electroweak baryogenesis~\cite{Morrissey:2012db}, the generations of a primordial
magnetic field~\cite{Yang:2021uid,Di:2020kbw,Zhang:2019vsb,Stevens:2012zz,Kahniashvili:2009qi,Hindmarsh:1997tj,Grasso:1997nx,Ahonen:1997wh} and detectable gravitational waves~\cite{Roshan:2024qnv,Athron:2023xlk,Caprini:2015zlo}. Such PTs commonly appear in models beyond the Standard Model(BSM). Once new physics models containing new particles are excluded, the choice remaining is the Standard Model effect field theory (SMEFT). The SMEFT can introduce a potential barrier between the ``symmetric'' and ``broken'' phase via high-dimensional operators, and contains only SM particles while respecting the SM gauge symmetry $SU(3)_C \allowbreak
\bigotimes SU(2)_L \bigotimes U(1)_Y$ . Previous studies based on the SMEFT have shown that there exists a range of parameter space capable of producing detectable gravitational waves from a strong first-order PT~\cite{Cai:2017tmh,Croon:2020cgk,Grojean:2004xa,Delaunay:2007wb,Chala:2018ari,Bodeker:2004ws,Hashino:2022ghd,Damgaard:2015con,Postma:2020toi,Gould:2021oba,Camargo-Molina:2021zgz,Kierkla:2023von,Gould:2024jjt,Ramsey-Musolf:2024ykk}.

The studies of the PT dynamics are based on the thermal effective potential~\cite{Athron:2023xlk,Schicho:2022wty,Ekstedt:2024etx}. The effective potential is gauge-dependent since the elementary fields are not invariant under gauge transformations~\cite{Hu:1996qa,Schimmoller:2020kvg}. Because the physical observables are gauge-independent, one needs to find a way to obtain the gauge-invariant results from a gauge-dependent theory. There are two different methods for obtaining gauge-invariant results, one is introduce an external source coupled to a gauge-invariant operator~\cite{Hu:1996qa,Buchmuller:1994vy,Hebecker:1995kd,Buchmuller:1995sf,Qin:2024dfp}, and the other is construct a theory satisfying the Nielsen identity~\cite{Nielsen:1975fs,Aitchison:1983ns,Metaxas:1995ab}. In this work, we choose the second method. The theory satisfying the Nielsen identity usually split the calculation into two different part, leading order(LO) and next leading order(NLO). To generate the potential barrier at LO, the contribution of one-loop gauge bosons should be included in the LO part by using the power counting, so that gauge-dependent terms will arise only at NLO. Previous studies have shown that this method will reduce the theoretical uncertainty caused by the gauge parameter~\cite{Hirvonen:2021zej,Metaxas:1995ab,Aitchison:1983ns,Fukuda:1975di,Espinosa:2016nld,Lofgren:2021ogg,Nielsen:1975fs,DiLuzio:2014bua,Garny:2012cg,Johnston:1986ib}.
These works prove gauge invariance under a $U(1)$ symmetry~\cite{Metaxas:1995ab,Hirvonen:2021zej,Lofgren:2021ogg,Garny:2012cg}, but they did not address phase transition parameters or gravitational wave studies, nor did they involve power counting beyond $\lambda \sim g^3$. Furthermore, the works based on 3d EFT only concerned the soft scale~\cite{Hirvonen:2021zej,Lofgren:2021ogg,Garny:2012cg}.

In this paper, we adopt the SMEFT with a single dimension-six operator $(\Phi^\dagger\Phi)^3/\Lambda^2$ and the 3d EFT related two-loop order dimensional reduction (DR), where $\Lambda$ is the only new physics (NP) scale. For the electroweak minimum being the global one, the NP scale should be
$\Lambda\geq v^2/m_h$ which is higher than that of the heavy scale $\pi T$ of the DR procedure since $T\lesssim \mathcal{O}(10^2)$ GeV, and the 4D studies of PT show that the
potential barrier for the first-order PT can be raised when $\Lambda < \sqrt{3}v^2/m_h$~\cite{Huang:2015tdv,Zhou:2019uzq,Grojean:2004xa}. Since the potential barrier can be directly produced by tree level potential, the power counting $\lambda\sim g^2$ firstly has been considered in the framework of the Nielsen identity. Additionally, we study the PT parameters and gravitational waves by this gauge-invariant method and compare predictions at the soft and ultra-soft scales. We found that the theory based on the power-counting $\lambda\sim g^2$ does not strictly satisfy the Nielsen identity, and the difference between the soft and ultrasoft scales is significant. We then show that one can indeed construct a gauge invariant bubble nucleation rate based on the power-counting $\lambda\sim g^3$, see the accompanied article~\cite{Liu:2025ipj} for the ultrasoft scale results. Our results show that a detectable gravitational waves can be produced at the range of $\Lambda\lesssim 570$GeV.

The structure of this paper is organized as follows.
In Section~\ref{sec:effective potential}, we begin by introducing the model under study and systematically reviewing the theoretical framework of dimensional reduction and the finite-temperature effective potential. We compute the potential at the two-loop level for both the soft and ultrasoft scales.
Section~\ref{sec:Nielsen Identity at Finite Temperature} addresses the challenge of maintaining explicit gauge invariance by adopting two distinct power-counting schemes and analyzing them as separate scenarios. This is achieved, in particular, through the application of the Nielsen identity to handle gauge dependence.
Section~\ref{sec:Power counting and perturbative expansion} classifies the relevant power-counting regimes and scale hierarchies, and then calculates the critical temperature $T_c$ through the corresponding perturbative expansions: a formal $\hbar$-expansion and a strict $g$-expansion.
In Section~\ref{sec:The bubble nucleation action}, based on the different cases established, we investigate their impact on the effective action. We perform a quantitative analysis of the effects due to gauge choice (’t~Hooft-Feynman gauge and Landau gauge) and the different scales (soft and ultrasoft).
Section~\ref{sec:First-order phase transition parameters} discusses the implications of our findings for the parameters of the phase transition and the associated gravitational wave signals.
Finally, Section~\ref{sec:Summary and outlook} summarizes the key findings and discusses their implications for future studies of phase transitions, including predictions for gravitational waves.Additionally, Appendix~\ref{appendix:SMEFT in four dimensions and dimensional reduction} presents the parameter choices in the pure 4D theory and the matching after dimensional reduction. Appendix~\ref{appdenix:zzfactor} details the calculation of the field renormalization factor, the $Z$-factor, for different cases. Appendix~\ref{two-loop effective} provides a detailed computation of the two-loop potential in the $R_\xi$ gauge. Appendix~\ref{appendix:Nielsen identity derivation} systematically derives the Nielsen identity within this model and calculates the involved $D$ and $\tilde{D}$ factors, as well as the calculation of the $C$-factor in the Nielsen identity at the two-loop level; for completeness, we also present the Nielsen identity for the 4D case in Appendix~\ref{appendix:4D Nilsen C}.
\section{The effective potential}
\label{sec:effective potential}

\subsection{The model}

Let us split the classical Lagrangian density of the electroweak sector of the SM into gauge, Higgs and fermion parts
\begin{equation}
\label{SMEFT}
\mathcal{L}=\mathcal{L}_{\mathrm{YM}}+\mathcal{L}_{\mathrm{H}}+\mathcal{L}_{\mathrm{F}}+\mathcal{L}_\mathrm{g.f}+\mathcal{L}_{\mathrm{ghost}}\;,
\end{equation}
where the Yang-Mills part is
\begin{equation}
    \mathcal{L}_\mathrm{YM}=-\frac{1}{4} W^a_{\mu \nu}W^{a \mu\nu}-\frac{1}{4} B_{\mu\nu}B^{\mu \nu}\;,
\end{equation}
with
\begin{equation}
\begin{aligned}
    W^a_{\mu\nu}&=\partial_\mu A^a_\nu-\partial_\nu A^a_\mu+g f^{abc}A^b_\mu A^c_\nu\,,\\
    B_{\mu\nu}&=\partial_\mu B_\nu-\partial_\nu B_\mu\;,
\end{aligned}
\end{equation}
where $A_{\mu}^{a}(a=1,2,3)$ and $B_{\mu}$ are the SU(2) and U(1) gauge fields, and $f^{abc}$ is the antisymmetric tensor.
The fermion part is
\begin{equation}
    \mathcal{L}_{\mathrm{F}}=\bar{Q}_L i\gamma^\mu D_\mu Q_L+ \bar{t}_R i\gamma^\mu D_\mu t_R+(-y_t \bar{Q}_L(i \sigma^2)H^* t_R+h.c.)\;,
\end{equation}
where $Q_{L}^{T}=(t_{L},b_{L})$ is the left-handed third generation quark doublet. Only the top quark is retained among the fermions and the QCD indices are suppressed in the quark sector.

The Higgs part is
\begin{equation}
    \mathcal{L}_\mathrm{H}=(D_\mu H)^\dagger (D_\mu H)-V(H)\;,
\end{equation}
with H being the SM Higgs doublet,
and the covariant derivative is defined as
\begin{equation}
	D_{\mu}=\partial_{\mu}-i g \frac{\sigma^{a}}{2}A_{\mu}^{a}-i g^{\prime} \frac{Y}{2}B_{\mu}\;,
\end{equation}
where $\sigma^{a}(a=1,2,3)$ are the Pauli matrices. The Higgs potential at tree-level and zero-temperature under study is
\begin{equation}
	V(H)=-m^{2}H^{\dagger}H+\lambda\lp H^{\dagger}H\rp^{2}+\frac{1}{\Lambda^2}\lp H^{\dagger}H\rp^{3}\;.
\end{equation}
Gauge invariance allows us to perform the shift of the Higgs doublet in a specific direction of the SU(2)$\otimes$U(1) space:
\begin{equation}
	H(x)=\frac{1}{\sqrt{2}} \lp \begin{matrix}
		\chi^1(x) + i \chi^2(x) \\
		\phi + h(x) + i \chi^3(x)
	\end{matrix} \rp\;,
\end{equation}
where h denotes the Higgs field and $\chi^{a}(a=1,2,3)$ are the Goldstone boson field. At the tree level, the effective potential reads
\begin{equation}\label{eq:Vtree}
	V_\mathrm{eff}^{(0)}(\phi)=-\frac{m^{2}}{2}\phi^{2}+\frac{\lambda}{4}\phi^{4}+\frac{c_6}{8}\phi^{6}\;,
\end{equation}
following, we will also use $c_6=1/\Lambda^2$ for the coefficient of the higher dimensional operator, as it is more convenient to work with $c_6$ when carrying out Feynman diagrammatic calculations, but $\Lambda$, being related to the energy scale of NP, aids intuition.
In 4D framework, if H acquires a background field, we define it,
\begin{equation}
	\Phi_0=\braket{H}=\frac{1}{\sqrt{2}}\begin{pmatrix}
		0\\
		\tilde{\phi}
	\end{pmatrix}\;.
\end{equation}
There we set the background field of gauge fixing item as dependent field structure, although one need not relate $\tilde{\phi}$ directly to $\phi$, but eventually identifies $\tilde{\phi}=\phi$ to eliminate the mixing between the gauge field and Goldstone mode.

We work in the $R_\xi$ gauge and choose the gauge-fixing item to be
\begin{equation}
	\begin{aligned}
		\mathcal{L}_\mathrm{g.f}=&-\frac{1}{2 \xi}\lp\partial_\mu A^{a \mu}+ig \xi \lp H^\dagger t^a \Phi_0-\Phi_0^\dagger t^a H\rp \rp^2\,\\
		&-\frac{1}{2 \xi}\lp \partial_\mu B^\mu+i\frac{g'}{2} \xi\lp H^\dagger \Phi_0-\Phi_0^\dagger H \rp\rp^2\;,
	\end{aligned}
\end{equation}
or
\begin{equation}
	\begin{aligned}
		\mathcal{L}_{\mathrm{g.f}}=&-\frac{1}{2 \xi}(\partial_\mu A^{1\mu}+\xi m_W \chi^2)^2-\frac{1}{2 \xi}(\partial_\mu A^{2\mu}+\xi m_W \chi^1)^2\,\\
		&-\frac{1}{2 \xi}(\partial_\mu A^{3\mu}-\xi m_W \chi^3)^2-\frac{1}{2 \xi}(\partial_\mu B^\mu+\xi m_B \chi^3)^2\,,
	\end{aligned}
\end{equation}
where, $t^a=\sigma^a/2$, $m_W=\frac{1}{2}g \phi$ and $m_B=\frac{1}{2}g' \phi$.
The Faddeev–Popov ghost-field relevant Lagrangian is given by
\begin{equation}
	\mathcal{L}_{\mathrm{ghost}}=-\begin{pmatrix}
		\bar{c}^a & \bar{c}^0
	\end{pmatrix} \begin{pmatrix}
		M^{ab} & M^a\\
		M^b & M
	\end{pmatrix}\begin{pmatrix}
		c^b\\
		c^0
	\end{pmatrix}\;,
\end{equation}
where
\begin{equation}
	\begin{aligned}
		M^{ab}&=(\partial^\mu D_\mu^{ab})+g^2\xi[(t^b H)^\dagger(t^a \Phi_0)+(t^a \Phi_0)^\dagger(t^b H)]\,,\\
		M^a&=\frac{g g'}{2}\xi [H^\dagger t^a \Phi_0+(t^a \Phi_0)^\dagger H]\,,\\
		M^b&=\frac{g g'}{2}\xi [(t^b H)^\dagger \Phi_0+\Phi_0^\dagger (t^b H)]=M^a\,,\\
		M&=\partial^2+\frac{g^{'2}}{4}(H^\dagger \Phi_0+\Phi_0^\dagger H)\;,
	\end{aligned}
\end{equation}
with $D^{ab}_\mu$ being the covariant derivative in the adjoint representation,
\begin{equation}
	D^{ab}_\mu=\partial_\mu \delta^{ab}-g f^{abc}A^c\;.
\end{equation}
For the ghost fields we can define combinations
\begin{equation}
	c^\pm =\frac{1}{\sqrt{2}}(c^1 \mp c^2)\;,
\end{equation}
to simplify expression further. Introducing the weak mixing angle $\theta_w$, we can change the basis from $(c^3,c^0)$ to $(c^Z,c^\gamma)$:
\begin{equation}
	\begin{pmatrix}
		c^Z\\
		c^\gamma
	\end{pmatrix}=\begin{pmatrix}
		\cos\theta_w & -\sin\theta_w\\
		\sin\theta_w & \cos\theta_w
	\end{pmatrix}\begin{pmatrix}
		c^3\\
		c^0
	\end{pmatrix}\;.
\end{equation}

\subsection{Dimensional reduction of SMEFT}
We adopt the dimensional reduction(DR) approach in this work, since it can effectively reduce the theoretical uncertainty caused by the renormalization scale~\cite{Croon:2020cgk,Qin:2024dfp}.
This approach splits the original theory into three different energy scales $\pi T, g T$, and $g^2T$, named heavy, soft and ultrasoft mode. To applied this approach, one needs a matching between the 3d and 4d fields and parameters. In general, the DR requires two steps to integrate out heavy(fermion and non-zero boson Matsubara modes) and soft (temporal gauge fields with Debye mass) modes, thereby obtaining a theory that contains only ultrasoft modes(spatial gauge and scalar fields).
By integrating out heavy modes, the $\mathcal{L}_\mathrm{YM}$ in soft theory includes
\begin{equation}
\begin{aligned}\label{gaugei0}
	W^a_{i 0}W^{a i 0}&=\lp \partial_i A^a_0-\partial_0 A^a_i+g f^{abc}A^b_i A^c_0 \rp^2 \\
	&=\lp \partial_i A^a_0+g f^{abc}A^b_i A^c_0 \rp^2\\
	&=(\partial_i A^a_0)(\partial^i A^a_0)+2 g f^{abc}(\partial_i A^a_0)A^{bi}A^c_0+g^2(f^{abc}A^b_i A^c_0)^2\;,
\end{aligned}
\end{equation}
and the $A^0$ and $B^0$ relevant parts can be collected as,
\begin{equation}
\begin{aligned}
\mathcal{L}^{0}_\mathrm{YM}&=\frac{1}{2}(\partial_i A^a_0)^2+g f^{abc}(\partial_i A^a_0)A^{bi}A^c_0+\frac{1}{2}g^2(f^{abc}A^b_i A^c_0)^2+\frac{1}{2}(\partial_i B_0)^2\,.
\end{aligned}
\end{equation}
The kinetic part of the Higgs doublet reads,
\begin{equation}
	\begin{aligned}\label{scalar0}
(D_0 H)^\dagger D_0 H=\frac{g^2}{4}A^a_0A^a_0H^\dagger H+\frac{g g'}{2}A^a_0B_0 H^\dagger \sigma^a H+\frac{g'}{4}B_0^2 H^\dagger H\,,
	\end{aligned}
\end{equation}
where
\begin{equation}
	D_0=-i g \frac{\sigma^{a}}{2}A_0^{a}-i g^{\prime} \frac{Y}{2}B_0\;.
\end{equation}
Then, the 3d effective Lagrangian at the soft scale takes the following form
\begin{equation}\label{soft lagrangian}
\mathcal{L}^{\mathrm{soft}}_{3d}=-\frac{1}{4}W^a_{ij}W^a_{ij}-\frac{1}{4}B_{ij}B_{ij}+\frac{1}{2}\lp D_i A^a_0\rp^2+\frac{1}{2}(\partial_i B_0)^2+(D_i H)^\dagger(D_i H)-V^{\mathrm{soft}}_{3d}\;,
\end{equation}
with
\begin{equation}
	\begin{aligned}
		W^a_{ij}&=\partial_i A^a_j-\partial_j A^a_i+g f^{abc}A^b_i A^c_j\;,\\
		B_{ij}&=\partial_i B_j-\partial_j B_i\;,\\
        D_i H&=\lp \partial_i-ig \frac{\sigma^a}{2} A^a_i-i g' \frac{Y}{2} B_i\rp H\;.
	\end{aligned}
\end{equation}
And, $D_i A^a_0$ originates from Eq.~(\ref{gaugei0})
\begin{equation}
D_i A^a_0=\partial_i A^a_0+g f^{abc}A^b_i A^c_0 \;,
\end{equation}

The scalar potential in 3d theory at the soft scale reads
\begin{equation}
\begin{aligned}\label{Vsoft3d}
V^{\mathrm{soft}}_{3d}&=-m_3^2 H^\dagger H+\lambda_3 (H^\dagger H)^2+c_{6,3} (H^\dagger H)^3\\
&+\frac{1}{2}m_D^2A^a_0A^a_0+\frac{1}{2}m_D^{'2}B_0^2\\
&+\frac{1}{4}\kappa_1(A^a_0A^a_0)^2+\frac{1}{4}\kappa_2B_0^4+\frac{1}{4}\kappa_3 A^a_0A^a_0B_0^2\\
&+h_1 A^a_0A^a_0H^\dagger H+h_2 B_0^2 H^\dagger H+h_3 B_0 H^\dagger A^a_0 \sigma^a H\;.
\end{aligned}
\end{equation}
The gauge fixing item at the soft scale is
\begin{equation}
\begin{aligned}\label{eq:soft g.f}
\mathcal{L}^{\mathrm{soft}}_{\mathrm{g.f}}=&-\frac{1}{2 \xi}\lp\partial_i A^{a}_i+i\frac{g_3}{2} \xi \lp H^\dagger \sigma^a \Phi_0-\Phi_0^\dagger \sigma^a H\rp \rp^2\,\\
		&-\frac{1}{2 \xi}\lp \partial_i B_i+i\frac{g'_3}{2} \xi\lp H^\dagger \Phi_0-\Phi_0^\dagger H \rp\rp^2\;,
\end{aligned}
\end{equation}
and the ghost parts is:
\begin{equation}\label{eq:soft ghost}
	\mathcal{L}^{\mathrm{soft}}_{ghost}=-\begin{pmatrix}
		\bar{c}^a & \bar{c}^0
	\end{pmatrix} \begin{pmatrix}
		M^{ab} & M^a\\
		M^b & M
	\end{pmatrix}\begin{pmatrix}
		c^b\\
		c^0
	\end{pmatrix}\;,
\end{equation}
with
\begin{equation}
	\begin{aligned}
		M^{ab}&=(\partial^i D_i^{ab})+g_3^2\xi[(t^b H)^\dagger(t^a \Phi_0)+(t^a \Phi_0)^\dagger(t^b H)]\,,\\
		M^a&=\frac{g_3 g'_3}{2}\xi [H^\dagger t^a \Phi_0+(t^a \Phi_0)^\dagger H]\,,\\
		M^b&=\frac{g_3 g'_3}{2}\xi [(t^b H)^\dagger \Phi_0+\Phi_0^\dagger (t^b H)]=M^a\,,\\
		M&=\partial^2+\frac{g^{'2}_3}{4}(H^\dagger \Phi_0+\Phi_0^\dagger H)\;,
	\end{aligned}
\end{equation}
where $ D_i^{ab}=\partial_i- g_3 f^{abc} A^c_i\,$.

When the temporal scalars $A^a_0$ and $B_0$ are heavy and could be integrated out at the ultrasoft scale~\cite{Kajantie:1995dw}, the ultrasoft Lagrangian density is,
\begin{equation}
\begin{aligned}
\label{ultrasoft lagrangian}
    \mathcal{L}^{\mathrm{ultrasoft}}_{3d}=-\frac{1}{4}W^a_{ij}W^a_{ij}-\frac{1}{4}B_{ij}B_{ij}+(D_i H)^\dagger(D_i H)+\mathcal{L}_\mathrm{g.f}+\mathcal{L}_{\mathrm{ghost}}
-V^{\mathrm{ultrasoft}}_{3d}\;.
\end{aligned}
\end{equation}

The implicit gauge couplings are $\bar{g}_3$ for the SU(2) and $\bar{g}^{\prime}_3$ for the U(1) sectors. The ultrasoft potential reads
\begin{equation}
   V^{\mathrm{ultrasoft}}_{3d}=-\bar{m}_3^2 H^\dagger H+\bar{\lambda}_3 (H^\dagger H)^2+\bar{c}_{6,3} (H^\dagger H)^3\;.
\end{equation}
The Lagrangian for the gauge fixing terms and the ghost terms is similar to Eqs.~(\ref{eq:soft g.f}) and~(\ref{eq:soft ghost}), only the parameters are changed to the corresponding ultrasoft parameters.

We here note that the temporal components of gauge fields can be integrated out when the Debye mass is well separated from the background-field dependent contribution for the weak phase transition that can be studied with lattice simulations~\cite{Kajantie:1995dw}.
Recent studies have emphasized that the soft EFT should be employed for stronger first-order phase transitions~\cite{Ekstedt:2020abj,Gould:2023ovu,Gould:2024jjt,Bernardo:2025vkz}, particularly when the hierarchy required for integrating out the temporal gauge fields is not parametrically satisfied.
Motivated by these developments, we investigate both the soft and ultrasoft EFTs throughout this work, rather than focusing on the latter one~\cite{Croon:2020cgk}. This enables us to compare their predictions and assess the residual gauge dependence in both descriptions.


Within the framework of 3D finite-temperature field theory, we need to compute the $Z$-factor, thermal effective pontential, as well as the $C,D,\tilde{D}$-factors for the justification of the Nielsen identities. The details are provided in Appendices.

\subsection{The calculation of the thermal effective potential}




The central object for studying the phase transition is the effective potential. As in generic gauge theories, the effective potential is gauge dependence. We work in the $R_\xi$ gauge and $\MSbar$ renormalization scheme for our computations.
The effective potential may be written as
\be\label{v:first}
V^{\mathrm{eff}}=V_{\mathrm{tree}}+V_{\mathrm{1 loop}}+V_{\mathrm{2 loop}}+\cdots\; .
\ee

At soft scale, the tree-level effective potential for the 3d theory is defined by~\eqref{soft lagrangian}
\be\label{eq:Vsofttree}
V^{\mathrm{soft}}_{\mathrm{tree}}=-\frac{1}{2}m_3^2\phi_3^2+\frac{\lambda_3 \phi_3^4}{4}+\frac{c_{6,3} \phi_3^6}{8}\;.
\ee
Then, the one-loop contribution ($V_{g^3}$) at the soft scale is given by
\begin{equation}
\begin{aligned}
V^{\mathrm{soft}}_{\rm{1loop}}=&J_{3d}(m_{h,3})+2 J_{3d}(m_{\chi^\pm,3})+J_{3d}(m_{\chi^0,3})+4 J_{3d}(m_{W,3})+2 J_{3d}(m_{Z,3})\,\\
	&-2 J_{3d}(m_{c_W,3})-J_{3d}(m_{c_Z,3})+3 J_{3d}( m_L)+J_{3d}(m_L^{\prime})\;,
\end{aligned}
\end{equation}
with
\begin{equation}
	J_{3d}(m)=-\frac{1}{12 \pi} m^3\;.
\end{equation}
where the mass parameters appearing in this expression can be computed by Eq.~\eqref{field dependent masses} and substituting $g \to g_3, g^\prime \to g^\prime_3, \lambda \to\lambda_3, c_6 \to c_{6,3}, m \to m_3$. For more details on the soft scale parameters, see Appendix~\ref{parameter matching}, and Eqs.~\eqref{soft couplings}-\eqref{soft masses}.
And, the two-loop contribution to the effective potential ($V_{g^4}$) at the soft scale reads
\begin{equation}
\begin{aligned}
 V^{\mathrm{soft}}_{\rm{2loop}}=&-\Big((SSS)+(SGG)+(VSS)+(VGG)+(VVS)+(VVV)\\
    &+(SS)+(SV)+(VV) \Big)+V^{(A^a_0,B_0)}_{2,3d}\;.
\end{aligned}
\end{equation}
The effective potential for the 3d theory at the ultrasoft scale is defined by~\eqref{ultrasoft lagrangian}. The tree-level effective potential reads
\begin{equation}\label{eq:Vultrasofttree}
    V^{\mathrm{ultrasoft}}_{\mathrm{tree}}=-\frac{1}{2}\bar{m}_3^2\phi_3^2+\frac{\bar{\lambda}_3 \phi_3^4}{4}+\frac{\bar{c}_{6,3} \phi_3^6}{8}\;.
\end{equation}
At the ultrasoft scale, the one-loop effective potential ($V_{g^3}$) reads
\begin{equation}
\begin{aligned}\label{eq:V1loop}
V^{\mathrm{ultrasoft}}_{\rm{1loop}}&=J_{3d}(\bar{m}_{h,3})+2 J_{3d}(\bar{m}_{\chi^\pm,3})+J_{3d}(\bar{m}_{\chi^0,3})+4 J_{3d}(\bar{m}_{W,3})+2 J_{3d}(\bar{m}_{Z,3})\,\\
	&-2 J_{3d}(\bar{m}_{c_W,3})-J_{3d}(\bar{m}_{c_Z,3})\;.
\end{aligned}
\end{equation}
These paramaters can be defined by Eq.~\eqref{field dependent masses} and substituting $g \to \bar{g}_3, g^\prime \to \bar{g}^\prime_3, \lambda \to\lambda_3, c_6 \to \bar{c}_{6,3}, m \to \bar{m}_3$.
Correspondingly, the two-loop effective potential at the ultrasoft scale is,
\begin{eqnarray}\label{v:end}
    V^{\mathrm{ultrasoft}}_{\rm{2loop}}=&-&\Big((SSS)+(SGG)+(VSS)+(VGG)+(VVS)+(VVV)\notag\\
    &+&(SS)+(SV)+(VV) \Big)\;.
\end{eqnarray}
When we calculate the effective potential, excluding the contributions of $A^a_0$ and $B_0$ at the soft scale, the 3d effective potential forms at the two different scales are similar, we only need to use the parameters corresponding to the scale being considered, see Appendix~\ref{parameter matching}. See Appendix~\ref{two-loop effective} for details on expressions for the 3d two-loop effective potential contributions in the SMEFT. The effective potential and the following wave function Z-factor for new physics models can be obtained also by utilizing the public code {\it DRalgo}~\cite{Ekstedt:2022bff} within the Landau gauge.


Having obtained the effective potential of the model, we now turn to its application to vacuum decay and phase transition dynamics, where a conceptual issue immediately arises. Owing to the gauge-fixing procedure inherent in perturbative calculations, the effective potential itself is generally gauge-dependent. Consequently, physical quantities extracted directly from the effective potential, such as the critical temperature read off from the location of the minimum, or the vacuum decay rate computed from the shape of the potential barrier --- may inherit spurious gauge dependence. In particular, while the vacuum decay rate is required to be a gauge-invariant physical observable, this property is contingent upon a proper treatment of the gauge dependence encoded in the effective potential. Similarly, the extraction of phase transition parameters, including the transition strength and the nucleation temperature, is physically meaningful only when performed within a gauge-invariant framework. This section systematically examines the gauge dependence of the effective potential, on the basis of which we will later address how to extract a gauge-invariant vacuum decay rate and reliable phase transition parameters.

\section{Nielsen Identity at Finite Temperature}
\label{sec:Nielsen Identity at Finite Temperature}
\begin{figure}[h!]
	\centering
	\includegraphics[width=0.8\linewidth]{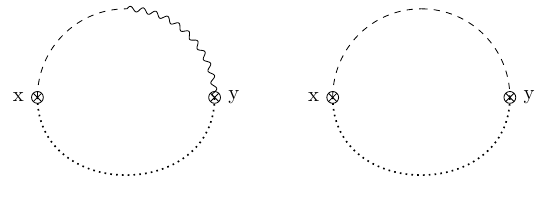}
	\caption{The two graphs that contribute to $C$ at one-loop order}
	\label{fig:C:one loop}
\end{figure}

The proof that the nucleation rate and PT parameters are gauge-independent requires the generalization of the Nielsen identity for the thermal EFT.
We expect that Nielsen identities still hold at finite temperature after suitably taking into account thermal effects, since  this ultimately follows from the fact that the partition function respects the BRST symmetry ensuring that Ward identities still hold at finite temperature. The same applies to the Nielsen identity, that can be regarded as a Ward identity with the thermal effective potential. Our approach is based on Ref.~\cite{Nielsen:1975fs}, which describes the gauge dependence of the effective action, and has been used to show that gauge-independent physical quantities can be obtained from a gauge-dependent effective potential~\cite{Fukuda:1975di}:
\be\label{Nielsenkey}
\xi\frac{\partial }{\partial \xi}S[\phi(x),\xi]=-\int_y K[\phi(y),\xi]\frac{\delta S}{\delta \phi(y)}\;,
\ee
wherein,
\be
K[\phi]=C(\phi)+D(\phi)(\partial_\mu \phi)^2-\partial_\mu(\tilde{D}(\phi)\partial_\mu \phi)+\mathcal{O}(\partial^4)\;,
\ee
which have a loop expansion~\cite{AITCHISON19841}, and the detailed derivation is provided in Appendix~\ref{appendix:Nielsen identity derivation}.
\begin{equation}
	\begin{aligned}
		K_j[\phi]=&-\int \diff^4 x i \hbar \bra{0}T \lp\frac{i}{\hbar}\rp^2\left[ \frac{1}{2} \bar{c}^a(x)\lp\partial_\mu W^{a \mu}+g v^a_i \varphi_i\rp igc^b(0) t^b_{jk}\varphi_k(0) \exp \frac{i}{\hbar}S_\mathrm{eff}\right]\ket{0}\,\\
		&-\int \diff^4 x i \hbar \bra{0}T \lp\frac{i}{\hbar}\rp^2\left[ \frac{1}{2} \bar{c}^0(x)\lp\partial_\mu B^\mu+g' v_i \varphi_i\rp ig' c^0(0) n'_{jk}\varphi_k(0) \exp \frac{i}{\hbar}S_\mathrm{eff}\right]\ket{0}\;,
	\end{aligned}
\end{equation}
we can extract the $C(\phi)$ part after expand $\exp\lp(i/\hbar)S_\mathrm{eff} \rp$ at one-loop,
\begin{equation}
	\begin{aligned}
		C_i[\phi]=&-\frac{g}{2}\int_y\braket{c^b(x) t^b_{ij}\varphi_j\bar{c}^a(x)\lp\partial_\mu W^{a \mu}(y)+g v^a_k \varphi_k(y)\rp}\,\\
		&-\frac{g'}{2}\int_y\braket{c^0(x) n'_{ij}\varphi_j\bar{c}^0(x)\lp\partial_\mu B^\mu(y)+g' v_k \varphi_k(y)\rp}\;,
	\end{aligned}
\end{equation}
The corresponding Feynman diagram is shown in Figure~\ref{fig:C:one loop}. For detailed calculations, see the following calculation of the contribution of one loop to C factor. We expand $\exp\lp(i/\hbar)S_\mathrm{eff} \rp$ and calculate the C factor for two loops, for detailed calculations see Appendix~\ref{appendix:C factor two loop}.

The Nielsen identity describes the $\xi$-dependence of the effective action and plays a central role in discussing how to obtain gauge-independent quantities. For cases with a Higgs background only, the identity reads:
\be
\label{Nielsen_1}
\xi\frac{\partial V}{\partial \xi}=-C \frac{\partial V}{\partial \phi}\;,
\ee

\begin{equation}
\label{Nielsen_2}
\xi\frac{\partial Z}{\partial \xi}=-C\frac{\partial Z}{\partial \phi}-2 Z \frac{\partial C}{\partial \phi}-2 D \frac{\partial V}{\partial \phi}-2 \tilde{D}\frac{\partial^2 V}{\partial \phi^2}\;.
\end{equation}

In perturbation theory, the coefficients C and $D,\tilde{D}$ appearing in Nielsen identities are expanded as
\begin{align}
	C=C_g+C_{g^2}+\cdots\;,\quad
	D, \tilde{D}=\mathcal{O}(g^{-1})\;.
\end{align}
The leading-order contributions to C are identical for the \(\lambda \sim g^2\) and \(\lambda \sim g^3\) power-counting schemes, and the $C_{g^2}$ is suppressed in the \(\lambda \sim g^3\) scheme. The terms D and \(\tilde{D}\) need only be evaluated in the \(\lambda \sim g^2\) scheme, as they are power-counting suppressed in the \(\lambda \sim g^3\) scheme, as will be detailed below.
When we calculate the field  renormalization factor (Z), we use
\begin{equation}
    \tilde{\phi} \to \tilde{\phi}+\tilde{h}
\end{equation}
to shift the gauge fixing background field and treat $\tilde{h}$ as an external auxiliary field that only appears on external legs and does not contribute to the propagator.
Where, the field renormalization factor for the scalar field can be computed as
\begin{equation}
	Z=\frac{\partial }{\partial k^2}\lp \Pi_{hh}+\Pi_{h\tilde{h}}+\Pi_{\tilde{h}h}+\Pi_{\tilde{h}\tilde{h}}\rp\;,
\end{equation}
Where $\Pi$ denotes the scalar two-point correlation function. The Feynman diagrams involved in the calculation of the $Z$-factor are shown in Figure~\ref{fig:Zfactor small}, with the detailed computational procedure provided in Appendix~\ref{appendix:3d Z-factor}.
Finally, calculating the field renormalization Z factor at a one-loop level on the soft scale, we have
\begin{equation}
    Z \to Z+Z^{(A^a_0,B_0)}_{1,3d}\;.
\end{equation}
The corresponding parameters should be adjusted to the soft level parameters.
And, the functions of $D,\tilde{D}$ are given by:
\begin{eqnarray}
    	D=\frac{\partial }{\partial k^2}\lp \Pi^D_{h,h}+\Pi^D_{h,\tilde{h}}+\Pi^D_{\tilde{h},\tilde{h}} \rp\;,\quad
	\tilde{D}=\frac{\partial }{\partial k^2}\lp \Pi^{\tilde{D}}_h +\Pi^{\tilde{D}}_{\tilde{h}} \rp\;.
\end{eqnarray}
The relevant calculations are detailed in Appendix~\ref{appendix:D}-\ref{appendix:Dtilde}.

\begin{figure}[!htp]
    \centering
    \includegraphics[width=0.8\linewidth]{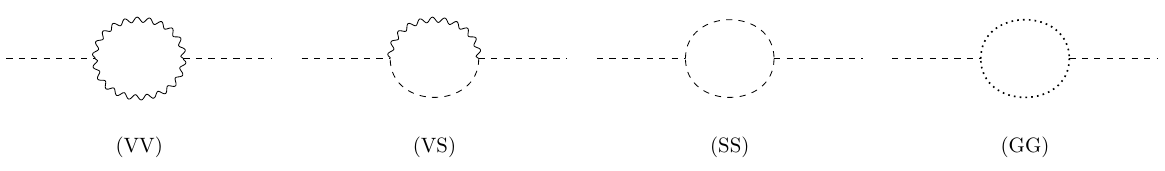}
    \caption{The diagrams needed for calculating the $Z$-factor.The dashed lines represent scalar propagators, the wavy lines represent gauge field propagators, and the dotted lines represent ghost field propagators.}
    \label{fig:Zfactor small}
\end{figure}

\subsection{$\lambda \sim g^2$ scenario}\label{sec:TLB}
Now, we organize the potential using the ultrasoft scale as an illustration (the soft scale can be treated analogously). With $c_6$ introduced, it is possible to appear a barrier at the tree-level potential, which requires a negative $\bar{\lambda}_3$. We can take $\bar{\lambda}_3\sim g^2 T$ (for brevity, denoted as $\lambda \sim g^2$) and split the the effective potential in powers of $g$:
\begin{equation}
    V^{\mathrm{eff}}=V_{g^2}+V_{g^3}+V_{g^4}+\cdots\,,
\end{equation}

In the following, we drop the subscript $``3"$ and the bar notation. And, here, we have the correspondence of $V_{g^2}=V_{\mathrm{tree}}, V_{g^3}=V_{\mathrm{1loop}}$ and $V_{g^4}=V_{\mathrm{2loop}}$ with Eqs.~\eqref{v:first}-\eqref{v:end}.

When we consider the power counting $\lambda\sim g^2$, the tree level and one-loop level effective potential contribute at the $g^2$ and $g^3$ orders, then the first Nielsen identity at $g^3$ order reads,
\begin{equation}\label{Nielsen_1a}
    \xi\frac{\partial V_{g^3}}{\partial \xi}=-C_g \frac{\partial V_{g^2}}{\partial \phi}\;,
\end{equation}
and the second Nielsen identity at the order of $g$ reads:
\begin{equation}
	\label{Nielsen_2_With D}
	\xi \frac{\partial Z_g}{\partial \xi}=-2\frac{\partial C_g}{\partial \phi}-2 D_{g^{-1}} \frac{\partial V_{g^2}}{\partial \phi}-2 \tilde{D}_{g^{-1}}\frac{\partial^2 V_{g^2}}{\partial \phi^2}\;.
\end{equation}
Here, at 1-loop level,
$C_{g}=C_{W}(\phi,T,\xi)+C_{Z}(\phi,T,\xi)$ with
\begin{align}
	C_{W}(\phi,T,\xi)&=\frac{1}{2} g \int_{p} \frac{\xi m_W}{\lp p^2+m^2_{\chi^\pm}\rp \lp p^2+m^2_{cW} \rp}
	\ , \\
	C_{Z}(\phi,T,\xi)&=\frac{1}{2} g' \int_{p} \frac{\frac{1}{2}\xi m_B}{\lp p^2+m^2_{\chi^0}\rp \lp p^2-m^2_{cZ} \rp} + \frac{1}{2} g \int_{p} \frac{\frac{1}{2}\xi m_W}{\lp p^2+m^2_{\chi^0}\rp \lp p^2+m^2_{c^Z} \rp}\;,
\end{align}
Wherein, the main loop integral reads
\be\label{eq:C1loop}
I_{3d}(m_1,m_2)=\int_{p}\frac{1}{(p^2+m^2_1)(p^2+m^2_2)}=\frac{1}{4 \pi(m_1+m_2)}\;.
\ee
Therefore, the $C_{g}$ is
\begin{equation}
    C_{g}=\frac{g^2 \xi \phi}{16 \pi(m_{\chi^\pm}+m_{c_W})}+\frac{(g^2+g^{\prime 2})\xi \phi}{32 \pi(m_{\chi^0}+m_{c_Z})}\;.
\end{equation}

For the case of $\lambda\sim g^2$, one can easily verify the first Nielsen identity at $g^3$ order (Eq.~(\ref{Nielsen_1a})) holds analytically.
If we consider the matching of power counting up to the two-loop potential, we will involve the two-loop calculation of the C factor. The corresponding calculation is given in Appendix~\ref{appendix:C factor two loop}. In this case, we have
\begin{equation}\label{Nielsen_1b}
    \xi \frac{\partial V_{g^4}}{\partial \xi}=-C_{g^2}\frac{\partial V_{g^2}}{\partial \phi}-C_{g}\frac{\partial V_{g^3}}{\partial \phi}\;,
\end{equation}

\begin{figure}[!htp]
    \centering
    \includegraphics[width=0.4\linewidth]{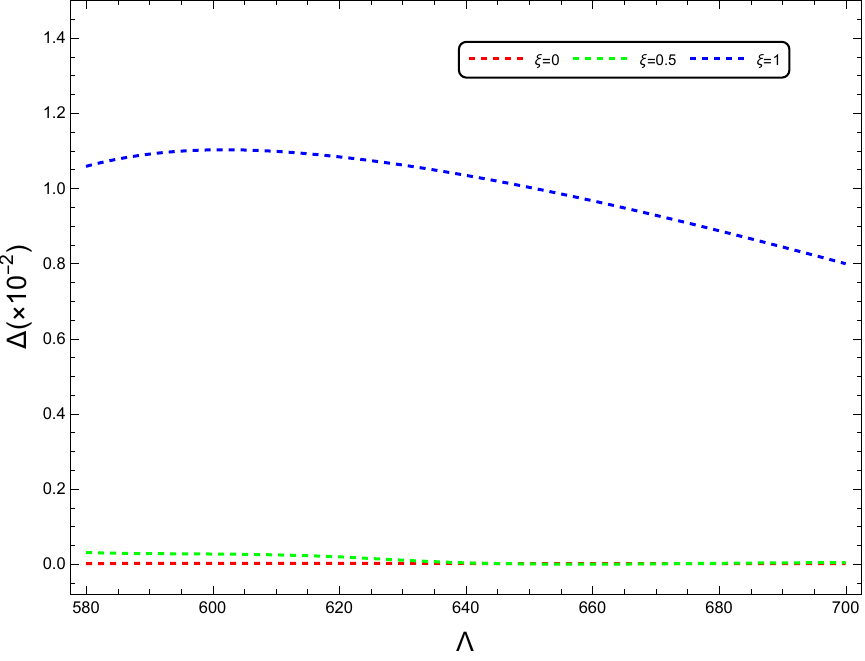}
    \quad
    \includegraphics[width=0.4\linewidth]{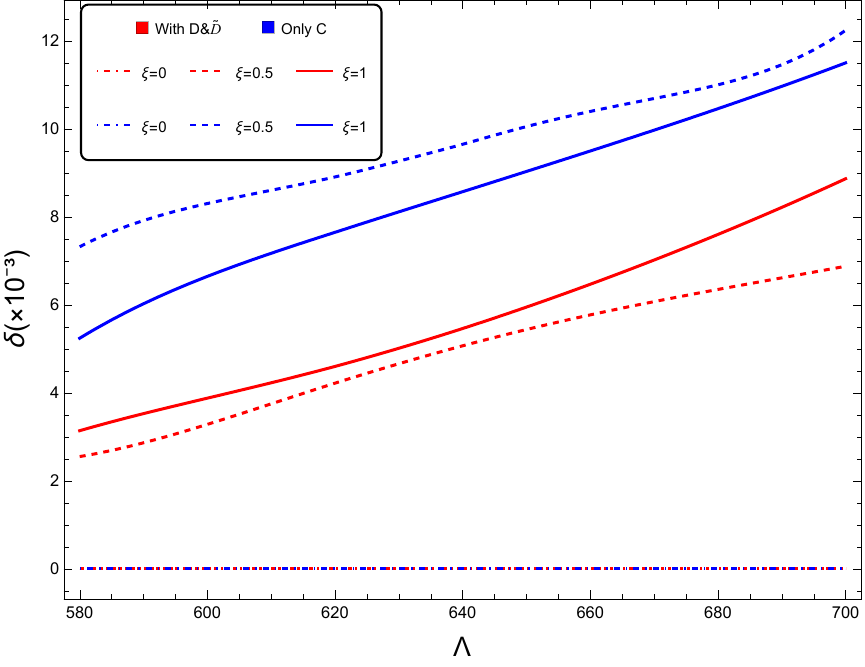}
    \caption{Left: the deviation $\Delta$ as function of $\Lambda$, where the Red, Green, and Blue curves respectively correspond to $\xi =0$, $\xi =0.5$, $\xi =1$. Right: the deviation $\delta$ as a function of $\Lambda$, the red curve represents the contribution from Eq.~(\ref{Nielsen_1b}), which includes the terms $D$ and $\tilde{D}$, while the blue curve corresponds to the contribution without $D$ and $\tilde{D}$, like Eq.~(\ref{like2}), different line styles denote different gauge parameters.}
    \label{fig:deviation analysis}
\end{figure}

Due to the complexity of calculating the two-loop level of the C factor and $D,\tilde{D}$ at one-loop level, we use numerical verification for Eqs.~(\ref{Nielsen_1b}) and~(\ref{Nielsen_2_With D}) as shown in Figure~\ref{fig:deviation analysis}.
We consider the deviations of $\Delta$ and $\delta$ being the differences between the left-hand side and the right-hand side of Eq.~(\ref{Nielsen_1b}) and Eq.~(\ref{Nielsen_2_With D}), respectively.
\begin{equation}
\begin{aligned}
\Delta&=\xi \frac{\partial V_{g^4}}{\partial \xi}-\Big(-C_{g^2}\frac{\partial V_{g^2}}{\partial \phi}-C_{g}\frac{\partial V_{g^3}}{\partial \phi}\Big)\;,\\
    \delta &=\xi \frac{\partial Z_g}{\partial \xi}-\Big(-2\frac{\partial C_g}{\partial \phi}-2 D_{g^{-1}} \frac{\partial V_{g^2}}{\partial \phi}-2 \tilde{D}_{g^{-1}}\frac{\partial^2 V_{g^2}}{\partial \phi^2}\Big)\;.
\end{aligned}
\end{equation}
The left panel of the Figure~\ref{fig:deviation analysis} shows the Nielsen identity about the effective potential is essentially valid, the deviation on the sides of the expression~(\ref{Nielsen_1b}) for the $\xi=1$ case becomes noticeable. The right panel of Figure~\ref{fig:deviation analysis} presents a numerical test of the Nielsen identity concerning the Z part. We find that including the terms $D$ and $\tilde{D}$ results in a better agreement between both sides of the identity. Furthermore, when the contributions from $D$ and $\tilde{D}$ are taken into account, the discrepancy between the two sides becomes more pronounced as the gauge parameter increases or $\Lambda$.

\subsection{$\lambda \sim g^3$ scenario}\label{sec:RB6}
If the gauge boson one-loop contributions play a role in the formation of a barrier, we can features the power-counting $\bar{\lambda}_3\sim g^3T$ (denoted as $\lambda \sim g^3$). For a more detailed discussion, see~\cite{Camargo-Molina:2024sde}.
In this scheme, we have
\be
V^{\mathrm{eff}}=V_{\rm LO}+V_{\rm NLO}+\cdots\;,
\ee
one can reorganize the effective potential of tree-level and part of one-loop contributions as:
\begin{equation}\label{eq.lo}
    V_{\mathrm{LO}}=-\frac{1}{2}m^2\phi^2+\frac{\lambda \phi^4}{4}+\frac{c_6 \phi^6}{8}-\frac{g^3 \phi^3}{24 \pi}-\frac{\lp g^2+g^{\prime 2}\rp^{3/2}\phi^3}{48 \pi}\,,
\end{equation}
and, after considering  $ (m_G^2+m_{c_W}^2)^\frac{3}{2}-m_{c_W}^3 \sim \frac{3}{2} m_G^2 m_{c_W}$ and $ (m_G^2+m_{c_Z}^2)^\frac{3}{2}-m_{c_Z}^3 \sim \frac{3}{2} m_G^2 m_{c_Z}\,$, we collect the rest of the one-loop potential as well as the two-loop contributions into the $V_{\rm NLO}$:
\begin{equation}\label{eq.nlo}
    V_{\mathrm{NLO}}=-\frac{\sqrt{\xi}\lp 2 g+\sqrt{g^2+g^{'2}}\rp \phi }{16 \pi}\lp-m^2+\lambda \phi^2+\frac{3}{4}c_6 \phi^4\rp+V^{\mathrm{2-loop}}_{g^4}\;,
\end{equation}
when we calculate the two-loop contribution to the effective potential, considering the power counting $\lambda\sim g^3$, one can set $\lambda,m_h \to 0, m_{\chi^\pm} \to m_{c_W},m_{\chi^0} \to m_{c_Z}$ similar to the Abelian Higgs scenarios~\cite{Garny:2012cg,Hirvonen:2021zej}, and
\begin{equation}
    V^{\mathrm{2-loop}}_{g^4}=V^{0}_{2,g^4}+V^{\xi}_{2,g^4}\;.
\end{equation}
Here, $V^{0}_{2,g^4}$ has no explicit $\xi$-dependence and is equivalent to the $V^0_{\mathrm{NLO}}$ term in~\cite{Liu:2025ipj}. We have
\begin{eqnarray}
    V^{\xi}_{2,g^4}&=&\frac{5 g^4 \sqrt{\xi } \phi ^2}{256 \pi ^2}+\frac{g^2 g^{\prime 2} \sqrt{\xi } \phi ^2}{128 \pi ^2}+\frac{g g^{\prime 4} \sqrt{\xi } \phi ^2}{128 \pi ^2 \sqrt{g^2+g^{\prime 2}}}+\frac{g^5 \sqrt{\xi } \phi ^2}{64 \pi ^2 \sqrt{g^2+g^{\prime 2}}}\notag\\
    &&+\frac{3 g^3 g^{\prime 2} \sqrt{\xi } \phi ^2}{128 \pi ^2 \sqrt{g^2+g^{\prime 2}}}+\frac{g^{\prime 4} \sqrt{\xi } \phi ^2}{256 \pi ^2}\;.
\end{eqnarray}
For the gauge-dependence analysis in high-temperature dimensional reduction, preserving the perturbativity of the 3D effective field theory requires the condition \(|\xi| \ll 1/g\)~\cite{Croon:2020cgk,Kripfganz:1995jx,Laine:1994bf,Garny:2012cg}.
The result here is indeed a consequence of the perturbative expansion in the gauge coupling $g$.
Within the $\lambda\sim g^3$ power-counting scheme, all terms with $\xi^{n\geq 1}$ and those of the form $\xi^{m>0}\log(\bar{\mu}/\sqrt{\xi})$ cancel exactly, leaving only the $\sqrt{\xi}$ term.


In this scenario, these coefficients related with the Nielsen identities are computed at the leading order. We do not need the explicit expression of $D$ and $\tilde{D}$ because the terms on the second line of Eq.~(\ref{Nielsen_2}) appear at $\mathcal{O}(g^2)$, and are hence suppressed relative to those on the first line at $\mathcal{O}(g)$, and at leading order we have $C \sim g$ when we consider the power counting $\lambda \sim g^3$. An explicit counting in powers of the weak gauge coupling $g$ in the identities of Eqs.~(\ref{Nielsen_1}) and (\ref{Nielsen_2}) yields the Nielsen identities for the effective potential:
\begin{equation}\label{likev}
	\xi\frac{\partial V_{\mathrm{NLO}}}{\partial \xi}=-C_{\mathrm{LO}} \frac{\partial V_{\mathrm{LO}}}{\partial \phi}\,,
\end{equation}
and the wave function:
\begin{equation}\label{like2}
	\xi \frac{\partial Z_{\mathrm{NLO}}}{\partial \xi}=-2\frac{\partial C_{\mathrm{LO}}}{\partial \phi}\;.
\end{equation}
In this situation, the first Nielsen identity establishes at the order of $\sim\mathcal{O}(g^3)$, and the second Nielsen identity holds at the order of $\sim\mathcal{O}(g)$.

At leading order in the power counting of $\lambda \sim g^3$ (where, one can set $m_{\chi^\pm} \to m_{c_W}, m_{\chi^0} \to m_{c_Z}$), we obtain
\begin{equation}
    C^{3d}_{\mathrm{LO}}=\frac{\lp 2 g+\sqrt{g^2+g^{'2}}\rp \sqrt{\xi}}{32 \pi}\,.
\end{equation}
With the gauge dependence of $V_{\mathrm{NLO}}$, we now find that the first Nielsen identity holds.
In 3d framework, the wave function of the Higgs field at the order of $g$ reads,
 \begin{equation}
    Z^{3d}_{\mathrm{NLO}}= -\frac{11 \left(\sqrt{g^2+g^{'2}}+2 g\right)}{48 \pi  \phi }\;,
\end{equation}
 therefore, we have
\begin{equation}
     \xi \frac{\partial Z_{\mathrm{NLO}}^{3d}}{\partial \xi}=-2\frac{\partial C_{\mathrm{LO}}^{3d}}{\partial \phi}=0\;,
\end{equation}
this justifies the second Nielsen identity.

\section{Power counting and perturbative expansion}
\label{sec:Power counting and perturbative expansion}
\subsection{Power counting regimes and scale hierarchies}
The perturbative organization of a thermal effective theory is determined not only by the loop order, but also by the relative scaling of its Wilson coefficients and background fields. In particular, loop order and perturbative order need not coincide when a loop-induced contribution participates in the formation of the leading-order potential barrier. It is therefore necessary to specify the relevant power-counting regime before separating the effective potential into leading- and  next leading-order contributions.

Following the classification of the three-dimensional effective theories given in Ref.~\cite{Camargo-Molina:2024sde}, we distinguish two power-counting regimes relevant for the present SMEFT setup. The first is a tree-level-barrier regime, denoted by TLB, in which the barrier is generated by the competition among the quadratic, quartic and sextic scalar interactions. The second is the RB6 regime, in which the transverse gauge-boson contribution and the dimension-six interaction both contribute to the barrier at leading order.

The four- and three-dimensional background fields are related by $\phi=\sqrt{T}\phi_3$.
The three dimensional gauge couplings are
\begin{equation}
    g_3^2\sim g^2T, \qquad g^{\prime 2}_3\sim g^2 T\;.
\end{equation}
Moreover, for the dimension-six operator normalized as $c_6=1/\Lambda^2$, its leading three-dimensional coefficient scales as
\begin{equation}
    c_{6,3}\sim c_6 T^2\sim \frac{T^2}{\Lambda^2}
\end{equation}

In the {\it TLB} regime which corresponds to the $\lambda\sim g^2$ scenario in this study,  the leading-order barrier is generated by the tree-level scalar potential (Eq.~\eqref{eq:Vsofttree} and Eq.~\eqref{eq:Vultrasofttree} for soft and ultrasoft scales). The three tree level terms must balance parameterically:
\begin{equation}\label{eq:TLB balance}
    m_3^2\phi_3^2 \sim \lambda_3\phi_3^4 \sim c_{6,3}\phi_3^6\;.
\end{equation}
The TLB regime considered here corresponds to $\lambda_3\sim g^2 T$. Taking the transition background to satisfy $\phi_3\sim \sqrt{T}$.
The balance condition in Eq.~\eqref{eq:TLB balance} therefore implies
\begin{equation}
    m_3^2\sim g^2 T^2, \qquad c_{6,3}\sim g^2\;.
\end{equation}
Therefore, the new-physics scale and the hard scale in the dimensional reduction scheme satisfying $\pi T/\Lambda\sim g$ in the $\lambda\sim g^2$ power-counting scenario.
The characteristic masses in the broken phase are
\begin{equation}
    m_W\sim g_3 \phi_3\sim gT, \qquad m_h^2\sim \frac{\partial^2 V_{\mathrm{tree}}}{\partial\phi^2}\sim g^2 T^2\;.
\end{equation}
Thus,
\begin{equation}
    m_h\sim m_W\sim gT\;.
\end{equation}
The scalar and gauge fields are consequently both associated with the soft scale of the DR procedure. This absence of a parametric separation between the Higgs and gauge-field masses will become important when discussing the derivative expansion of the nucleation action in Sec~\ref{sec:TLB} and~Sec.\ref{sec:The bubble nucleation action}.

We next discuss the {\it RB6} regime, which corresponds to $\lambda\sim g^3$ scenario under study. In this case, the gauge-bosons' contributions are promoted to leading order because it participates directly in the formation of the barrier. The leading-order potential is given by Eq.~\eqref{eq.lo}.
In this regime, the balance condition reads
\begin{equation}\label{eq:RB6 balance}
    m_3^2\phi_3^2 \sim g^3 \phi_3^3 \sim \lambda_3\phi_3^4 \sim c_{6,3}\phi_3^6\;.
\end{equation}
For $\phi_3\sim\sqrt{T}$, the cubic contribution scales as $g^3 \phi_3^3 \sim g^3T^3$. Requiring the remaining terms in Eq.~\eqref{eq:RB6 balance} to be of the same order yields
\begin{equation}
    m_3^2 \sim g^3 T^2,\qquad \lambda_3 \sim g^3T, \qquad c_{6,3}\sim g^3\;.
\end{equation}
Consequently, the heavy scale is $\pi T/\Lambda \sim g^{3/2}$ in the $\lambda\sim g^3$ power-counting scenario. Since
\begin{equation}
    m_h^2 \sim \frac{\partial^2 V_{\rm LO}}{\partial\phi^2}\sim g^3 T^2\;,
\end{equation}
the resulting scale hierarchy yields
\begin{equation}
    m_h \sim g^{3/2} T\ll m_W\sim g T\ll \pi T\;.
\end{equation}

\subsection{Strict perturbative expansion and critical temperature}

The Nielsen identity dictates the gauge dependence of the effective potential~\cite{Nielsen:1975fs}
\begin{equation}\label{funneq}
\xi\frac{\partial V_{\rm eff}(\phi,\xi)}{\partial \xi}=C(\phi,\xi)\frac{\partial V_{\rm eff}(\phi,\xi)}{\partial \phi},
\end{equation}
where $\xi$ is the gauge parameter, and $C(\phi,\xi)$ is independent of $V_{\rm eff}$ and can be expanded order by order, see Sec~\ref{sec:Nielsen Identity at Finite Temperature} for details. To analyze the gauge dependence of the effective potential, we now formulate the strict perturbative expansion in a unified manner. We introduce a general parameter $\epsilon$ and write~\cite{Lofgren:2023sep,Ekstedt:2020abj}.
\begin{equation}\label{veffnloop}
  V_{\rm eff}= V_0+\epsilon V_1+\epsilon^2 V_2+\cdots.
\end{equation}
Its meaning, as well as the identification of the coefficients $V_n$, is fixed by the power-counting regime under consideration. In particular, $\epsilon$ should not in general be identified with the number of loops.

In the TLB regime, where $\lambda\sim g^2$, the strict expansion is aligned with the loop expansion within the 3d EFT. In this case, the formal parameter can be identified with the conventional loop counting parameter $\hbar$,
\begin{equation}
    \epsilon :=\hbar, \quad V_0=V_{\mathrm{tree}},\quad V_1=V_{\mathrm{1loop}},\quad V_2=V_{\mathrm{2loop}}
\end{equation}
By contrast, in the RB6 regime, where $\lambda\sim g^3$, the loop order and strict perturbative order are not aligned. The expansion parameter should in this case be understand as a power counting parameter $g$, we have
\begin{equation}
    \epsilon :=g, \quad V_0=V_{\mathrm{LO}},\qquad V_1=V_{\mathrm{NLO}}\;.
\end{equation}
Hence, the the same formal expansion in $\epsilon$ describes two different expansion:a $\hbar$-expasion in the $\lambda\sim g^2$ scheme and a $g$-expansion in the $\lambda\sim g^3$ scheme.

At the VEV, we have
\begin{equation}
  \left.\frac{\partial}{\partial \phi}( V_0+\epsilon V_1+\epsilon^2 V_2+\cdots+\epsilon^n V_n)\right|_{v_n}=0,
\end{equation}
where $v_n=v_{(0)}+\epsilon v_{(1)}+\epsilon^2 v_{(2)}+\cdots+\epsilon^n v_{n}$ is the $n$ order VEV. Then the Nielsen identity implies
\begin{align}\label{neq1}
\xi &\left.\frac{\partial}{\partial\xi}( V_1+\epsilon V_2+\epsilon^2 V_2+\cdots)\right|_{v_n}\nn\\
=&\left.C(\phi,\xi)\frac{\partial}{\partial \phi}( V_0+\epsilon V_1+\epsilon^2 V_2+\cdots)\right|_{v_n}\nonumber\\
=&C(\phi,\xi)\biggr[\left.\frac{\partial}{\partial\phi}(V_0+\epsilon V_1+\epsilon^2 V_2+\cdots+\epsilon^n V_n)\right|_{v_n}\nonumber\\
&+ \left.\frac{\partial}{\partial\phi}(\epsilon^{n+1} V_{n+1}+\epsilon^{n+2} V_{n+2}+\cdots)\right|_{v_n}  \biggr]\nonumber\\
=&C(\phi,\xi)\left(\left.\frac{\partial}{\partial\phi}(\epsilon^{n+1} V_{n+1}+\epsilon^{n+2} V_{n+2}+\cdots)\right|_{v_n}\right),
\end{align}
and
\begin{align}\label{neq2}
\xi &\left.\frac{\partial}{\partial\xi}( V_1+\epsilon V_2+\epsilon^2 V_2+\cdots)\right|_{v_n}\nn\\
=&\xi\biggr[\left.\frac{\partial}{\partial\xi}(V_0+\epsilon V_1+\epsilon^2 V_2+\cdots+\epsilon^n V_n)\right|_{v_n}\nonumber\\
&+ \left.\frac{\partial}{\partial\xi}(\epsilon^{n+1} V_{n+1}+\epsilon^{n+2} V_{n+2}+\cdots)\right|_{v_n}  \biggr].
\end{align}
Combining the Eq.\eqref{neq1} and Eq.\eqref{neq2}, we obtain
\begin{align}\label{resneq}
\xi&\left.\frac{\partial}{\partial\xi}(V_0+\epsilon V_1+\epsilon^2 V_2+\cdots+\epsilon^n V_n)\right|_{v_n}\nonumber\\
=&\left.\left(C(\phi,\xi)\frac{\partial}{\partial\phi}-\xi\frac{\partial}{\partial\xi}\right)
(\epsilon^{n+1} V_{n+1}+\epsilon^{n+2} V_{n+2}+\cdots)\right|_{v_n}\nonumber\\
=&\mathcal{O}(\epsilon^{n+1}).
\end{align}
The Eq.\eqref{resneq} shows that, without considering higher-order effective potential, the residual gauge dependence of the effective potential is at order $\epsilon^{n+1}$. This implies that thermodynamic quantities possess a residual gauge dependence associated with higher-order correction~\cite{Biekotter:2025npc}.

At the critical temperature, the broken phase and symmetric phase have equal free energy density, which can be obtained by solving
\begin{equation}\label{tcfun1}
\Delta V_{\rm eff}=V_{\rm eff}(v_3)-V_{\rm eff}(0)=0,
\end{equation}
where the $V_{\rm eff}$ is the effective potential which defined in Eqs.~\eqref{v:first}-\eqref{v:end} . We solve the Eq.\eqref{tcfun1} using the $\epsilon-$expansion, which is a perturbative expansion carried out order by order around the leading-order minimum of the potential, thereby ensuring gauge invariance at each order~\cite{Farakos:1994xh,Gould:2023ovu,Biekotter:2025npc,Croon:2020cgk,Patel:2011th}. Since the effective potential does not depend explicitly on temperature $T$, we solve the Eq.\eqref{tcfun1} by calculate the free energy of effective potential $\Delta V_{\rm eff}$ to carry out the $\epsilon-$expansion. In the TLB scenario, the effective potential is retained through $\mathcal{O}(\epsilon^2)$, whereas in the RB6 scenario it is computed through $\mathcal{O}(\epsilon)$. The free energy at $\epsilon^2$ order read as
\begin{equation}\label{veffmin}
\Delta V_{\rm eff}=\Delta V_{0}(v^2)+\epsilon \Delta V_{1}(v^2)+\epsilon^2 \Delta V_{2}(v^2)+\mathcal{O}(\epsilon^3),
\end{equation}
and the definition of $v$ is
\begin{equation}\label{vevhbar}
 v^2=v_{(0)}^2+\epsilon v_{(1)}^2+\epsilon^2 v_{(2)}^2+\mathcal{O}(\epsilon^3).
\end{equation}
Insert this in $\Delta V_{\rm eff}^\prime=0$, we can obtained the following equations~\cite{Farakos:1994xh,Gould:2023ovu,Biekotter:2025npc,Croon:2020cgk}:
\begin{equation}\label{vevhbar2}
\begin{aligned}
&\Delta V_0^{\prime}(v_{(0)}^2)=0\;,\quad
v_{(1)}^2=-\frac{\Delta V_{\rm 1}^\prime}{\Delta V_{\rm 0}^{\prime\prime}},\\
&v_{(2)}^2=-\frac{\Delta V_{\rm 2}^\prime}{\Delta V_{\rm 0}^{\prime\prime}}+\frac{\Delta V_{\rm 1}^\prime \Delta V_{\rm 1}^{\prime\prime}}{(\Delta V_{\rm 0}^{\prime\prime})^2}-\frac{(\Delta V_{\rm 1}^\prime)^2 \Delta V_{\rm 0}^{\prime\prime\prime}}{2(\Delta V_{\rm 0}^{\prime\prime})^3}.
\end{aligned}
\end{equation}
The prime in this part means derivative with respect to $\phi^2$, and the functions $\Delta V_i$ are evaluated at $v_{(0)}^2$. The parameters $m, \lambda, c_6,g$ in this part denoted the 3d parameters $m_3, \lambda_3, c_{6,3}, g_3$($\overline{m}_3, \overline{\lambda}_3, \overline{c}_{6,3}, \overline{g}_3$) in soft(ultrasoft) scale, See Appendix~\ref{parameter matching}. With these definitions, the free energy Eq.\eqref{veffmin} can be rewritten as in Refs.~\cite{Farakos:1994xh,Gould:2023ovu,Biekotter:2025npc,Croon:2020cgk}
\begin{equation}\label{veffmin2}
\begin{aligned}
\Delta &V_{\rm eff}(v^2)\\
=&\Delta V_{\rm 0}(v_{(0)}^2+\epsilon v_{(1)}^2+\epsilon^2 v_{(2)}^2)+\epsilon \Delta V_{\rm 1}(v_{(0)}^2+\epsilon v_{(1)}^2)+\epsilon^2 \Delta V_{\rm 2}(v_{(0)}^2)\\
=&\Delta V_{\rm 0}(v_{(0)}^2)+\epsilon\biggr[\Delta V_{\rm 1}(v_{(0)}^2)+v_{(1)}^2\Delta V_{\rm 0}^\prime(v_{(0)}^2)\biggr]+\epsilon^2\biggr[\Delta V_{\rm 2}(v_{(0)}^2)\\
&+v_{(2)}^2 V_{\rm 0}^\prime(v_{(0)}^2)+v_{(1)}^2\Delta V_{\rm 1}^\prime(v_{(0)}^2)+\frac{1}{2}v_{(1)}^2\Delta V_{\rm 0}^{\prime\prime}(v_{(0)}^2)\biggr]\;\\
=&\Delta V_{\rm 0}(v_{(0)}^2)+\epsilon \Delta V_1(v_{(0)}^2)+\epsilon^2\biggr[\Delta V_2(v_{(0)}^2)-\frac{\Delta V_1^{\prime 2}(v_{(0)}^2)}{2 \Delta V_0^{\prime\prime}(v_{(0)}^2)} \biggl]\;.
\end{aligned}
\end{equation}
Then the critical temperature $T_c$ can be solved by $\Delta V_{\rm eff}(v^2,T_c)=0$. This method is the same as the method described in Ref.~\cite{Gould:2023ovu}, which calculates the critical temperature $T_c$ by solving the equation $\Delta p(T_c)=-T\Delta V_{\rm eff}(v^2,T_c) =0$.
\begin{figure}[!htp]
    \centering
    \includegraphics[width=0.4\linewidth]{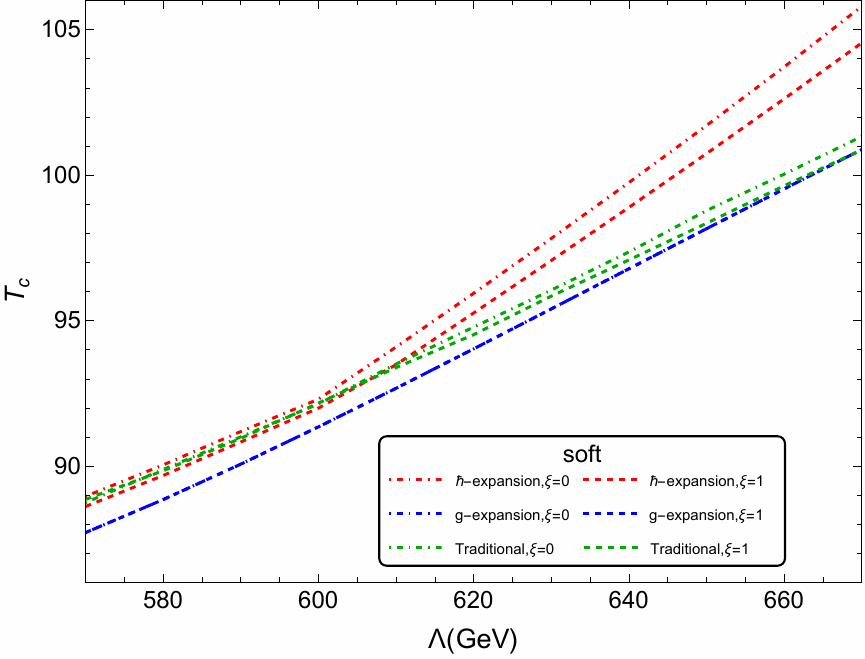}
    \quad
    \includegraphics[width=0.4\linewidth]{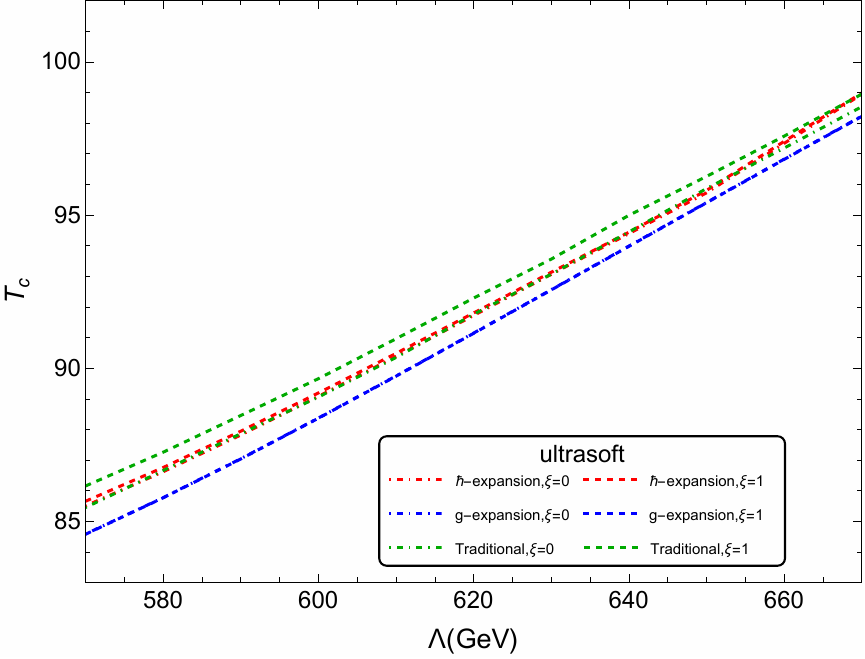}
    \caption{The critical temperature $T_c$ as a function of the cutoff scale $\Lambda$.The left (right) panels show the results obtained in the soft (ultrasoft) effective theory. For each panel, the predictions from diffenent expansions are compared with those from the traditional approach for the gauge-fixing parameters, $\xi=0,1$.}
    \label{fig:TC}
\end{figure}

As shown in Figure~\ref{fig:TC}, we compare the critical temperatures obtained from the formal $\hbar$-expansion, the $g$-expansion, and the traditional direct method. In both the soft and ultrasoft descriptions, $T_c$ increases monotonically with the new physical scale $\Lambda$, indicating that the transition occurs at a higher temperature as the dimension-six interaction becomes weaker. At the soft scale, the three computational methods yield similar results in the region of small $\Lambda$, but as $\Lambda$ increases, their predictions gradually diverge. In particular, the $\hbar$-expansion results exhibit significant gauge dependence in the large $\Lambda$ region and yield higher values of $T_c$. This behavior should be interpreted as indicating that the $\hbar$-expansion still suffers from gauge uncertainty. In contrast, the curves for the gauge parameters in the $g$-expansion coincide perfectly, consistent with the expectation that the physical quantities are independent of the gauge order. At the ultrasoft scale, the numerical differences among the three methods are significantly reduced; the curves exhibit an approximately common linear growth trend and gradually converge in the large $\Lambda$ region. Therefore, the $g$-expansion yields gauge-independent results, whereas the $\hbar$-expansion does not.

\begin{figure}[!htp]
  \centering
  \includegraphics[width=0.4\textwidth]{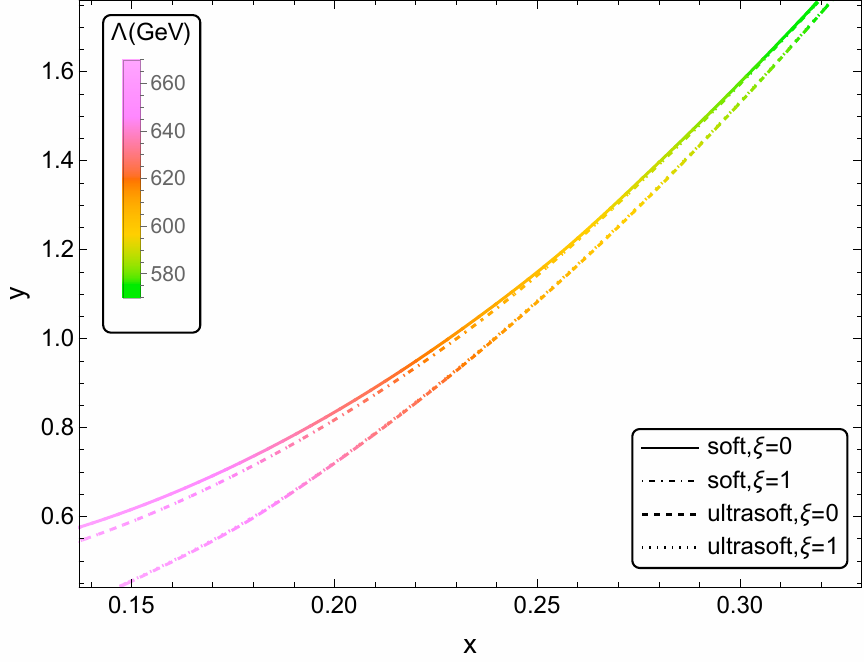}
  \quad
  \includegraphics[width=0.4\textwidth]{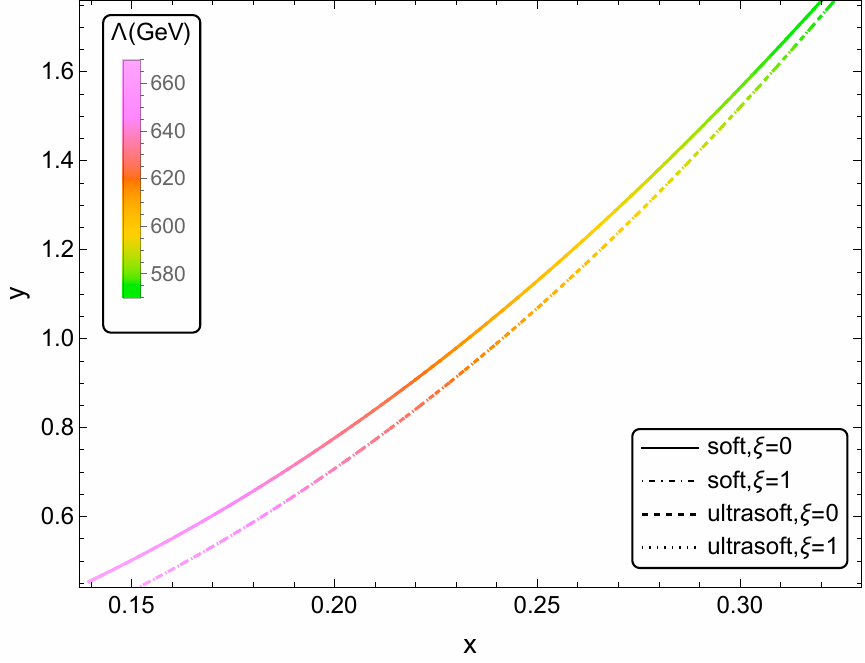}
  \caption{The parameters $x$ and $y$ as function of $\Lambda$ in soft and ultrasoft scale at $\hbar-$expansion (Left) and at $g$-expamsion (Right). The color denote the values of $\Lambda$, and the lines denote the different scale and gauge choice.}\label{figxyLambda}
\end{figure}

The Figure~\ref{figxyLambda} shows the relation between the dimensionless dynamical parameters $x=-\lambda/g^2$ and $y=-m^2/g^4$ at critical temperature $T_c$ in the range of $570~\text{GeV}\leq\Lambda\leq 670~\text{GeV}$. The parameters $x$ and $y$ are commonly used to investigate the phase structure of theories in lattice~\cite{Chala:2025xlk,Kajantie:1996mn,Chala:2025xlk,Gould:2022ran,Ekstedt:2022zro}. As shown in this figure, both $x$ and $y$ are decreases as $\Lambda$ increases. In the formal $\hbar$-expansion shown in the left panel, there is a visible separation between different scale descriptions and different gauge choices, particularly in the region of large $\Lambda$ corresponding to small $x$. This separation reflects the fact that if loop expansions are performed around tree-level minima and thermodynamic quantities are calculated based on them, the results will be gauge-dependent. In contrast, the $g$-expansion shown in the right panel yields strictly gauge-independent results, while the differences between the soft and ultrasoft descriptions are significantly reduced. This result indicates that the g-expansion scheme ensures $T_c$ as well as its matching parameters $x$ and $y$ gauge invariant.

\section{The bubble nucleation action}
\label{sec:The bubble nucleation action}
The Universe first lives in the “symmetric" phase, and as the Universe cools down, the “symmetric” and “broken" phases have the same free energy at the critical temperature.
In particular, when the Universe cools further, the vacuum bubbles of the true phase nucleated in the symmetric phase.
The bubble nucleation rate, $\Gamma$, has the semiclassical approximation
\be
\Gamma=A e^{-B}\; .
\ee
Here, the exponent $B=S_3$, with $S_3$ being the three-dimensional Euclidean effective action evaluated at the "bounce" solution that solves the classical Euclidean field equations. A is an expression involving functional determinants that is generally equal to a numerical factor of order unity times a dimensionful quantity determined by the characteristic mass scales of the theory.
 The nucleation temperature $T_n$ is obtained when the bubble nucleation rate $\Gamma=A \exp[-S_c]$ is equal to Hubble parameter $\Gamma\sim H$, i.e., $S_c\approx 140$~\cite{Linde:1981zj}.

At 3d EFT, the Euclidean effective action can be expressed as:
\be\label{effective Euclidean action}
S^{\mathrm{eff}}=\int d^3x\left[ V^{\mathrm{eff}}+\frac{1}{2} Z(\phi) (\partial_\mu \phi)^2+\mathcal{O}(\partial^4)\right]\;,
\ee
with the last represent terms containing higher order derivatives that do not enter the calculation we work. From the 3d Euclidean action, and consider the $d Z/d\phi$ factor at vacuum expectation value(VEV) is close to zero, the bounce function can be recast as~\cite{Qin:2024dfp}:
\begin{equation}\label{bouncefunction}
\frac{d^2 \phi_b}{d\rho^2}+\frac{2}{\rho}\frac{d\phi_b}{d\rho}=\frac{1}{Z}\frac{d V^{\mathrm{eff}}}{d\phi},
\end{equation}
with the boundary conditions:
\begin{equation}
\phi(\rho\rightarrow \infty)=0,\quad \left.\frac{d\phi}{d\rho}\right|_{\rho=0}=0\;.
\end{equation}
Numerically, we utilize ``FindBounce'' to obtain the field configuration of the bounce solution at the nucleation temperature $T_n$~\cite{Guada:2020xnz}.

 \begin{figure}[!htp]
     \centering
    \includegraphics[width=0.4\linewidth]{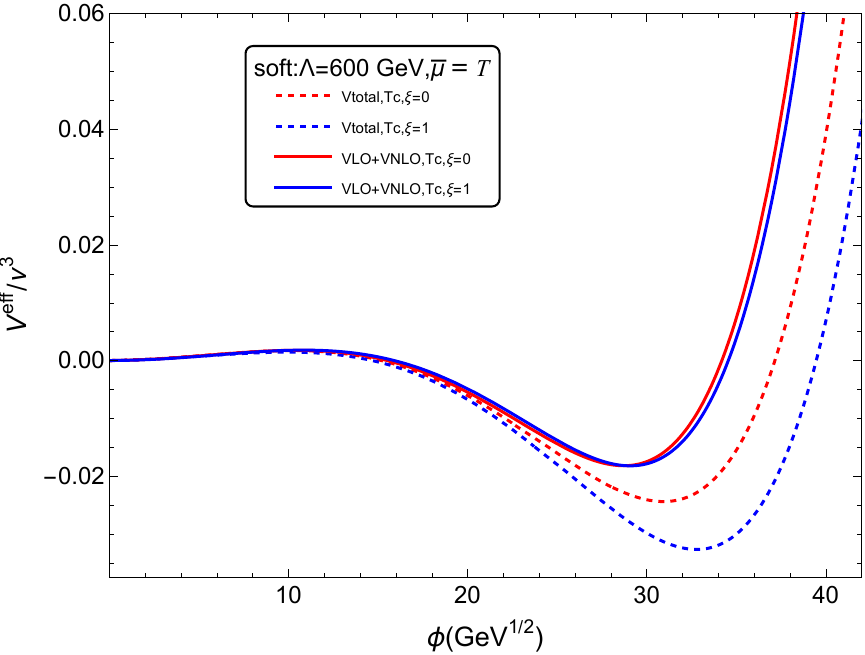}
     \quad
     \includegraphics[width=0.4\linewidth]{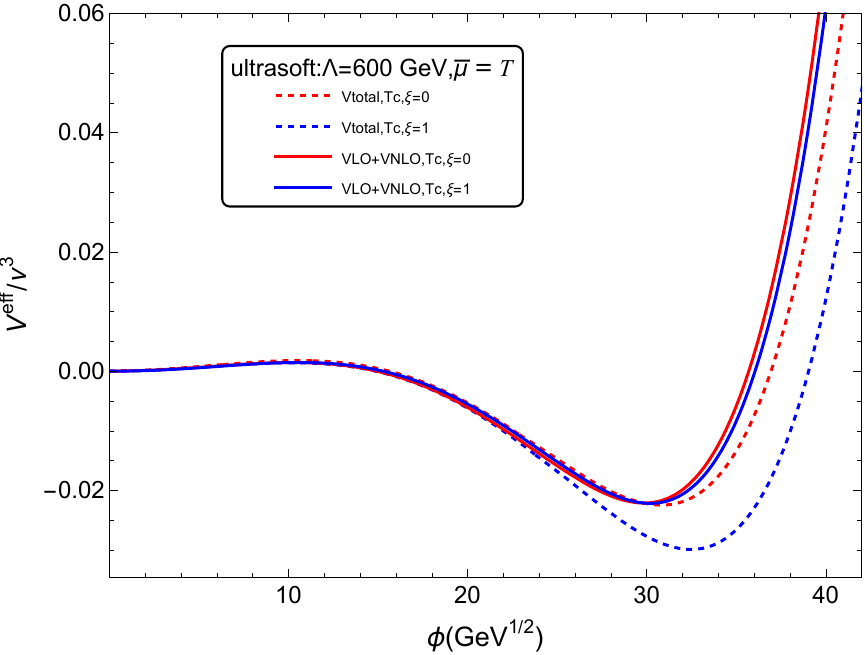}
     \caption{The effective potential at $\Lambda = 600$ GeV for $T_n$ with different gauge parameters in the soft scale (left) and ultrasoft scale (right).}
     \label{fig:softv&ultrasoftV}
 \end{figure}

In Figure~\ref{fig:softv&ultrasoftV}, we present the effective potential shapes for $\Lambda=600$ GeV at the nucleation temperature $T_n$ (when vacuum bubbles start to nucleate). We find that, in comparison with the traditional calculation based on the $V_{\rm total}$, the gauge dependences of the effective potentials (calculated with $V_{\rm LO}+V_{\rm NLO}$) has been greatly reduced in the gauge-independent scheme at both soft and ultrasoft scales. In this work, given that the renormalization scale $\bar{\mu}$ dependence of the results is tiny~\cite{Croon:2020cgk,Qin:2024idc}, all numerical results presented are computed with the choice $\bar{\mu} = T$.

Meanwhile, one can expand the wave function of the background field as
\be
Z=1+Z(g)+\mathcal{O}(g^2)+\cdots\;,
\ee
and the $S^{\rm eff}(\phi_b)$ as
\be\label{b00}
B_0 = \int d^3 x \left[ V_{\rm LO}(\phi_b)+\frac{1}{2}(\partial_\mu \phi_b)^2 \right]\;,
\ee
while
\be\label{b11}
B_1 = \int d^3 x \left[ V_{\rm NLO}(\phi_b)+\frac{1}{2}Z_{\rm NLO}(\phi_b)(\partial_\mu \phi_b)^2 \right]\;,
\ee
The first step in this approach is to use the leading approximation of the effective action to determine the bounce solution $\phi_b(x)$ through the equation
\be\label{eq:bounce solution}
\Box \phi_b =\frac{\partial V_{\rm LO}}{\partial \phi}\;.
\ee
In this perturbative method, the desired nucleation rate can be given by
\be
\Gamma=A' e^{-(B_0+B_1)}\;,
\ee
Here, the prefactor $A'$ encodes high-order corrections to the effective action as well as functional determinant from quantum fields and fluctuation effects at finite temperature~\cite{Athron:2023xlk}. Like any physically measurable quantity, the nucleation rate should be gauge independent. Since the leading terms in the effective potential are gauge independent for the cases of $\lambda\sim g^2$ and $\lambda\sim g^3$, there is no difficulty in this regard with respect to either $B_0$ or the bounce solution itself. However, both of the functions that enter in $B_1$ are known to depend on gauge. Our goal is to show, if these combine would give a gauge-independent contribution to the nucleation rate. Although we do not explicitly examine the prefactor $A'$, we expect that our methods could be extended---albeit with considerably more technical complication---to show that it too is independent of gauge.


For scenario with the power-counting of $\lambda\sim g^2$, we consider the $B_0$ and $B_1$ at the order of weak gauge coupling $\sim g^2$ and $\sim g^3$ respectively. Then, the $B_1$ recast the form of,
\begin{equation}
	\begin{aligned}
\xi \frac{\partial }{\partial \xi}B_1&=\xi \frac{\partial }{\partial \xi}\int \diff^d x \left[ V_{g^3}(\phi_b)+\frac{1}{2}Z_{g}(\phi_b)(\partial_\mu \phi_b)^2 \right]\,\\
&=\int \diff^d x \left[-C_g\frac{\partial V_{g^2}}{\partial \phi}-\frac{\partial C_g}{\partial \phi}(\partial_\mu \phi_b)^2-\lp D\frac{\partial V_{g^2}}{\partial \phi}+\tilde{D}\frac{\partial^2 V_{g^2} }{\partial \phi^2}\rp (\partial_\mu \phi_b)^2 \right]\\
&=-\int \diff^d x \left[\lp D\frac{\partial V_{g^2}}{\partial \phi}+\tilde{D}\frac{\partial^2 V_{g^2} }{\partial \phi^2}\rp (\partial_\mu \phi_b)^2 \right]\\
&=-\int \diff^d x \left[\lp D \partial^\mu V_{g^2}+\tilde{D}\partial^\mu(\frac{\partial V_{g^2}}{\partial \phi}) \rp(\partial_\mu \phi_b)\right]\\
&=\int \diff^d x \left[\lp D V_{g^2}+\tilde{D} \frac{\partial V_{g^2}}{\partial \phi} \rp \Box \phi_b \right]\\
&=\int \diff^d x  \left[D V_{g^2}\frac{\partial V_{g^2}}{\partial \phi}+\tilde{D}\lp \frac{\partial V_{g^2}}{\partial \phi} \rp^2  \right]\;.
	\end{aligned}
\end{equation}
Here, in the second equality we utilize the first Nielsen identity given by Eq.~(\ref{Nielsen_1a}), and we assume Eq.~(\ref{Nielsen_2_With D}) holds based on the results of Figure~\ref{fig:deviation analysis}. And, in the last equality we consider the bounce solution can be obtained at the order of $g^2$. From above derivation, we find that the $B_1$ indeed depends on the gauge choice since the $D,\tilde{D}$ coefficients rely on the gauge parameter.

The accumulation of infinitely many higher-order terms in the derivative expansion renders the expansion invalid~\cite{Croon:2020cgk}. In general, a well-defined derivative expansion requires a separation of scales, leading to an expansion in powers of $P/M$ with formal  and M being a mass scale $P\ll M$~\cite{Hirvonen:2021zej}, The calculation contributing to action involves propagating vector bosons, so that $M=m_{W,Z} \sim g \sigma$. However, In the scheme with $\lambda\sim g^2$,
\begin{equation}
    P^2 \sim \frac{1}{\phi}\frac{\partial V_{g^2}}{\partial \phi} \sim g^2 \sigma^2\;, \quad \frac{P^2}{M^2} \sim 1\;.
\end{equation}
where $\sigma$ is a characteristic value of $\phi_b(\rho =0)\sim \sigma$.
We can determine the expected sizes of $B_0$ and $B_1$ in these two schemes by using power counting and the characteristic size of the critical bubble, with radius $R \sim m_h^{-1}$.

For the power-counting of $\lambda \sim g^2$, we have
\begin{align}
    R &\sim m_h^{-1} \sim g^{-1}T^{-1}\nn\;,\\
    \int\diff^3 x &\sim g^{-3}T^{-3} \Longrightarrow B_0\sim g^{-1}, B_1\sim g^{0}
\end{align}
Therefore, the derivative expansion breaks down in this scheme, the $\lambda \sim g^2$ is not a good choice.
In the scheme with $\lambda\sim g^3$,
\begin{equation}
    P^2 \sim \frac{1}{\phi}\frac{\partial V_{\rm{LO}}}{\partial \phi} \sim g^3 \sigma^2\;, \quad \frac{P^2}{M^2} \sim g\;,
\end{equation}
and
\begin{align}
    R &\sim m_h^{-1} \sim g^{-3/2}T^{-1}\nn\;,\\
    \int\diff^3 x &\sim g^{-9/2}T^{-3} \Longrightarrow B_0\sim g^{-3/2}, B_1\sim g^{-1/2}
\end{align}
The leading order is evidently gauge invariance , while the gauge invariance of $B_1$ is not immediately obvious since  $V_{\rm{NLO}}$ in Eq.~\eqref{eq.nlo} explicitly depends on $\xi$. However the gauge-invariance of $B_1$ can be proved by using the Nieslen identity:
\begin{align}\label{eq:B1xi}
\xi \frac{\partial }{\partial \xi}B_1&=\xi \frac{\partial }{\partial \xi}\int \diff^d x \left[ V_{\rm NLO}(\phi_b)+\frac{1}{2}Z_{\rm NLO}(\phi_b)(\partial_\mu \phi_b)^2 \right]\,\nn\\
&=\int \diff^d x \left[-C\frac{\partial V_{\rm LO}}{\partial \phi}-\frac{\partial C}{\partial \phi}(\partial_\mu \phi_b)^2\right]\,\nn\\
&=-\int \diff^d x \left[C\frac{\partial V_{\rm LO}}{\partial \phi}+\partial^\mu C (\partial_\mu \phi_b)\right]\,\nn\\
&=-\int \diff^d x \left[C (\frac{\partial V_{\rm LO}}{\partial \phi}-\Box \phi_b)\right]\,\nn\\
&=0\;.
\end{align}

Here, we use the relations given by Eq.~(\ref{likev}) and Eq.~(\ref{like2}) in the second equality, and in last step we utilize the Eq.~(\ref{eq:bounce solution}).

\begin{figure}[!htp]
    \centering
    \includegraphics[width=0.8\linewidth]{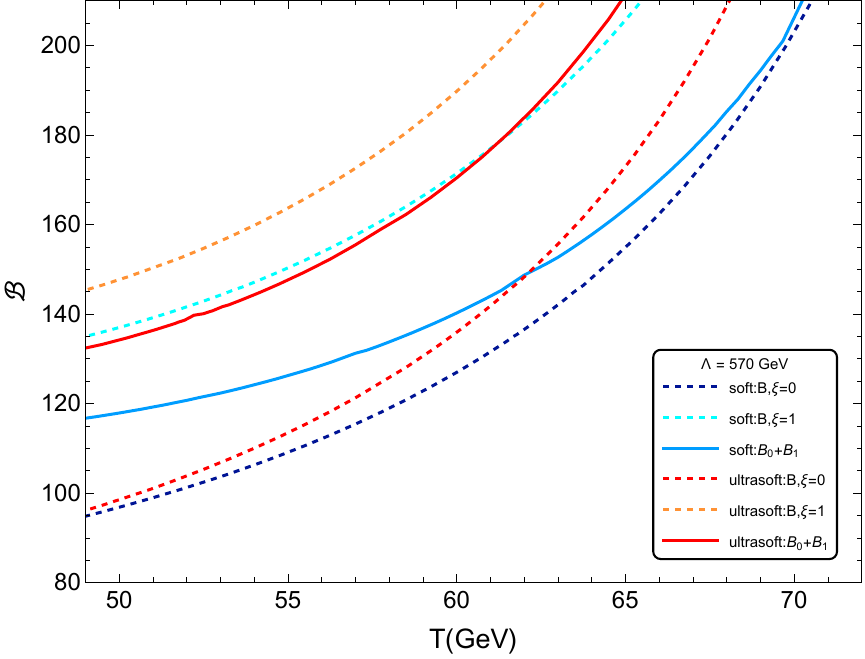}
    \caption{The effective action $\mathcal{B}$ as functions of temperature T at the soft scale and ultra-soft scale  with $\Lambda$ =570 GeV, the scenarios of $\xi=1$ and $\xi=0$ correspond to 't Hooft-Feynman gauge and the Landau gauge. The $B_0+B_1$ denotes the results obtained through the gauge-invariant method, and the $B$ indicating the results of the traditional method. }
    \label{fig:s3}
\end{figure}

As shown in Figure~\ref{fig:s3}, the effective action increases with the rising temperature.
The gauge-invariant method yields results with negligible residual gauge dependence; as is evident from the figures, predictions for different values of the gauge-fixing parameter are visually indistinguishable. This result confirms the theoretical prediction obtained in Eq.~\eqref{eq:B1xi}. The gauge-invariant result of the bounce action calculated at the ultra-soft scale is slightly larger than that of the soft scale.

\begin{figure}[!htp]
  \centering
  \includegraphics[width=0.8\textwidth]{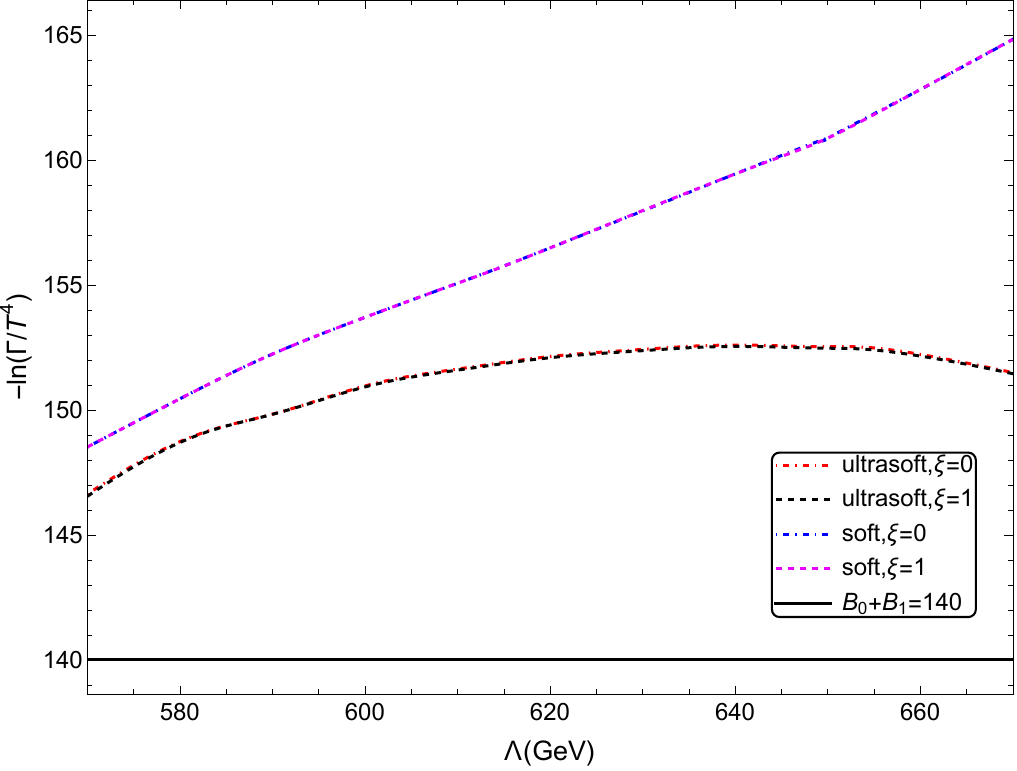}
  \caption{The nucleation rate $\Gamma$ as function of $\Lambda$ with $\xi=0,1$ at soft and ultrasoft scale at temperature $T_n$. The relation between the $-\ln(\Gamma/T^4)$ and $\Lambda$ at nucleation temperature $T_n$. The Dashed and Dotted lines denotes the result of the code ``BubbleDet'' by using $V_{LO}$, and the solid line denote the result of $\Gamma=T^4e^{-(B_0+B_1)}$.}\label{figgamma}
\end{figure}

Since the effective action is spilt to LO and NLO part, the nucleation rate $\Gamma$ can be rewritten as
\begin{equation}\label{gamma1}
\Gamma=A e^{-B},
\end{equation}
The factor A in the finite temperature theory includes the dynamic and statistical parts~\cite{Gould:2021ccf,Hirvonen:2022jba,Ekstedt:2023sqc,Kierkla:2025qyz}:
\begin{equation}\label{factorA1}
A=A_{\text{dyn}}\times A_{\text{stat}},
\end{equation}
with
\begin{align}\label{factorA2}
  &A_{\text{dyn}}=\frac{1}{2\pi}\left(\sqrt{|\lambda_-|+\frac{1}{4}\eta^2}-\frac{\eta}{2}\right),\\
  &A_{\text{stat}}=\left(\frac{B_0}{2\pi}\right)^{3/2}
  \left|\frac{\text{det}^{'}(-\nabla^2+V_{\rm LO}^{''}(\phi_b))}{\text{det}(-\nabla^2+V_{\rm LO}^{''}(\phi_F))}\right|^{-1/2}\;,
\end{align}
where $\phi_F$ is the value of false vacuum, and $\phi_b$ is the bounce solution which can be solving by bounce function Eq.~(\ref{eq:bounce solution}). We calculate the nucleation rate $\Gamma$ and the prefactor $A$ utilizing the public code ``BubbleDet''~\cite{Ekstedt:2023sqc}. The Figure~\ref{figgamma} shows the nucleation rate calculated by the code ``BubbleDet'' and the one directly obtained through $\Gamma=T^4e^{-(B_0+B_1)}$ at temperature $T_n$. The nucleation rate obtained through $B_0+B_1$ yields horizontal lines since the calculation of temperature $T_n$ requires $B_0 + B_1 = 140$. The difference between the ``BubbleDet" and ``$B_0+B_1$" is significant.

\begin{figure}[!htp]
    \centering
    \includegraphics[width=0.8\linewidth]{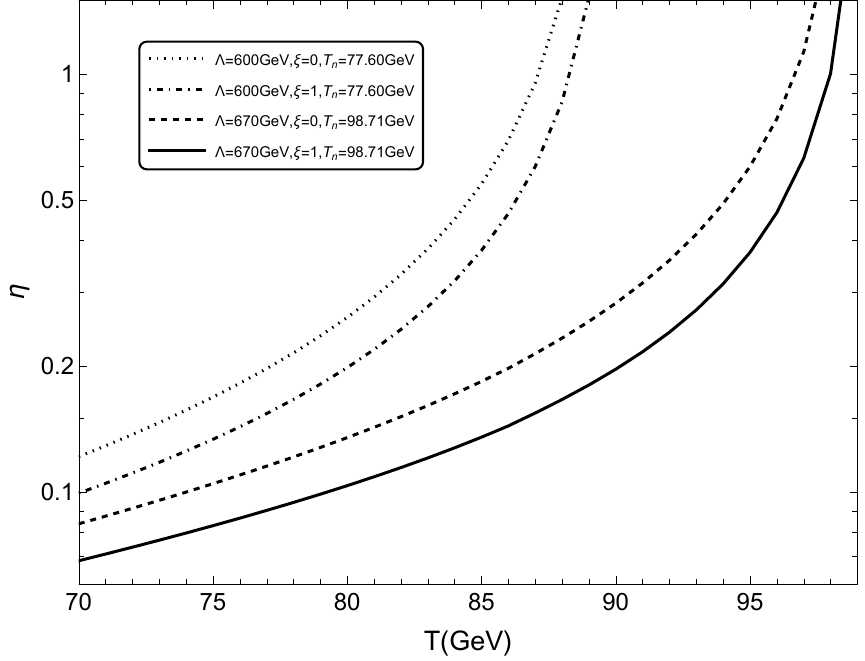}
    \caption{The difference ratio $\eta$ as a function of temperature $T$ with $\Lambda=600,670$ GeV and $\xi=0,1$ around the nucleation temperature $T_n$ below the critical temperature $T_c$. Here, $T_c \approx 92$ GeV for $\Lambda=670$ GeV and $T_c \approx 101$ GeV for $\Lambda=670$ GeV. This figure highlights the influence of different $\Lambda$ and $\xi$ on the behavior of $\eta$ near $T_n$.}
    \label{figsoftA0inf}
\end{figure}

Figure~\ref{figsoftA0inf} show the impact of contributions from the temporal scalars and their sensitivity to temperature at the soft scale. We define difference ratio $\eta = (V_1 - V_2)/V_2$, where $V_1$ is the minimum of the full effective potential, and $V_2$ is the minimum of the potential excluding contributions from the temporal scalars.
On one hand, as shown in Figure~\ref{figsoftA0inf}, for smaller values of $\Lambda$ (e.g., $\Lambda = 600$ GeV), the contribution of temporal scalars is negligible near the corresponding, relatively lower nucleation temperature $T_n$. As $\Lambda$ increases (e.g.$\Lambda = 670$ GeV), the associated $T_n$ also increases, and the contribution from temporal scalars becomes significant.
On the other hand, the tree-level contribution dominates at $\Lambda = 600$ GeV. With increasing $\Lambda$, the tree-level contribution diminishes while the loop-level contribution grows, which further amplifies the influence of the temporal scalar contributions.



\begin{figure}[!htp]
     \centering
     \includegraphics[width=0.4\linewidth]{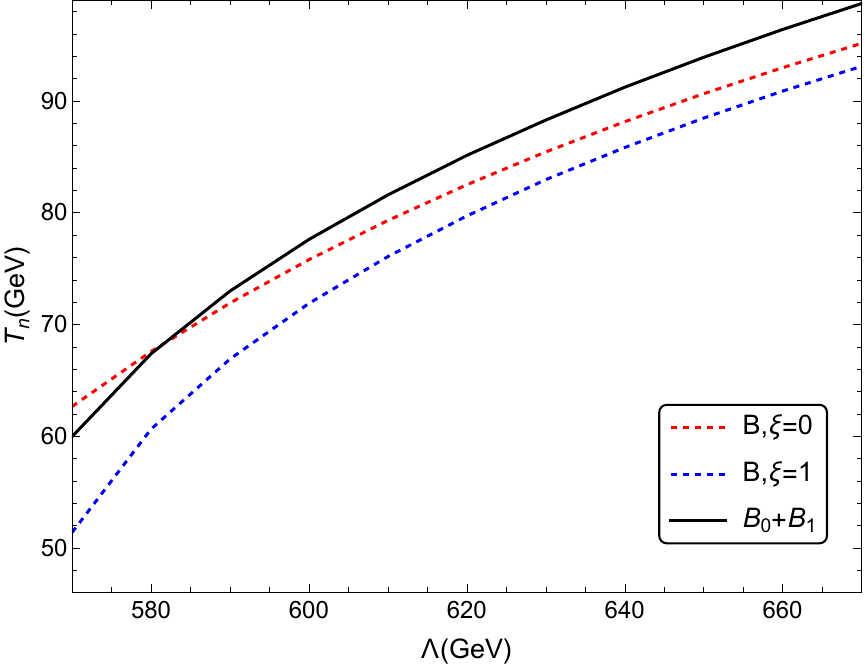}
     \quad
     \includegraphics[width=0.4\linewidth]{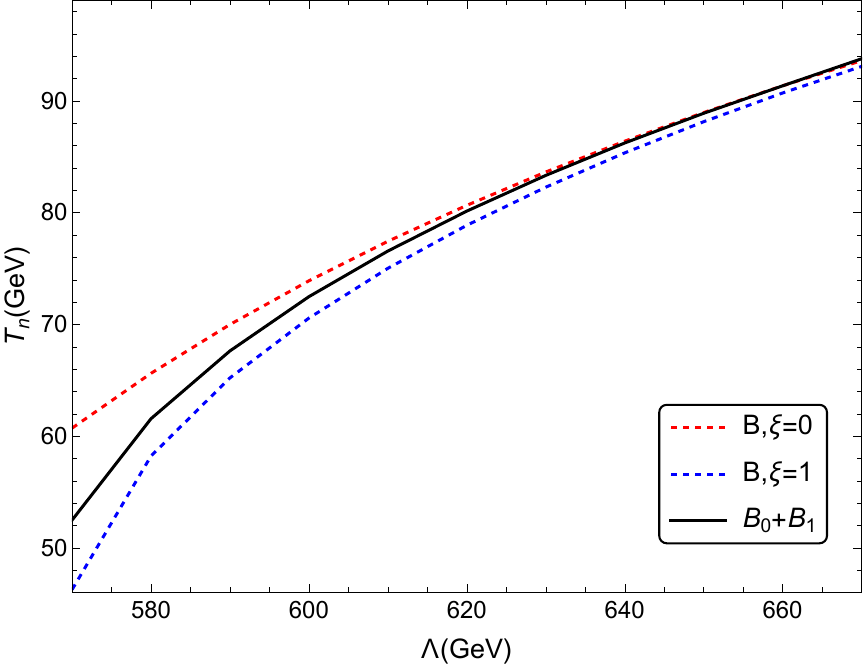}
     \quad
     \includegraphics[width=0.4\linewidth]{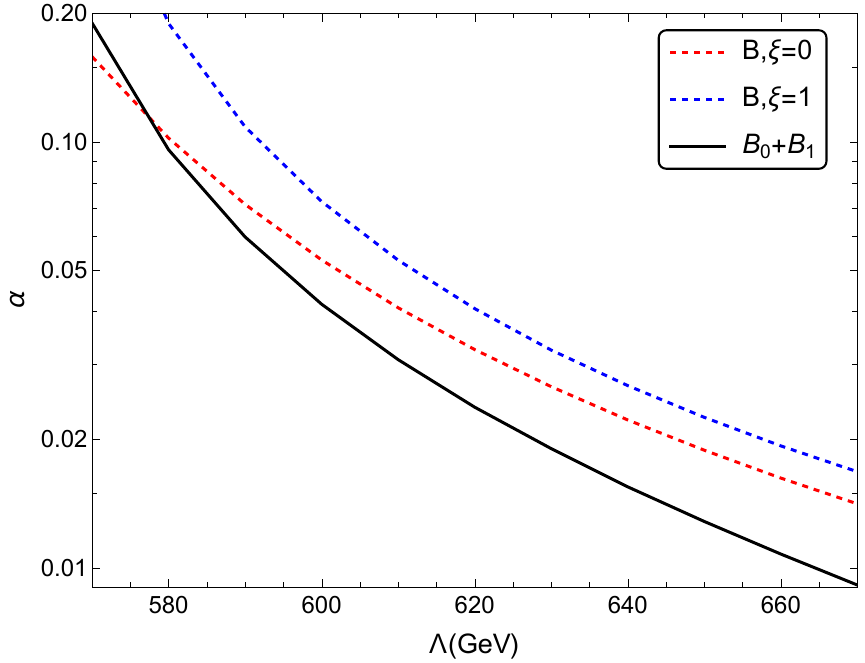}
     \quad
     \includegraphics[width=0.4\linewidth]{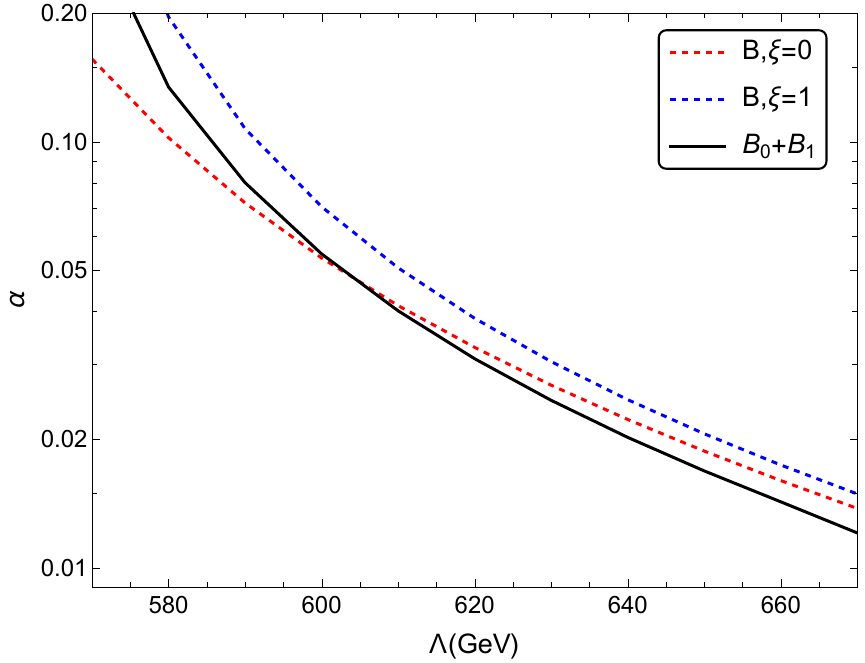}
     \quad
     \includegraphics[width=0.4\linewidth]{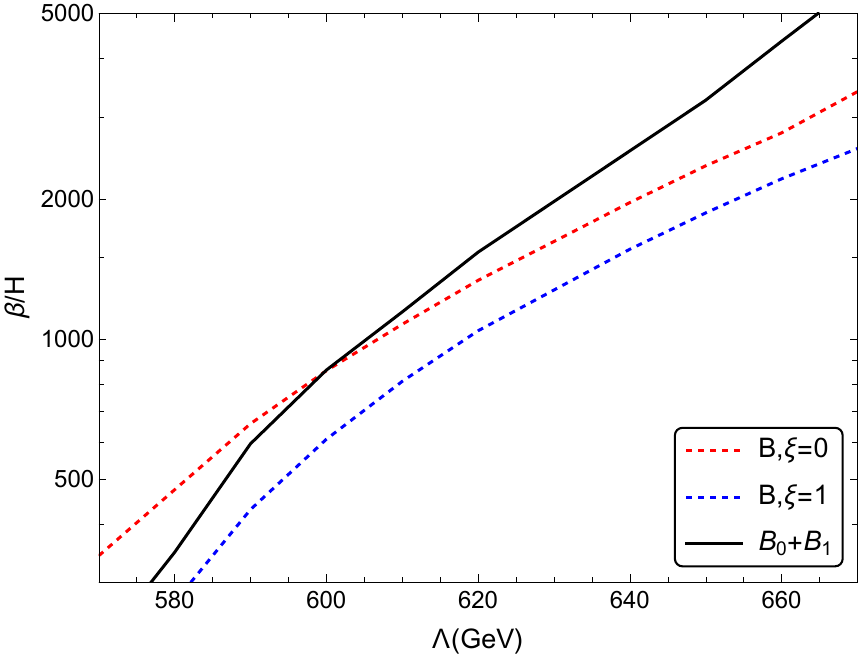}
     \quad
     \includegraphics[width=0.4\linewidth]{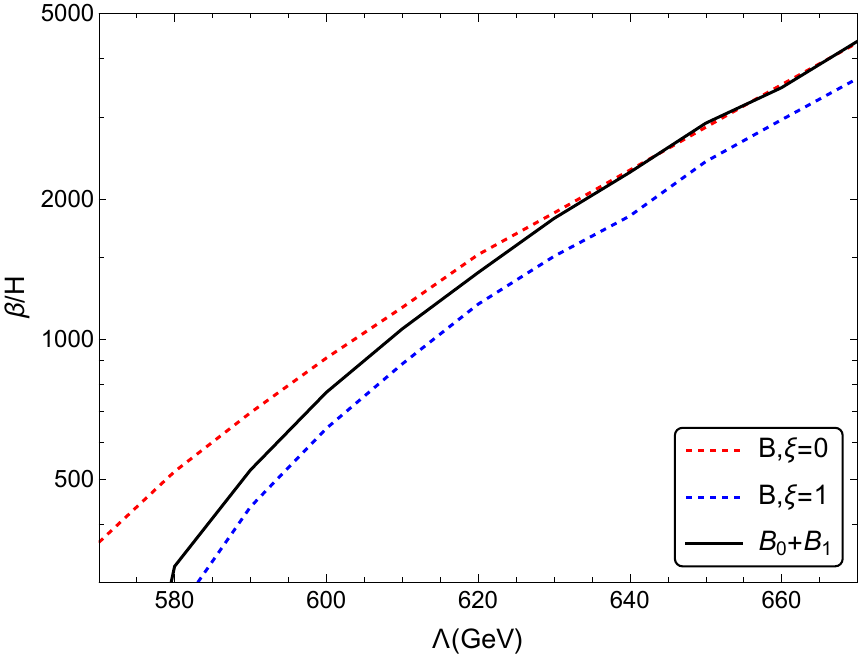}
     \caption{The nucleation temperature $T_n$,the phase parameter $\alpha$ and $\beta /H$ as functions of $\Lambda$ at the soft scale (left) and ultrasoft scale (right). Solid lines ($B_0+B_1$) represent the results obtained by gauge-invariant method (with the results of $\xi=0$ and $\xi=1$ indistinguishable), and dashed lines ($B$) represent the conventional method predictions.}
     \label{fig:Tn&alpha&betah}
 \end{figure}

\section{First-order phase transition parameters}
\label{sec:First-order phase transition parameters}

The trace anomaly ($\alpha$) usually determines the amplitude of the generated GW. The PT temperature and the duration determine the peak frequency of the produced GW from PT~\cite{Huber:2008hg,Caprini:2015zlo,Caprini:2009yp}, the inverse PT duration is defined as:
$\beta/H_n=T_n(dS_c/dT)|_{T_n}$.
After apply the relation between $4d$ and $3d$ potential $V_{4d}\approx T V_{\mathrm{eff}}$, we have $\alpha=T (\Delta\rho/\rho_{rad})$ with
\begin{equation}
\Delta\rho=-\frac{3}{4}\Delta V_{\mathrm{eff}}(\phi_n,T_n)+\frac{1}{4}\left. T_n \frac{d \Delta V_{\mathrm{eff}}(\phi_n,T)}{d T}\right|_{T=T_n}\;,
\end{equation}
where
$
\Delta V_{\mathrm{eff}}(\phi,T)=V_{\mathrm{eff}}(\phi,T)-V_{\mathrm{eff}}(0,T)
$ and $\rho_{rad}=\pi^2g_* T_n^4/30$, $g_*=106.75$ is the effective number of relativistic degrees of freedom~\cite{Croon:2020cgk}.

The Figure~\ref{fig:Tn&alpha&betah} presents the gauge dependence of PT paramater in the conventional method and the gauge-invariant formalism at the soft scale and ultrasoft scale. At two different scales, as $\Lambda$ increases, the temperature $T_n$ and $\beta /H$ increase, while $\alpha$ decreases with the increase of $\Lambda$.
At the soft scale, the gauge-invariant treatment yields quantitatively noticeable shifts in all three key phase transition observables relative to the conventional approach: the nucleation temperature \(T_n\) is elevated, the transition strength \(\alpha\) is suppressed, and the inverse duration parameter \(\beta/H\) is enhanced. This indicates that gauge-invariant calculations systematically predict a weaker, shorter-lived phase transition at the soft scale. The magnitude of this correction grows monotonically with \(\Lambda\), implying that the contributions encoded in the \(B_0 + B_1\) terms become progressively dominant in this parameter regime. At the ultrasoft scale, by contrast, the gauge-invariant method produces predictions for \(T_n\), \(\alpha\), and \(\beta/H\) that are closely aligned with those of the conventional approach, and generally fall within the scatter of the latter’s gauge-dependent results.
As \(\Lambda\) increases, the discrepancies between the different calculations diminish steadily. This behavior confirms that gauge-dependent artifacts are suppressed to some extent at the ultrasoft scale in comparison with that of the soft scale. Taken together, the results at both scales demonstrate that gauge-invariant effects can introduce sizable quantitative corrections to phase transition observables, most prominently at the soft scale. Gauge-invariant formulation at both scales produce gauge-independent results, as we have explicitly demonstrated in the preceding section.

\begin{figure}[!htp]
    \centering
    \includegraphics[width=0.8\linewidth]{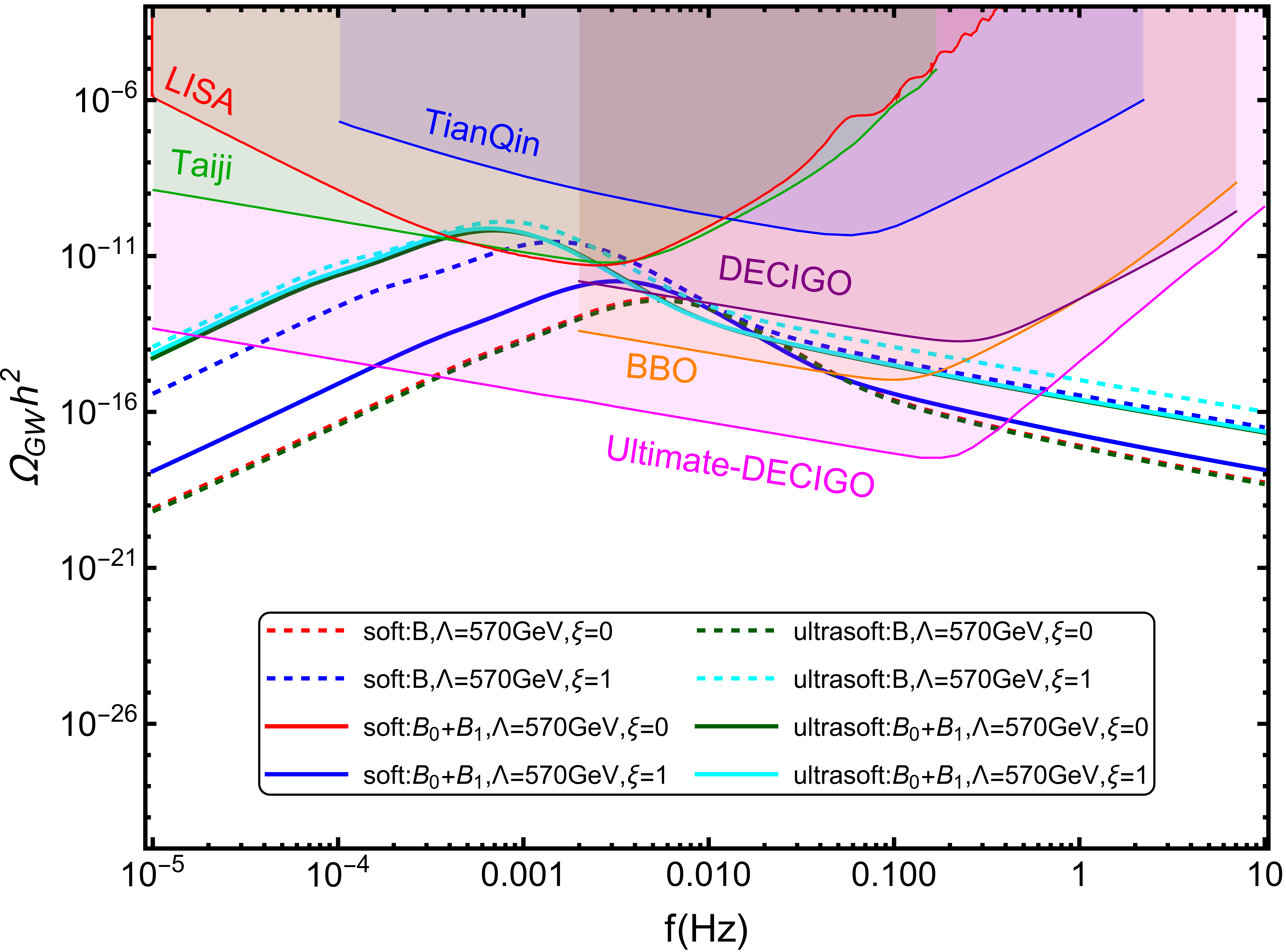}
    \caption{The effect of gauge parameter on GW prediction at the soft scale and ultrasoft scale at $\Lambda$= 570. The color region dentes the sensitivity of detectors: LISA~\cite{LISA:2017pwj,Baker:2019nia}, Taiji~\cite{Hu:2017mde,Ruan:2018tsw}, Tianqin~\cite{TianQin:2015yph,Zhou:2023rop},  BBO~\cite{Crowder:2005nr,Corbin:2005ny,Harry:2006fi}, and DECIGO~\cite{Seto:2001qf,Kawamura:2006up,Yagi:2011wg,Isoyama:2018rjb}. The $B$ and $B_0+B_1$ denote the results obtained with conventional and gauge-invariant methods.}
    \label{fig:gw}
\end{figure}

For the GW prediction from the first-order phase transition, there are three main contributions, i.e., bubble collisions~\cite{Huber:2008hg,Caprini:2015zlo,Kamionkowski:1993fg}, sound waves~\cite{Hindmarsh:2013xza,Hindmarsh:2015qta,Hindmarsh:2017gnf,Caprini:2019egz,Ellis:2019oqb}, and MHD turbulence~\cite{Caprini:2009yp}. With the formula listed in Ref.~\cite{Qin:2024idc}, we give the GW spectra predictions for $\Lambda=570$ GeV in Fig.~\ref{fig:gw}, which shows that, in comparison with the traditional method, the effect of the gauge parameter in the gauge-independent method becomes negligible at the ultra-soft scale and the soft scale. And, the ultrasoft scale result give rise to a much higher magnitude of the GW spectra than that of the soft scale.
\section{Summary and outlook}
\label{sec:Summary and outlook}
In this work, we study the gauge dependence of the Electroweak vacuum decay through the first-order phase transition. Utilizing the SMEFT as a representative framework to capture the new physics accounting for first-order electroweak phase transition, we construct the 3d EFT in the general $R_\xi$ gauge, and analyze the gauge dependences of the nucleation rate and PT parameters. We show that, in comparison with that of the $\lambda\sim g^2$, both the bubble nucleation rate and the PT parameters are gauge invariant for the power-counting scenario of $\lambda\sim g^3$ at both soft scale and ultra-soft scale up to two-loop level. We further present the GW predictions for soft scale and ultrasoft scale,
both the two scale results show null gauge dependence. The predicted GW at the ultra-soft scale are slightly stronger than that of the soft scale results.

We note that some parts of effective potential and all the matching from 4d to 3d theory in the framework of DR are conducted with high-temperature expansion, which reduce the prediction ability of the 3d thermal EFT for the strong phase transitions~\cite{Qin:2024idc}, see Refs.~\cite{Bernardo:2025vkz,Chala:2024xll,Chala:2025oul} also for the study within the Abelian Higgs model where high dimensional operators effect on the GW are addressed. Precise perturbative predictions of the phase structure in strong phase transition scenarios motivated by new physics models and the associated GW signals requires to go beyond the limitations of high-temperature approximation~\cite{Navarrete:2025yxy}. We expect the constructed gauge-invariant framework can be generalized to study the electroweak sphaleron rate and increase the baryon number washout condition for the baryon asymmetry generation~\cite{Zhou:2019uzq,Qin:2024idc,Qin:2024dfp,Li:2025kyo}. The gauge invariant prediction of the GW reduce the uncertainty in the complementary search for new physics with colliders~\cite{Jahedi:2025yjz,Bian:2019bsn,Chala:2025aiz,Hashino:2022ghd}.

The gauge-invariant results obtained in the three-dimensional SMEFT model for first-order phase transitions, such as the phase transition parameter and latent heat, can be compared with the lattice simulation
results as in the models of SM extended by scalar singlet~\cite{Niemi:2024axp}, scalar doublet~\cite{Andersen:2017ika}, and scalar triplet~\cite{Niemi:2020hto}. The reliability of perturbative gauge-independent nucleation rates arrived here can also be tested against non-perturbative calculations~\cite{Moore:2000jw,Moore:2001vf}, though thermal fluctuation effects might make physical picture more complete~\cite{Hirvonen:2024rfg,Hirvonen:2025hqn,Wang:2025ooq,Bian:2025twi}.
With the nucleation rate at hand, one can obtain more realistic real-time simulations of the generation of primordial magnetic fields~\cite{Di:2020kbw,Di:2024gsl,Di:2025ncl}, and gravitational waves spectra~\cite{Hindmarsh:2015qta,Hindmarsh:2013xza,Hindmarsh:2017gnf,Cutting:2018tjt,Cutting:2019zws,Cutting:2020nla}.

\section*{Acknowledgments}
We thank Michael J. Ramsey-Musolf and Philipp Schicho for helpful discussions on the gauge dependence problem of the Abelian Higgs scenario, and Long-Bin Chen, Wen Chen, Feng Feng, Hai Tao Li, Yan-Qing Ma, Wen-Long Sang, and Jian Wang for helpful discussions on the calculation of two-loop Feynmann diagrams, and Joonas Hirvonen for clarify the issue on bubble nucleation rate.
This work is supported by the National Natural Science Foundation of China (NSFC) under Grants Nos.12322505, 12547101. We also acknowledge Chongqing Natural Science Foundation under Grant No. CSTB2024NSCQJQX0022 and Chongqing Talents: Exceptional Young Talents Project No. cstc2024ycjhbgzxm0020.

\appendix
\section{SMEFT in four dimensions and dimensional reduction}
\label{appendix:SMEFT in four dimensions and dimensional reduction}
This appendix collects multiple technical details of matching between parameter in 4D and 3D SMEFT.

\subsection{Relations between $\MSbar$-parameters and physical observables}
\label{appendix:setparameters}
We relate the $\MSbar$-parameters of the Lagrangian
to physical observables, that serve as input parameters
\begin{equation}
    (M_h, M_W, M_Z, M_t, G_F,) \mapsto (m,\lambda,g,g',g_Y) \;.
\end{equation}
Note that the physically observed masses are the pole masses. These relations also depend on the new $\MSbar$-scheme BSM parameters $c_6$, which we also treat as input parameters. For the values of the physical observables used in this work, we set these parameters from~\cite{ParticleDataGroup:2024cfk}. We define the shorthand notation
$g^2_0 \equiv 4 \sqrt{2}\,G_F M^2_W$ for the tree-level coupling and
$v^2_0 \equiv 4 M^2_W/g^2_0 \approx (246.22~ {\rm GeV})^2$ for the tree-level minimum.

At tree and one-loop level only the Higgs mass parameter and self-coupling are affected by $c_6$, and the tree-level relations can be solved from ($V_{\rmi{tree}}$ is defined in Eq.~\ref{eq:Vtree})
\begin{align}
\left.\frac{\partial^2 V_{\rmi{tree}}(\phi)}{\partial \phi^2} \right|_{\phi=v} = M^2_h
\;, \qquad
\left.\frac{\partial V_{\rmi{tree}}(\phi)}{\partial \phi} \right|_{\phi=v} = 0
\;,
\end{align}
resulting in
\begin{align}
m^2 =\frac{1}{2} M^2_h - \frac{3}{4} c_6 v_0^4\;, \qquad
\lambda =\frac{1}{2} \frac{M^2_h}{v^2_0}- \frac{3}{2} c_6 v_0^2\;.
\end{align}
At tree-level, the relations for gauge and Yukawa couplings are unaffected by $c_6$ and read
\begin{align}
g^{2} &= g^2_0\;, \\
g'^{2} &= g^2_0 \Big(\frac{M^2_Z}{M^2_W} - 1 \Big)\;, \\
g_Y^{2} &= \frac{1}{2} g^2_0 \frac{M^2_t}{M^2_W}\;.
\end{align}

For an accurate numerical analysis of the thermodynamics, the above tree-level relations
can be improved by their one-loop corrections (Refs.~\cite{Kajantie:1995dw,Kainulainen:2019kyp,Laine:2017hdk}).
These corrections are necessary for the complete $\mathcal{O}(g^4)$ accuracy of our 3d approach. Regarding the masses, this can be achieved with a standard one-loop pole mass renormalization at zero temperature.
For experimentally measured physical parameters we will use the central values given in~\cite{ParticleDataGroup:2024cfk} throughout the paper, as taken from Ref.~\cite{Karsch:1997gj}. Our perturbative calculations use the $\MSbar$ scheme for renormalization, with the 4-dimensional renormalization scale denoted by $\bmu$. We match experimental results to $\MSbar$ parameters at 1-loop order, matching pole masses using the full 1-loop self-energies. This includes momentum-dependent terms in addition to those from evaluating the second derivative of the 1-loop effective potential at the minimum.





The field-dependent masses are defined by
\begin{align}\label{field dependent masses}
&h:m_h^2=-m^2+3 \lambda \phi^2+\frac{15}{4}c_6 \phi^4,\quad m_G^2=-m^2+\lambda \phi^2+\frac{3}{4}c_6\phi^4 \,\nn\\
&\chi^\pm:m_{\chi^\pm}^2=m_G^2+\xi m_W^2,\qquad \chi:m_{\chi^0}^2=m_G^2+\xi m_Z^2,\,\nn\\
&W^\pm:m_W^2=\frac{1}{4}g^2\phi^2,\qquad Z:m_Z^2=\frac{1}{4}(g^2+g^{\prime 2})\phi^2,\quad A:m_\gamma^2=0,\,\nn\\
&c^\pm:m_{cW}^2=\xi m_W^2,\quad\qquad c_Z:m_{cZ}^2=\xi m_Z^2,\qquad\qquad c_A:m_{cA}^2=0.
\end{align}

The above are the field-dependent masses in the 4D theory before dimensional reduction. The 4D field-dependent masses also include the top quark mass $(T:m_t^2=\frac{1}{2}y_t^2\phi^2)$,however, the fermionic contributions are absorbed into the parameters of the 3D effective theory.This 3D theory is associated with the soft scale.After dimensional reduction the couplings and masses are replaced by the corresponding quantities in the 3D effective theory, conventionally denoted by a subscript “3”, like $g \to g_3, \lambda \to \lambda_3,m_h^2\to m^2_{h,3}$. We have further integrated out certain soft modes to reach the ultrasoft scale,where the effective couplings are denoted by barred quantities, e.g.$g_3 \to \bar{g}_3$.

Throghout this work, the symbol g (Without subscript or overbar) denotes the gauge coupling of the 3D effective. Unless a distinction between 4D and 3D quantities is explicitly needed, all parameters, couplings and masses refer to those of the 3D effective theory, and the subscript “3” and overbar are suppressed for notational simplicity.


\subsection{Parameter matching}
\label{parameter matching}
In the general context of low-energy effective field theories, Ref.~\cite{Vepsalainen:2007ji} reviews the rationale for dimensional reduction. It discusses the required resummations to remove the high-temperature infrared divergences by matching the correlation functions at the higher scale and lower scale EFT. Ref.~\cite{Schicho:2020xaf} presents a practical tutorial for the matching procedure for a real scalar field. The automated package DRalgo~\cite{Ekstedt:2022bff} can be utilized for dimensional reduction for generic models was in Landau gauge. Below, we present a review of the formal recipe for this matching procedure by following Ref.~\cite{Croon:2020cgk}. For a generic field $\psi$, we denote the n-point correlation functions by

\begin{equation}
	\Gamma_{\psi^n}\equiv \braket{\psi^n}\,,\quad \Pi_{\psi^2}\equiv \braket{\psi^2}\,,
\end{equation}
where $n \geq 2$. We distinguish the 2-point function $\Gamma$ and expand in soft external momenta $K=(0,\bm{k})$ with $|\bm{k}|\sim gT$ :
\begin{align}
	\Gamma_{\psi^n}&=G_{\psi^n}+\mathcal{O}(K^2)\,,\\
	\Pi_{\psi^2}&=G_{\psi^2}+K^2 \Pi_{\psi^2}^\prime+\mathcal{O}(K^4)\,.
\end{align}
$G$ denotes the correlator at zero external momenta and $\Pi^\prime$ is the quadratic-momenta correction that contributes to the field renormalization factor $Z$
\begin{equation}
	Z_{\psi^2}=1+\Pi_{\psi^2}^\prime\,.
\end{equation}

By matching the effective actions in both theories, the leading (quadratic) kinetic terms yield the relation between 3d and 4d fields
\begin{equation}\label{eq:field4dto3d}
	\begin{aligned}
		\phi_{3d}^2 Z_{3d} &=\frac{1}{T} \phi_{4d}^2 Z_{4d}\,,\\
		\phi_{3d}^2 (1+\Pi_{3d}^\prime) &=\frac{1}{T} \phi_{4d}^2 (1+\Pi_{\mathrm{soft}}^\prime+\Pi_{\mathrm{hard}}^\prime)\,,\\
		\phi_{3d}^2 &= \frac{1}{T} \phi_{4d}^2 (1+\Pi_{\mathrm{hard}}^\prime)\,,
	\end{aligned}
\end{equation}
where we denote the scalar background fields by $\phi$ and illustrate the separation into soft $(k \sim g T)$ and hard $(K \sim \pi T)$ modes. For simplicity, we omit the field subscript from $z$ and $\Pi^\prime$ for a moment. By construction of the 3d EFT, contributions $\Pi_{3d}^\prime=\Pi_{\mathrm{soft}}^\prime$ cancel, this is a requirement that the theories are mutually valid in the IR. Therefore, only the hard modes contribute to the last line in Eq.~(\ref{eq:field4dto3d}). By equating the quartic terms of the effective actions, we get
\begin{equation}\label{eq:four v}
	\frac{1}{4}\lp -2 \lambda+\Gamma_{\mathrm{hard}}+\Gamma_{\mathrm{soft}}\rp \phi_{4d}^4 = T \frac{1}{4}\lp -2 \lambda_3+\Gamma_{3d}\rp \phi_{3d}^4 \,.
\end{equation}
where $\Gamma$ denotes four-point vertex correction, again by virtue of the EFT construction, terms $\Gamma_{\mathrm{soft}}=\Gamma_{3d}$ cancel. After inserting Eq.~(\ref{eq:four v}) for the field normalization into eq.~(\ref{eq:field4dto3d}) one can solve for the 3d quartic coupling $\lambda_3$:
\begin{equation}
		\lambda_3=T \Gamma_{\phi^4} Z_\phi^{-2}\simeq T\lp \lambda-\frac{1}{2}\Gamma_{\phi^4}-2 \lambda \Pi_{\phi^2}^{\prime} \rp\,.	
\end{equation}

The 3d mass parameter related with the 4D $\MSbar$ parameters at the soft scale as~\cite{Croon:2020cgk}:
\begin{align}
    m_3^2 =&
     m^2(\bmu)
    - \frac{T^2}{16}\Big(3\g^{2}(\bmu) + {\gp}^{2}(\bmu) + 4 \gY^{2}(\bmu) + 8 \lambda(\bmu) \Big)
    - \frac{1}{4} T^4 c_6
    \nonumber \\ &
    - \frac{1}{(4\pi)^2} \bigg\{ -m^2 \bigg[ \Big(\frac{3}{4}(3\g^{2} + {\gp}^{2}) - 6 \lambda \Big)L_b - 3 \gY^{2} L_f \bigg]
    \nonumber\\ &
    + T^2 \bigg[ \frac{167}{96}\g^{4} + \frac{1}{288}{\gp}^{4} - \frac{3}{16}\g^{2} {\gp}^{2} + \frac{1}{4}\lambda (3\g^{2}+{\gp}^{2})
    \nonumber \\ &
    + L_b \Big( \frac{17}{16}\g^{4} - \frac{5}{48}{\gp}^{4} - \frac{3}{16}\g^{2}{\gp}^{2} + \frac{3}{4}\lambda(3\g^{2}+{\gp}^{2}) - 6 \lambda^2 \Big)
    \nonumber \\ &
    + \Big( c + \ln\left(\frac{3T}{\bmu_{3}}\right) \Big)\Big(  \frac{81}{16}\g^{4} +  3\lambda ( 3\g^{2} + {\gp}^{2})  - 12\lambda^2
     -\frac{7}{16} {\gp}^{4} - \frac{15}{8}\g^{2} {\gp}^{2} \Big)
    \nonumber \\ &
    - \gY^{2} \Big(\frac{3}{16}\g^{2} + \frac{11}{48}{\gp}^{2} + 2 \gs^{2} \Big)
    + \Big(\frac{1}{12}\g^{4} + \frac{5}{108}{\gp}^{4}\Big)\Nf
    \nonumber \\ &
    + L_f \Big( \gY^{2} \Big(\frac{9}{16}\g^{2} + \frac{17}{48}{\gp}^{2} + 2 \gs^{2} - 3 \lambda \Big) +\frac{3}{8}\gY^{4}
    - \Big(\frac{1}{4}\g^{4} + \frac{5}{36}{\gp}^{4}\Big) \Nf \Big)
    \nonumber \\ &
    + \ln(2) \Big( \gY^{2} \Big(-\frac{21}{8}\g^{2} - \frac{47}{72}{\gp}^{2} + \frac{8}{3} \gs^{2} + 9 \lambda \Big) -\frac{3}{2}\gY^{4}
  + \Big(\frac{3}{2}\g^{4} + \frac{5}{6}{\gp}^{4}\Big) \Nf \Big) \bigg] \bigg\}
    \;,
\end{align}
with
\begin{equation}
L_b =2\ln\Big(\frac{\bmu}{T}\Big)-2[\ln(4\pi)-\gammaE],\quad
L_f = L_b + 4\ln2,\quad c=\frac{1}{2}\bigg(\ln\Big(\frac{8\pi}{9}\Big) + \frac{\zeta'(2)}{\zeta(2)} - 2 \gammaE \bigg),
\end{equation}
where $\gammaE$ is the Euler-Mascheroni constant and $\zeta(x)$ is the Riemann zeta function. See also Ref.~\cite{Chala:2025aiz} for other contributions of high-dimensional operators. And, Debye masses read
\begin{align}
    \mD^{2} =&
    \g^{2}(\bmu) T^2\bigg(\frac{5}{6}+\frac{\Nf}{3}\bigg)
    \;,\\
\mD'^{2} =&
    {\gp}^{2}(\bmu)  T^2\bigg(\frac{1}{6}+\frac{5\Nf}{9}\bigg)\;,
\end{align}
Other 3d couplings at the soft scale can be obtained with the 4D couplings~\cite{Qin:2024idc,Croon:2020cgk}:
\begin{align}
\g_{3}^2 =&
    \g^{2}(\bmu)T\bigg[1 +\frac{\g^{2}}{(4\pi)^2}\bigg(\frac{43}{6}L_b+\frac{2}{3}-\frac{4\Nf}{3}L_f\bigg)\bigg]
    \;,\\
{\gp_{3}}^{2} =&
    {\gp}^{2}(\bmu)T\bigg[1 +\frac{{\gp}^{2}}{(4\pi)^2}\bigg(-\frac{1}{6}L_b-\frac{20\Nf}{9}L_f\bigg)\bigg]
    \;,\\
\lambda_{3} =&
    T\Big(\lambda(\bmu) + \frac{1}{(4\pi)^2}\bigg[\frac{1}{8}\Big(3\g^{4} + {\gp}^{4} +2 \g^{2} {\gp}^2 \Big)+ 3 L_f \Big(\gY^{4} - 2\lambda \gY^{2} \Big)\nonumber\\ &
    - L_b \bigg(\frac{3}{16}\Big(3\g^{4} + {\gp}^{4} + 2 \g^{2} {\gp}^{2} \Big) -\frac{3}{2}\Big(3\g^{2}+{\gp}^{2} -8 \lambda \Big) \lambda \bigg) \bigg]\Big)
    \nonumber \\ &
    + T^3 c_6
    - \frac{\mh^2}{(4\pi)^2} 12 T c_{6}L_{b}
    \;,
    \end{align}
    \begin{align}
c_{6,3} =&
    T^2 c_6(\bmu) \bigg (1 + \frac{1}{(4\pi)^2} \bigg[ \Big( -54 \lambda + \frac{9}{4} (3 \g^{2} + {\gp}^{2}) \Big) L_b- 9 \gY^{2} L_f  \bigg] \bigg)
    \nonumber \\ &
    - \frac{\zeta(3)}{768 \pi^4} \bigg( -\frac{3}{8} \Big( 3 \g^{6} + 3 \g^{4} {\gp}^{2} + 3 \g^{2} {\gp}^{4} + {\gp}^{6} \Big)  - 240 \lambda^3
 +84 \gY^{6}\bigg)
    \;, \\
h_{1} =&
    \frac{\g^{2}(\bmu)T}{4}\bigg(1+\frac{1}{(4\pi)^2}\bigg\{\bigg[\frac{43}{6}L_b+\frac{17}{2}-\frac{4\Nf}{3}(L_f-1)\bigg]\g^{2}+\frac{{\gp}^{2}}{2}-6\gY^{2}+12\lambda\bigg\} \bigg)\;,\nonumber\\
h_{2} =&
    \frac{{\gp}^{2}(\bmu)T}{4}\bigg(1 +\frac{1}{(4\pi)^2}\bigg\{\frac{3\g^{2}}{2}
    - \bigg[\frac{(L_b-1)}{6}  + \frac{20\Nf(L_f-1)}{9}\bigg]{\gp}^{2}- \frac{34}{3} \gY^{2}+12\lambda\bigg\} \bigg)
    \;,\\
h_{3} =&
    \frac{\g(\bmu){\gp}(\bmu)T}{2}\bigg\{1+\frac{1}{(4\pi)^2}\bigg[-\g^{2}+ \frac{1}{3}{\gp}^{2}+L_b\bigg(\frac{43}{12}\g^{2} -\frac{1}{12}{\gp}^{2}\bigg)\nonumber \\ &
    - \Nf(L_f-1)\bigg(\frac{2}{3}\g^{2}+\frac{10}{9}{\gp}^{2}\bigg)+4\lambda+ 2\gY^{2}\bigg] \bigg\}
    \;, \\
\kappa_{1} =& T\frac{\g^{4}}{(4\pi)^2} \bigg(\frac{17-4\Nf}{3}\bigg)
    \;,\\
\kappa_{2} =& T\frac{{\gp}^{4}}{(4\pi)^2} \bigg(\frac{1}{3}-\frac{380}{81} \Nf\bigg)
    \;,\\
\kappa_{3} =& T\frac{\g^{2} {\gp}^{2}}{(4\pi)^2}\bigg(2-\frac{8}{3}\Nf\bigg)\;,
\end{align}

The second step is to build the ultrasoft theory after integrating out the soft temporal scale. The Lagrangian at ultrasoft scale is given in Eq.~{(\ref{ultrasoft lagrangian})}. The parameters of the ultrasoft 3D EFT read
\begin{align}\label{lightparameters}
\bar{g}^2_3 =& \g_{3}^{2} \Big( 1 - \frac{\g_{3}^{2}}{6 (4\pi) \mD} \Big)
\;, \\
\bar{g}'^2_3 =& {\gp_{3}}^2
\;, \\
\bar{m}_3^2 =& m_3^2
    + \frac{1}{4\pi}\Big(3 h_{1}\mD +  h_{2}\mD' \Big)
    - \frac{1}{(4\pi)^2} \bigg( 3\g_{3}^{2}h_{1} - 3 h_{1}^2 - h_{2}^2 - \frac{3}{2} h_{3}^{2}
    \nonumber \\ &
    + \Big(-\frac{3}{4}\g_{3}^{4} + 12\g_{3}^{2}h_{1} \Big) \ln\Big(\frac{\bmu_{3}}{2\mD} \Big)
    - 6 h_{1}^{2} \ln\Big(\frac{\bmu_{3}}{2\mD} \Big)
    - 2 h_{2}^{2} \ln\Big(\frac{\bmu_{3}}{2\mD'} \Big)
    \nonumber \\&
    - 3 h_{3}^{2} \ln\Big(\frac{\bmu_{3}}{\mD+\mD'} \Big)
    \bigg)
\;, \\
\bar{\lambda}_{3} =& \lambda_{3} - \frac{1}{2(4\pi)}\Big(
      \frac{3 h_{1}^{2}}{\mD}
    + \frac{h_{2}^{2}}{\mD'}
    + \frac{h_{3}^{2}}{\mD+\mD'}
    \Big)
    \;,\\
\bar{c}_{6,3} =&
    c_{6,3} + \frac{1}{2(4\pi)}\frac{h_{1}^{3}}{\mD^3}
\;,
\end{align}
The ultrasoft renormalization scale $\bmu_{3}$ has less effect on the results, and we set $\overline{\mu}_3=g^2 T$ is this paper.
\section{$Z_\phi$ Factor}
\label{appdenix:zzfactor}

We use dimensional regularisation in $D = d+1 = 4-2\epsilon$ dimensions and the $\MSbar$-scheme with renormalisation scale $\bmu$. We define the notation $P \equiv (\omega_n,\vec{p})$ for Euclidean four-momenta where the bosonic Matsubara frequency is $\omega_n = 2\pi n T$ and the fermionic Matsubara frequency is $\omega_n = (2\pi n+1) T$,
\begin{align}
\label{eq:Tint}
\Tint{P} &\equiv T \sum_{\omega_n} \int_p\;,\quad\quad
\int_p \equiv \Big( \frac{\bmu^{2}e^\gammaE}{4\pi} \Big)^\epsilon \int \frac{{\rm d}^{d}p}{(2\pi)^d}\;,\\
\Tint{P}' &\equiv T \sum_{\omega_n \neq 0} \int_p,  \quad \quad
n_{\rmii{B/F}}(E_p,T) \equiv \frac{1}{e^{E_p/T}\mp 1}\;.
\end{align}
This last definition is the Bose(Fermi)-distribution with $E_p = \sqrt{p^2 + m^2}$.
We here present the results for 4d and 3d cases about the calculation of Z-factor.

\subsection{The 4d case}
\label{appendix:4D Z-factor}
We here consider the wave function renormalization at one-loop order in 4d case, as given in Fig.~\ref{fig:Z factor}. The contributions of each Feynmann diagrams is:
\begin{figure}[!htp]
	\centering
	\includegraphics[width=0.8\linewidth]{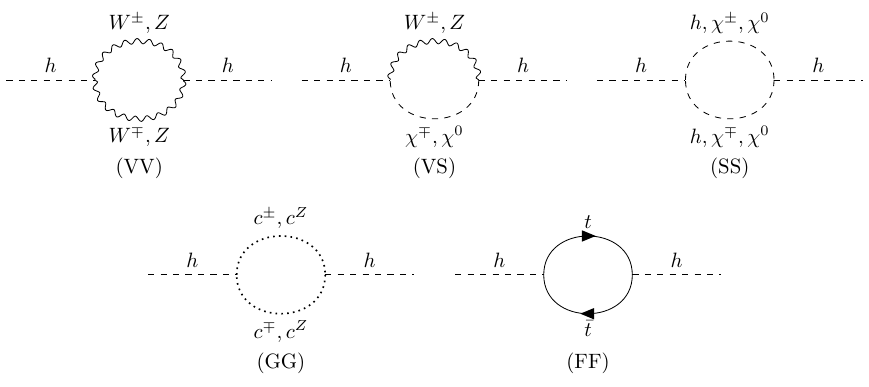}
	\caption{One-loop contributions to the wave function correction factors.}
	\label{fig:Z factor}
\end{figure}

The corresponding loop integrals are in Figure.~\ref{fig:Z factor}


\begin{equation}
    \begin{aligned}
\mathcal{D}_\mathrm{VV}(m)&=\Tint{p_b}'\dfrac{\left[g_{\mu\nu}-(1-\xi)\frac{p_\mu p_\nu}{p^2-\xi m^2}\right]\left[g^{\mu\nu}-(1-\xi)\frac{(p+k)^\mu (p+k)^\nu}{(p+k)^2-\xi m^2}\right]}{\lp p^2-m^2+i\epsilon\rp[(p+k)^2-m^2+i\epsilon]}   \\
&=i\biggl\{ \frac{\xi ^2+3}{16 \pi ^2 \epsilon _b}+\frac{23 k^2 \xi ^2 \zeta (3)}{2560 \pi ^4 T^2}-\frac{49 k^2 \xi  \zeta (3)}{3840 \pi ^4 T^2}+\frac{109 k^2 \zeta (3)}{7680 \pi ^4 T^2}-\frac{ (\xi ^3+3) \zeta (3)}{64 \pi ^4 T^2}m^2 \biggr\}\;,
    \end{aligned}
\end{equation}

\be
\begin{aligned}
\mathcal{D}_\mathrm{VS}(m_1,m_2)&=\Tint{p_b}'\dfrac{(2k^\mu+p^\mu)\left[g_{\mu\nu}-(1-\xi)\frac{p_\mu p_\nu}{p^2-\xi m_1^2}\right](2k^\nu+p^\nu)}{\lp p^2-m_1^2+i\epsilon\rp\lp(p+k)^2-m_2^2+i\epsilon\rp}\;\\
&=i \lp-\frac{k^2 \xi }{16 \pi ^2 \epsilon _b}+\frac{3 k^2}{16 \pi ^2 \epsilon _b}+\frac{m_1^2 \xi ^2}{16 \pi ^2 \epsilon _b}+\frac{m_2^2 \xi }{16 \pi ^2 \epsilon _b}-\frac{7 k^4 \xi  \zeta (3)}{960 \pi ^4 T^2}+\frac{k^4 \zeta (3)}{80 \pi ^4 T^2}\right.\\
	&+\frac{k^2 m_1^2 \xi ^2 \zeta (3)}{160 \pi ^4 T^2}+\frac{19 k^2 m_2^2 \xi  \zeta (3)}{1280 \pi ^4 T^2}-\frac{47 k^2 m_1^2 \zeta (3)}{1920 \pi ^4 T^2}-\frac{97 k^2 m_2^2 \zeta (3)}{3840 \pi ^4 T^2}\\
	& \left. -\frac{m_1^4 \xi ^3 \zeta (3)}{128 \pi ^4 T^2}-\frac{m_1^2 m_2^2 \xi ^2 \zeta (3)}{128 \pi ^4 T^2}-\frac{m_2^4 \xi  \zeta (3)}{128 \pi ^4 T^2}-\frac{\xi  T^2}{12} \rp\;,
\end{aligned}
\ee

\be
\begin{aligned}
    \mathcal{D}_\mathrm{SS}(m)
    =&\Tint{p_b}'\frac{1}{(p^2-m^2+i\epsilon)[(p+k)^2-{m^2}+i\epsilon]}\;\\
    =&i\lp\frac{1}{16 \pi ^2 \epsilon _b}+\frac{k^4 \zeta (5)}{10240 \pi ^6 T^4}-\frac{k^2 m^2 \zeta (5)}{1024 \pi ^6 T^4}+\frac{k^2 \zeta (3)}{384 \pi ^4 T^2}\right.\\
    &\left.+\frac{3 m^4 \zeta (5)}{1024 \pi ^6 T^4}-\frac{m^2 \zeta (3)}{64 \pi ^4 T^2}\rp\;,
\end{aligned}
\ee

because $GG$ characteristic integral is the same as that of $SS$, that is
\be
\mathcal{D}_\mathrm{GG}(m)=\mathcal{D}_\mathrm{SS}(m)\;,
\ee

\be
\begin{aligned}
\mathcal{D}_\mathrm{FF}(m)&=\Tint{p_f}\dfrac{\mathrm{Tr}[(\slashed{p}+m)(\slashed{p}+\slashed{k}+m)]}{(p^2- m^2+i\epsilon)[(p+k)^2-m^2+i\epsilon]}\;\\
&=i\lp-\frac{k^2}{8 \pi ^2 \epsilon _f}+\frac{3 m^2}{4 \pi ^2 \epsilon _f}+\frac{31 k^4 m^2 \zeta (5)}{2560 \pi ^6 T^4}-\frac{7 k^4 \zeta (3)}{192 \pi ^4 T^2}-\frac{31 k^2 m^4 \zeta (5)}{256 \pi ^6 T^4}\right. \\
	&\left.+\frac{35 k^2 m^2 \zeta (3)}{96 \pi ^4 T^2}+\frac{93 m^6 \zeta (5)}{256 \pi ^6 T^4}-\frac{35 m^4 \zeta (3)}{32 \pi ^4 T^2}+\frac{T^2}{6}\rp\;,
\end{aligned}
\ee

In total,we have
\be
\begin{aligned}
	-iM_{hh}=&1\times\lp\frac{2 m_W^2}{\phi}\rp^2 \mathcal{D}_\mathrm{VV}(m_W)+\frac{1}{2}\times\lp\frac{2 m_Z^2}{\phi}\rp^2 \mathcal{D}_\mathrm{VV}(m_Z)\\
	&+2\times\lp\frac{i g}{2}\rp^2\mathcal{D}_\mathrm{VS}(m_W,m_{\chi^\pm})-1\times\lp\frac{g}{2\cos\theta}\rp^2\mathcal{D}_\mathrm{VS}(m_Z,m_{\chi^0})\\
	&+\frac{1}{2}\times\frac{9m_h^4}{\phi^2}\mathcal{D}_\mathrm{SS}(m_h)+\frac{m_h^4}{\phi^2}\mathcal{D}_\mathrm{SS}(m_{\chi^+})+\frac{1}{2}\times\frac{m_h^4}{\phi^2}\mathcal{D}_\mathrm{SS}(m_{\chi^0})\\
	&-2\times\frac{\xi^2 m_W^4}{\phi^2}\mathcal{D}_\mathrm{GG}(m_{cW})-\frac{\xi^2 m_Z^4}{\phi^2}\mathcal{D}_\mathrm{GG}(m_{cZ}) -n_c\frac{m_t^2}{\phi^2}\mathcal{D}_\mathrm{FF}(m_t)\;,
\end{aligned}
\ee
where $n_c$ represents the number of colors of the quark, $n_c=3$. Finally,
\begin{equation}
	\begin{aligned}\label{4D Z}
		Z&=\frac{\partial M_{hh}}{\partial k^2}\\
		&=-\frac{g^2 \xi }{32 \pi ^2 \epsilon _b}-\frac{g^2 \xi }{64 \pi ^2 \cos^2\theta \epsilon _b}+\frac{3 g^2}{32 \pi ^2 \epsilon _b}+\frac{3 g^2}{64 \pi ^2 \cos^2\theta \epsilon _b}-\frac{3 m_t^2}{8 \pi ^2 \phi ^2 \epsilon _f}\\
		&+\frac{g^2 \xi ^2 \zeta (3) m_W^2}{320 \pi ^4 T^2}-\frac{47 g^2 \zeta (3) m_W^2}{3840 \pi ^4 T^2}+\frac{g^2 \xi ^2 \zeta (3) m_Z^2}{640 \pi ^4 T^2 \cos^2\theta}-\frac{47 g^2 \zeta (3) m_Z^2}{7680 \pi ^4 T^2 \cos^2\theta}\\
		&+\frac{19 g^2 \xi  \zeta (3) m_{\text{$\chi $0}}^2}{5120 \pi ^4 T^2 \cos^2\theta}+\frac{19 g^2 \xi  \zeta (3) m_{\text{$\chi $1}}^2}{2560 \pi ^4 T^2}-\frac{97 g^2 \zeta (3) m_{\text{$\chi $0}}^2}{15360 \pi ^4 T^2 \cos^2\theta}-\frac{97 g^2 \zeta (3) m_{\text{$\chi $1}}^2}{7680 \pi ^4 T^2}\\
		&+\frac{\zeta (5) m_h^4 m_{\text{$\chi $0}}^2}{2048 \pi ^6 T^4 \phi ^2}+\frac{\zeta (5) m_h^4 m_{\text{$\chi $1}}^2}{1024 \pi ^6 T^4 \phi ^2}+\frac{9 \zeta (5) m_h^6}{2048 \pi ^6 T^4 \phi ^2}-\frac{\zeta (3) m_h^4}{64 \pi ^4 T^2 \phi ^2}-\frac{93 \zeta (5) m_t^6}{256 \pi ^6 T^4 \phi ^2}\\
		&+\frac{35 \zeta (3) m_t^4}{32 \pi ^4 T^2 \phi ^2}-\frac{\xi ^4 \zeta (5) m_W^2 m_Z^4}{1024 \pi ^6 T^4 \phi ^2}-\frac{\xi ^4 \zeta (5) m_W^6}{512 \pi ^6 T^4 \phi ^2}-\frac{59 \xi ^2 \zeta (3) m_W^4}{1920 \pi ^4 T^2 \phi ^2}+\frac{49 \xi  \zeta (3) m_W^4}{960 \pi ^4 T^2 \phi ^2}\\
		&-\frac{109 \zeta (3) m_W^4}{1920 \pi ^4 T^2 \phi ^2}-\frac{59 \xi ^2 \zeta (3) m_Z^4}{3840 \pi ^4 T^2 \phi ^2}+\frac{49 \xi  \zeta (3) m_Z^4}{1920 \pi ^4 T^2 \phi ^2}-\frac{109 \zeta (3) m_Z^4}{3840 \pi ^4 T^2 \phi ^2}\;.
	\end{aligned}
\end{equation}

\subsection{The 3d case}
\label{appendix:3d Z-factor}
Based on 3d DR technology, the heavier fields are absorbed into the corresponding coupling parameters. The graph we consider is basically the same as the above one, except for the contribution of the Fermi fields which have been integrated out.
The field renormalization factor Z composes of the diagrams in Figure~\ref{fig:Z factor in 3d with hbar}, whose contributions to the self-energies read:

\begin{figure}[!htp]
    \centering
    \includegraphics[width=0.8\linewidth]{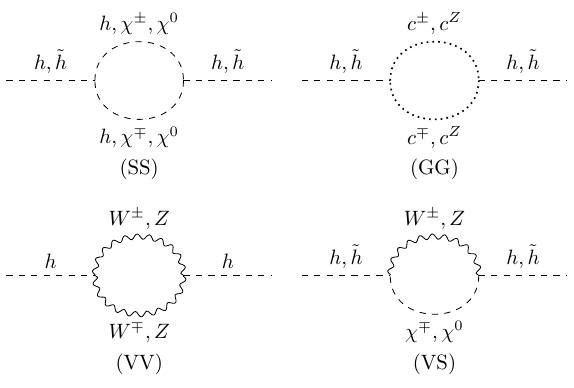}
    \caption{The diagrams contributing to the kinetic tern of effective action.}
    \label{fig:Z factor in 3d with hbar}
\end{figure}

\begin{equation}
	\begin{aligned}
		-\Pi_{hh}&=\frac{1}{2}C_{hhh}^2 I_{SS}(m_h)+\frac{2}{2}C_{H\chi^+\chi^-}^2 I_{SS}(m_{\chi^\pm})+\frac{1}{2}C_{h\chi\chi}^2I_{SS}(m_{\chi^0})\,\\
		&+\frac{2}{2}C_{hW^+W^-}^2 I_{VV}(m_W)+\frac{1}{2}C_{hZZ}^2 I_{VV}(m_Z)\,\\
		&-2 C_{hW^+\chi^-} C_{hW^-\chi^+} I_{VS}^{HH}(m_W,m_{\chi^\pm})-C_{hZ\chi}^2 I_{VS}(m_Z,m_{\chi^0})\,\\
		&-2 C_{hc^+c^-}^2 I_{GG}(m_{c_W})-C_{hc_Zc_Z}^2 I_{GG}(m_{c_Z})\;,
	\end{aligned}
\end{equation}

\begin{equation}
    \begin{aligned}
        -\Pi_{h\tilde{h}}&=\frac{2}{2}C_{h\chi^+\chi^-}C_{\tilde{h}\chi^+\chi^-}I_{SS}(m_{\chi^\pm})+\frac{1}{2}C_{h\chi\chi}C_{\tilde{h}\chi\chi}I_{SS}(m_{\chi^0})\\
        &+2 C_{h W^\pm \chi^\pm} C_{\tilde{h}W^\pm \chi^\pm}I_{VS}^{H\tilde{H}}(m_W,m_{\chi^\pm})+C_{h Z \chi}C_{\tilde{h}Z \chi}I_{VS}^{H\tilde{H}}(m_Z,m_{\chi^0})\\
        &-2 C_{hc^+c^-} C_{\tilde{h}c^+c^-} I_{GG}(m_{c_W})-C_{hc_Zc_Z} C_{\tilde{h}c_Zc_Z}I_{GG}(m_{c_Z})\;,
    \end{aligned}
\end{equation}

\begin{equation}
    \begin{aligned}
        -\Pi_{\tilde{h}\tilde{h}}&=\frac{2}{2}C_{\tilde{h}\chi^+\chi^-}^2I_{SS}(m_{\chi^\pm})+\frac{1}{2}C_{\tilde{h}\chi\chi}^2I_{SS}(m_{\chi^0})+2 C_{\tilde{h}W^\pm \chi^\pm}^2I_{VS}^{\tilde{H}\tilde{H}}(m_W,m_{\chi^\pm})\\
        &+C_{\tilde{h}Z \chi}^2I_{VS}^{\tilde{H}\tilde{H}}(m_Z,m_{\chi^0})-2C_{\tilde{h}c^+c^-}^2 I_{GG}(m_{c_W})-C_{\tilde{h}c_Zc_Z}^2I_{GG}(m_{c_Z})\;,
    \end{aligned}
\end{equation}
where we have the symmetry $\Pi_{h\tilde{h}}=\Pi_{\tilde{h}h}$. These $I_{xy}$ functions are listed bellow:
\begin{equation}
    I_{SS}(m)=\int_{p}\frac{1}{(p^2+m^2)[(p+k)^2+m^2]}=\frac{1}{8 \pi m}+k^2\lp-\frac{1}{96 \pi m^3} \rp+\mathcal{O}(k^4)\;,
\end{equation}
\begin{align}
    I_{GG}(m)&=I_{SS}(m)\;,\\
    I_{VV}(m)&=\int_{p}\frac{\left[\delta_{ij}-(1-\xi)\frac{p_i p_j}{p^2+\xi m^2}\right]\left[\delta_{ij}-(1-\xi)\frac{(p+k)_i (p+k)_j}{(p+k)^2+\xi m^2}\right]}{(p^2+m^2)[(p+k)^2+m^2]}\nn\,\\
	&=\frac{\xi^{3/2}+2}{8 \pi m}+k^2 \lp\frac{-9\xi+13 \sqrt{\xi}-10}{96 \pi m^3\lp\sqrt{\xi}+1\rp} \rp+\mathcal{O}(k^4)\;,\\
    I_{VS}^{H \tilde{H}}(m_1,m_2)&=\int_p \frac{p_i (p+2 k)_j \left[\delta_{ij}-(1-\xi)\frac{p_i p_j}{p^2+\xi m_1^2}\right]}{(p^2+m_1^2)[(p+k)^2+m_2^2]}\nn\\
	&=-\frac{\xi  \left(m_1^2 \xi +m_2 m_1 \sqrt{\xi }+m_2^2\right)}{4 \pi  \left(m_1 \sqrt{\xi }+m_2\right)}+k^2\lp \frac{ -\xi  \left(3 m_1^2 \xi +6 m_2 m_1 \sqrt{\xi }+2 m_2^2\right)}{12 \pi  \left(m_1 \sqrt{\xi }+m_2\right)^3}\rp+\mathcal{O}(k^4)\;,\\
	I_{VS}^{\tilde{H}\tilde{H}}(m_1,m_2)&=\int_p \frac{p_i p_j \left[\delta_{ij}-(1-\xi)\frac{p_i p_j}{p^2+\xi m_1^2}\right]}{(p^2+m_1^2)[(p+k)^2+m_2^2]}\nn\\
	&=-\frac{\xi  \left(m_1^2 \xi +m_2 m_1 \sqrt{\xi }+m_2^2\right)}{4 \pi  \left(m_1 \sqrt{\xi }+m_2\right)}+k^2\lp \frac{ m_1^2 \xi ^2}{12 \pi  \left(m_1 \sqrt{\xi }+m_2\right)^3}\rp+\mathcal{O}(k^4)\;,\\
    I_{VS}^{HH}(m_1,m_2)&=\int_{p}\frac{(p+2k)_i (p+2k)_j\left[\delta_{ij}-(1-\xi)\frac{p_i p_j}{p^2+\xi m_1^2}\right]}{(p^2+m_1^2)[(p+k)^2+m_2^2]}\nn\\
	&=-\frac{\xi\lp m_1 m_2 \sqrt{\xi}+m_1^2 \xi+m_2^2\rp}{4 \pi \lp m_1 \sqrt{\xi}+m_2\rp}\nn\\
	&+k^2 \lp -\frac{m_1 m_2 \xi^{3/2}}{3 \pi \lp m_1 \sqrt{\xi}+m_2\rp^3}-\frac{m_1^2 \xi^2}{4 \pi \lp m_1 \sqrt{\xi}+m_2\rp^3}+\frac{2}{3 \pi (m_1+m_2)}\rp\nn\\
	&+\mathcal{O}(k^4)\;.
\end{align}
The overall result for Z-factor reads
\begin{eqnarray}
Z&=&1-\frac{1}{768 \pi}\Bigg[  g^2 \Bigg(8 \xi  \phi ^2 \left(-\frac{1}{m_{\chi^0}^3}-\frac{2}{m_{\chi^\pm}^3}\right) \left(3 c_6 \phi ^2+\lambda \right)\nn\\
&+&2 g^{'2} \phi ^2 \left(\xi ^2 \left(-\frac{7}{m_{c_Z}^3}-\frac{1}{m_{\chi^0}^3}\right)+\frac{32 \xi }{m_{c_Z} m_Z (m_{c_Z}+m_Z)}-\frac{10}{m_Z^3}\right)\nn\\
&+&64 \left(\frac{\xi  (2 m_{c_Z}+m_{c_W}+2 m_{\chi^0}+m_{\chi^\pm})}{(m_{c_Z}+m_{\chi^0}) (m_{c_W}+m_{\chi^\pm})}+\frac{4}{m_W+m_{\chi^\pm}}+\frac{2}{m_{\chi^0}+m_Z}\right)\Bigg)\nn\\
&+&g^{'2} \left(64 \left(\frac{\xi }{m_{c_Z}+m_{\chi^0}}+\frac{2}{m_{\chi^0}+m_Z}\right)-\frac{8 \xi  \phi ^2 \left(3 c_6 \phi ^2+\lambda \right)}{m_{\chi^0}^3}\right)\nn\\
&+&16 \phi ^2 \left(-\frac{1}{m_{\chi^0}^3}-\frac{2}{m_{\chi^\pm}^3}\right) \left(3 c_6 \phi ^2+\lambda \right)^2-\frac{36 \left(5 c_6 \phi ^3+2 \lambda  \phi \right)^2}{m_h^3}\nn\\
&+&g^4 \phi ^2 \Bigg(\frac{64 \xi }{m_{c_W} m_W (m_{c_W}+m_W)}+\frac{32 \xi }{m_{c_Z} m_Z (m_{c_Z}+m_Z)}\nn\\
&+&\xi ^2 \left(-\frac{7}{m_{c_Z}^3}-\frac{14}{m_{c_W}^3}-\frac{1}{m_{\chi^0}^3}-\frac{2}{m_{\chi^\pm}^3}\right)-\frac{20}{m_W^3}-\frac{10}{m_Z^3}\Bigg)\nn\\
&+&g^{'4} \phi ^2 \left(\xi ^2 \left(-\frac{7}{m_{c_Z}^3}-\frac{1}{m_{\chi^0}^3}\right)+\frac{32 \xi }{m_{c_Z} m_Z (m_{c_Z}+m_Z)}-\frac{10}{m_Z^3}\right) \Bigg]\;.
\end{eqnarray}

At leading order in $\lambda\sim g^3$ (one can  set $\lambda,c_6 \to 0,m_{\chi^\pm} \to m_{c_W}, m_{\chi^0} \to m_{c_Z}$)
\begin{equation}
    Z^{3d}_{\mathrm{NLO}}= -\frac{11 \left(\sqrt{g^2+g^{'2}}+2 g\right)}{48 \pi  \phi }\;.
\end{equation}

\subsection{The contribution of $A^a_0$ and $B_0$ at soft scale}
\begin{figure}[h!]
    \centering
    \includegraphics[width=0.5\linewidth]{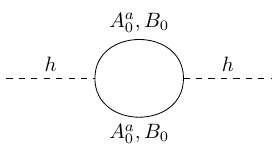}
    \caption{Contributions of $A^a_0$ and $B_0$ to the Z factor.}
    \label{fig:Z_A0B0}
\end{figure}

At soft scale, the contributions of $A^a_0,B_0$ will similarly be included in the calculation of the Z-factor. The corresponding figure is shown in Figure~\ref{fig:Z_A0B0}.

That contributions are
\begin{equation}
\begin{aligned}
-\Pi^{(A^a_0,B_0)}_{hh}&=6 (\phi h_1)^2 I_{SS}(m_L,m_L)+2 (\phi h_2)^2 I_{SS}(m'_L,m'_L)+\frac{1}{2}(\phi h_3)^2 I_{SS}(m_L,m'_L)\;,
\end{aligned}
\end{equation}
where
\begin{equation}
\begin{aligned}
     I_{SS}(m,M)&=\int_p \frac{1}{(p^2+m^2)[(p+k)^2+M^2]}\\
     &=\frac{1}{4\pi(m+M)}+k^2\lp -\frac{1}{12 \pi(m+M)^3}\rp+\mathcal{O}(k^4)\;,
\end{aligned}
\end{equation}

we have
\begin{equation}
    Z^{(A^a_0,B_0)}_{1,3d}=\frac{\partial \Pi^{(A^a_0,B_0)}_{hh}}{\partial k^2}=\frac{h_1^2 \phi ^2}{16 \pi  m_L^3}+\frac{h_2^2 \phi ^2}{48 \pi  m_L^{\prime 3}}+\frac{h_3^2 \phi ^2}{24 \pi  (m_L+m^\prime_L)^3}\;.
\end{equation}

\section{The calculation of two-loop effective potebtial in 3d approach}
\label{two-loop effective}
We define
\begin{equation}
	\int_{p} \equiv \mu^{2 \epsilon}\int \frac{\diff^d p}{(2 \pi)^d}\;,
\end{equation}
Where $\mu$ is the regularization scale, which has the $\MSbar$ renormalization scale $\bar{\mu}$,
\begin{equation}
	4 \pi \mu^2=e^\gammaE \bar{\mu}^2\;,
\end{equation}
for $d=3-2\epsilon, D=d+1=4-2\epsilon$.
In 3d framework,
\begin{equation}
	A(m)=\int_p\frac{1}{p^2+m^2}=\frac{m}{4 \pi}\;,
\end{equation}
\begin{equation}
	\begin{aligned}
		H(m_1,m_2,m_3)&=\int_{p,q}\frac{1}{(p^2+m_1^2)(q^2+m_2^2)[(p+q)^2+m_3^2]}\,\\
		&=\frac{1}{(4 \pi)^2}\lp \frac{1}{4 \epsilon}+\ln\lp\frac{\bmu}{m_1+m_2+m_3}\rp+\frac{1}{2}\rp\;.
	\end{aligned}
\end{equation}
In 4d framework,
\begin{equation}
	A(m)=\Tint{p} \frac{1}{p^2+m^2}=\frac{T^2}{12}-\frac{m T}{4 \pi}-\frac{2 m^2}{(4 \pi)^2}\ln \lp\frac{\bmu e^\gammaE}{4 \pi T} \rp\;,
\end{equation}

\begin{equation}
	\begin{aligned}
		H(m_1,m_2,m_3)&=\Tint{p,q}\frac{1}{(p^2+m_1^2)(q^2+m_2^2)[(p+q)^2+m_3^2]}\,\\
		&=\frac{T^2}{(4 \pi)^2}\lp \frac{1}{4 \epsilon}+\ln\lp\frac{\bmu}{m_1+m_2+m_3}\rp+\frac{1}{2}\rp\;.
	\end{aligned}
\end{equation}
In multi-loop calculations, we used the integration-by-parts (IBP) reduction algorithm with FIRE6\cite{Nishimura:2012ee,Smirnov:2019qkx}, transforming their characteristic integrals into functions of $H$ and $A$.

\begin{figure}[h!]
	\centering
	\includegraphics[width=0.8\linewidth]{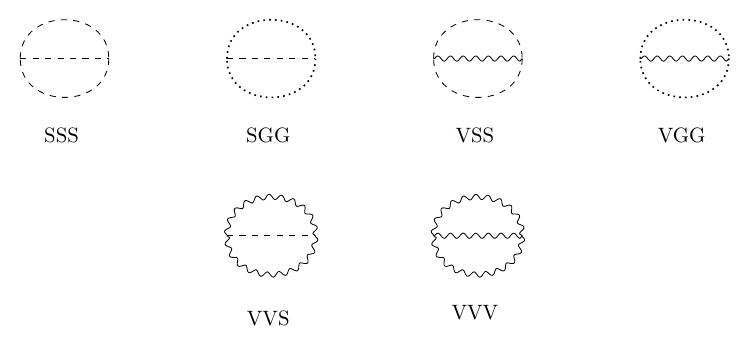}
	\caption{Sunset contributions to the effective potential.}
	\label{fig:Sunset}
\end{figure}

\begin{figure}[h!]
	\centering
	\includegraphics[width=0.8\linewidth]{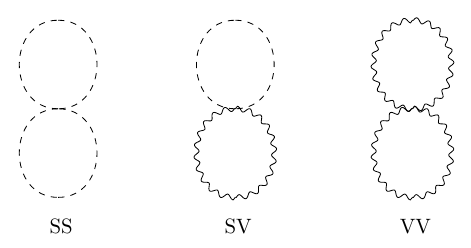}
	\caption{Double bubble contributions to the effective potential.}
	\label{fig:double bubble}
\end{figure}

The two-loop contribution to effective potential is obtained from the digrams in Figure~\ref{fig:Sunset} and Figure~\ref{fig:double bubble}:
\begin{equation}
    V_2=-\Big((SSS)+(SGG)+(VSS)+(VGG)+(VVS)+(VVV)+(SS)+(SV)+(VV) \Big).
\end{equation}

\subsection{Feynman diagram---sunset}

For SSS,
we have
\begin{equation}
\begin{aligned}
(\mathrm{SSS})&=\frac{1}{12} C_{h h h}^2 \mathcal{D}_{\mathrm{SSS}}(m_h, m_h, m_h)+\frac{1}{4} C_{h \chi \chi}^2 \mathcal{D}_{\mathrm{SSS}}(m_h, m_{\chi^0}, m_{\chi^0}) \\
&+\frac{1}{2} C_{h \chi^{+} \chi^{-}}^2 \mathcal{D}_{\mathrm{SSS}}\left(m_h, m_{\chi^\pm}, m_{\chi^\pm}\right),
\end{aligned}
\end{equation}
where
\be
\begin{aligned}
    \mathcal{D}_{\mathrm{SSS}}=\int_{p,q}\frac{1}{(p^2+m_1^2)(q^2+m_2^2)[(p+q)^2+m_3^2]}=H(m_1,m_2,m_3)\;.
\end{aligned}
\ee
For SGG,
we have
\be
\begin{aligned}
(\mathrm{SGG})&=\frac{1}{2}C_{h c_z \bar{c}_Z}^2 \mathcal{D}_{SGG}(m_h,m_{c_Z},m_{c_Z})+\frac{1}{2}C_{h c^+ \bar{c}^-}^2 \mathcal{D}_{SGG}(m_h,m_{c_W},m_{c_W})\,\\
&+\frac{1}{2}C_{h c^- \bar{c}^+}^2 \mathcal{D}_{SGG}(m_h,m_{c_W},m_{c_W})+\frac{1}{2}C_{\chi c^+ \bar{c}^-}^2 \mathcal{D}_{SGG}(m_{\chi^0},m_{c_W},m_{c_W})\,\\
&+\frac{1}{2}C_{\chi c^- \bar{c}^+}^2 \mathcal{D}_{SGG}(m_{\chi^0},m_{c_W},m_{c_W})-C_{\chi^+ c^- \bar{c}_Z} C_{\chi^- c_Z \bar{c}^+}\mathcal{D}_{SGG}(m_{\chi^\pm},m_{c_W},m_{c_Z})\,\\
&-C_{\chi^+ c_Z \bar{c}^-} C_{\chi^- c^+ \bar{c}_Z}\mathcal{D}_{SGG}(m_{\chi^\pm},m_{c_W},m_{c_Z})
-C_{\chi^+ c^- \bar{c}_A} C_{\chi^- c_A \bar{c}^+}\mathcal{D}_{SGG}(m_{\chi^\pm},m_{c_W},m_{c_A})\\
&-C_{\chi^+ c_A \bar{c}^-} C_{\chi^- c^+ \bar{c}_A}\mathcal{D}_{SGG}(m_{\chi^\pm},m_{c_W},m_{c_A})\,
,
\end{aligned}
\ee
where
\begin{equation}
   \mathcal{D}_{SGG}(m_1,m_2,m_3)=-\mathcal{D}_{SSS}(m_1,m_2,m_3)=-H(m_1,m_2,m_3)\;.
\end{equation}

For VSS, we have
\begin{equation}
\begin{aligned}
(\mathrm{VSS})= & -\frac{1}{2} C_{h \chi Z}^2 \mathcal{D}_{\mathrm{VSS}}(m_Z,m_h, m_{\chi^0}) \\
& +\frac{1}{2} C_{\chi^{+} \chi^{-} Z}^2 \mathcal{D}_{\mathrm{VSS}}(m_Z, m_{\chi^\pm}, m_{\chi^\pm})+\frac{1}{2} C_{\chi^{+} \chi^{-} A}^2 \mathcal{D}_{\mathrm{VSS}}(0, m_{\chi^\pm}, m_{\chi^\pm}) \\
& -C_{h \chi^{+} W^{-}} \times C_{h \chi^{-} \chi^{+}} \mathcal{D}_{\mathrm{VSS}}(m_W, m_h, m_{\chi^\pm}) \\
& -C_{\chi \chi^{+} W^{-}} \times C_{\chi \chi^{-} W^{+}} \mathcal{D}_{\mathrm{VSS}}(m_W, m_{\chi^0}, m_{\chi^\pm}),
\end{aligned}
\end{equation}
its characteristic integral function is
\begin{equation}
    \mathcal{D}_{\mathrm{VSS}}(m_1, m_2, m_3)=\int_{p,q} \frac{(2 p_i+q_i)(2 p_j+q_j)\lp \delta_{i j}-(1-\xi)\frac{q_i q_j}{q^2+\xi m_1^2}\rp}{(p^2+m_2^2)(q^2+m_1^2)[(p+q)^2+m_3^2]}\;,
\end{equation}

\be
\begin{aligned}\label{VSS}
   \mathcal{D}_{\mathrm{VSS}}(m_1, m_2, m_3)&=-\frac{\left(m_2^2-m_3^2\right)^2 }{m_1^2}H\lp m_2,m_1 \sqrt{\xi },m_3\rp\,\\
   &+\frac{m_2^4-2 m_2^2 \left(m_1^2+m_3^2\right)+\left(m_1^2-m_3^2\right)^2 }{m_1^2}H(m_2,m_1,m_3)\,\\
   &+\frac{-m_2^2+m_1^2 \xi +m_3^2}{m_1^2}A(m_2)A\lp m_1 \sqrt{\xi }\rp+\frac{m_2^2+m_1^2-m_3^2}{m_1^2}A(m_2)A(m_1)\\
   &-A(m_2)A(m_3)+\frac{-m_2^2+m_1^2+m_3^2}{m_1^2}A(m_1) A(m_3)\,\\
   &+\frac{m_2^2+m_1^2 \xi -m_3^2}{m_1^2}A\lp m_1 \sqrt{\xi }\rp A(m_3)\;,
\end{aligned}
\ee
for $m_1=0$, Eq.~(\ref{VSS}) then transforms to
\begin{equation}
	\mathcal{D}_{\mathrm{VSS}}(0, m_2, m_3)=(d (\xi -1)-3 \xi +1) \left(m_2^2+m_3^2\right) H(m_2,0,m_3)+ (d (\xi -1)-2 \xi +1)A(m_2) A(m_3)\;.
\end{equation}

For VGG,we have
\be
\begin{aligned}
(\mathrm{VGG})= & -C_{W^{+} \bar{c}^{-} c_{Z}} \times C_{W^{-} \bar{c}_{Z} c^{+}} \mathcal{D}_\mathrm{VGG}(m_W,m_{c_W},m_{c_Z}) \\
& -C_{W^{+} \bar{c}_{Z} c^{-}} \times C_{W^{-} c^{+} c_{Z}} \mathcal{D}_\mathrm{VGG}(m_W,m_{c_W},m_{c_Z}) \\
& -C_{W^{+} \bar{c}^{-} c_{A}} \times C_{W^{-} \bar{c}_{A} c^{c}} \mathcal{D}_\mathrm{VGG}(m_W,m_{c_W},m_{c_A}) \\
& -C_{W^{+} \bar{c}_{A} c^{-}} \times C_{W^{-} \bar{c}^{+} c_{A}} \mathcal{D}_\mathrm{VGG}(m_W,m_{c_W},m_{c_A}) \\
& -\frac{1}{2} C_{Z \bar{c}^{+} c^{-}}^{2} \mathcal{D}_\mathrm{VGG}(m_Z,m_{c_W},m_{c_W})-\frac{1}{2} C_{Z \bar{c}^{-} c^{+}}^{2} \mathcal{D}_\mathrm{VGG}(m_Z,m_{c_W},m_{c_W})\\
&-\frac{1}{2} C_{A \bar{c}^{+} c^{-}}^{2}\mathcal{D}_\mathrm{VGG}(0,m_{c_W},m_{c_W})-\frac{1}{2} C_{A \bar{c}^{-} c^{+}}^{2}\mathcal{D}_\mathrm{VGG}(0,m_{c_W},m_{c_W}),
\end{aligned}
\ee
the characteristic integral function is
\be
\begin{aligned}\label{eq:VGG}
    \mathcal{D}_{\mathrm{VGG}}(m_1, m_2, m_3)=\int_{p,q}\frac{p_i (p+q)_j \lp \delta_{i j}-(1-\xi)\frac{q_i q_j}{q^2+\xi m_1^2}  \rp}{(p^2+m_2^2)(q^2+m_1^2)[(p+q)^2+m_3^2]}\;,
\end{aligned}
\ee

\begin{align}
    \mathcal{D}_{\mathrm{VGG}}(m_1, m_2, m_3)&=\frac{m_1^4 \xi ^2-\left(m_2^2-m_3^2\right)^2}{4 m_1^2}H\left(m_2,m_1 \sqrt{\xi },m_3\right)\nn\\
    &+\frac{m_2^4-2 m_2^2 \left(m_1^2+m_3^2\right)+\left(m_1^2-m_3^2\right)^2 }{4 m_1^2}H(m_2,m_1,m_3)\nn\\
    &+\frac{\lp -m_2^2+m_1^2 \xi +m_3^2 \rp}{4 m_1^2}A(m_2) A\lp m_1 \sqrt{\xi }\rp +\frac{m_2^2+m_1^2-m_3^2}{4 m_1^2}A(m_2)A(m_1)\nn\\
    &-\frac{\xi+1}{4}A(m_2)A(m_3)+\frac{m_2^2+m_1^2 \xi -m_3^2}{4 m_1^2}A\left(m_1 \sqrt{\xi }\right)A(m_3)\nn\\
    &+\frac{-m_2^2+m_1^2+m_3^2}{4 m_1^2}A(m_1) A(m_3)\;.
\end{align}
For $m_1=0$, Eq.~\eqref{eq:VGG} becomes
\begin{equation}
\mathcal{D}_{\mathrm{VGG}}(0,m_2,m_3)=\frac{1+d(\xi-1)-3\xi}{4}\Big[A(m_2)A(m_3)+(m_2^2+m_3^2)H(0,m_2,m_3)\Big]\,.
\end{equation}
For VVS, we have
\begin{equation}
\begin{aligned}
(\mathrm{VVS})= & \frac{1}{4} C_{Z Z h}^2 \mathcal{D}_{\mathrm{VVS}}(m_Z, m_Z, m_h)+\frac{1}{2} C_{W^{+} W^{-} h}^2 \mathcal{D}_{\mathrm{VVS}}(m_W, m_W, m_h) \\
& +C_{W^{-} Z \chi^{+}} \times C_{W^{+} Z \chi^{-}} \mathcal{D}_{\mathrm{VVS}}(m_W, m_Z, m_{\chi^\pm}) \\
& +C_{W^{-} A \chi^{+}} \times C_{W^{+} A \chi^{-}} \mathcal{D}_{\mathrm{VVS}}(m_W, 0, m_{\chi^\pm}),
\end{aligned}
\end{equation}

its characteristic integral function is
\begin{equation}
\mathcal{D}_{\mathrm{VVS}}(m_1, m_2, m_3)=\int_{p,q} \frac{\lp \delta_{i j}-(1-\xi)\frac{p_i p_j}{p^2+\xi m_1^2}\rp \lp \delta_{i j}-(1-\xi)\frac{q_i q_j}{q^2+\xi m_2^2}\rp}{(p^2+m_1^2)(q^2+m_2^2)[(p+q)^2+m_3^2]}\;,
\end{equation}

\begin{align}
    \mathcal{D}_{\mathrm{VVS}}(m_1,m_2,m_3)&=\left(d+\frac{\left(m_1^2-m_3^2+m_2^2\right)^2}{4 m_1^2 m_2^2}-2\right) H(m_1,m_2,m_3)\nn\\
    &+\frac{\left(m_1^2 \xi -m_3^2+m_2^2 \xi \right)^2 }{4 m_1^2 m_2^2}H\left(m_1 \sqrt{\xi },m_2 \sqrt{\xi },m_3\right)\nn\\
    &+\left(\xi -\frac{\left(m_1^2-m_3^2+m_2^2 \xi \right)^2}{4 m_1^2 m_2^2}\right) H\left(m_1,m_2 \sqrt{\xi },m_3\right)\nn\\
    &+\left(\xi -\frac{\left(m_1^2 \xi -m_3^2+m_2^2\right)^2}{4 m_1^2 m_2^2}\right) H\left(m_1 \sqrt{\xi },m_2,m_3\right)\nn\\
    &+\frac{\left(m_1^2 \xi -m_3^2+m_2^2 \xi \right)}{4 m_1^2 m_2^2}A\left(m_1 \sqrt{\xi }\right) A\left(m_2 \sqrt{\xi }\right) \nn\\
    &+\frac{(\xi-1)m_2^2}{4 m_1^2 m_2^2}A(m_1) A(m_3)-\frac{m_1^2-m_3^2+m_2^2 \xi}{4 m_1^2 m_2^2} A(m_1) A\lp m_2 \sqrt{\xi}\rp  \nn\\
    &+\frac{m_1^2-m_3^2+m_2^2}{4 m_1^2 m_2^2}A(m_1) A(m_2)+\frac{(\xi-1) m_1^2}{4 m_1^2 m_2^2}A(m_2) A(m_3)\nn\\
    &+\frac{(1-\xi)m_2^2}{4 m_1^2 m_2^2}A\lp m_1 \sqrt{\xi}\rp A(m_3)+\frac{(1-\xi)m_1^2}{4 m_1^2 m_2^2} A(m_3) A\lp m_2 \sqrt{\xi}\rp\nn\\
    &-\frac{\left(m_1^2 \xi -m_3^2+m_2^2\right)}{4 m_1^2 m_2^2} A\left(m_1 \sqrt{\xi }\right) A(m_2)\;.
\end{align}
Specially, if $m_2=0$,we have
\begin{equation}
\mathcal{D}_{\mathrm{VVS}}(m_1, 0, m_3)=\int_{p,q} \frac{\lp \delta_{i j}-(1-\xi)\frac{p_i p_j}{p^2+\xi m_1^2}\rp \lp \delta_{i j}-(1-\xi)\frac{q_i q_j}{q^2}\rp}{(p^2+m_1^2) (q^2) [(p+q)^2+m_3^2]}\;,
\end{equation}

\begin{equation}
	\begin{aligned}
		\mathcal{D}_{\mathrm{VVS}}(m_1, 0, m_3)&=\frac{(d-1) \left(m_1^2 (\xi +3)+m_3^2 (\xi -1)\right) }{4 m_1^2}H(m_1,0,m_3)\,\\
		&+\frac{\left(m_1^2 \xi  (d-d \xi+5 \xi -1)+m_3^2 (d-1)(1-\xi)\right)}{4 m_1^2} H\left(m_1 \sqrt{\xi },0,m_3\right)\,\\
		&+\frac{(d-1) (\xi -1) }{4 m_1^2}A(m_1) A(m_3)-\frac{(d-1) (\xi -1)  }{4 m_1^2}A\left(m_1 \sqrt{\xi }\right)A(m_3)\;.
	\end{aligned}
\end{equation}

For VVV, we have

\begin{equation}
(\mathrm{VVV})=\frac{1}{2} C_{W^{+} W^{-} Z}^2 \mathcal{D}_{\mathrm{VVV}}(m_W, m_W, m_Z)+\frac{1}{2} C_{W^{+} W^{-} A}^2 \mathcal{D}_{\mathrm{VVV}}(m_W, m_W, 0)\;,
\end{equation}

the characteristic integral function is
\be
\begin{aligned}
    \mathcal{D}_{\mathrm{VVV}}(m_1, m_2, m_3)&=\int_{p,q} \frac{1}{(p^2+m_1^2)(q^2+m_2^2)[(p+q)^2+m_3^2]}\lp g_{\mu \nu}-(1-\xi)\frac{p_\mu p_\nu}{p^2+\xi m_1^2}\rp\\
    &\times \lp g_{\sigma \rho}-(1-\xi)\frac{q_\sigma q_\rho}{q^2+\xi m_2^2}\rp \lp g_{r s}-(1-\xi)\frac{(p+q)_r (p+q)_s}{(p+q)^2+\xi m_3^2}\rp\,\\
    &\times \lp g^{\mu \sigma} (q-p)^r-g^{\sigma r}(2q+p)^\mu +g^{\mu r} (2p+q)^{\sigma}  \rp\,\\
    &\times \lp g^{\rho \nu} (q-p)^s+g^{\nu s}(2p+q)^\rho -g^{\rho s} (2q+p)^{\nu} \rp\;.
\end{aligned}
\ee

Since the result of $\mathcal{D}_{\mathrm{VVV}}(m_1, m_2, m_3)$ is lengthy, we directly take the specific quality relationship, We consider two scenarios:

(a) $m_1=m_2=m, m_3=M$:
\begin{align}
&\mathcal{D}_{\mathrm{VVV}}(m, m, M)\nn\\
&=-\frac{\left(m^2-M^2\right)^2 \left(2 m^2 M^2 (2 d-\xi -3)+m^4 (\xi -1)^2+M^4\right) }{4 m^4 M^2}H\left(m,m \sqrt{\xi },M\right)\nn\\
&-\frac{\left(m^2-M^2\right)^2 \left(2 m^2 M^2 (2 d-\xi -3)+m^4 (\xi -1)^2+M^4\right) }{4 m^4 M^2}H\left(m \sqrt{\xi },m,M\right)\nn\\
&-\frac{(2 m-M) (2 m+M) \left(4 (d-1) m^4+4 (2 d-3) m^2 M^2+M^4\right) }{4 m^4}H(m,m,M)\nn\\
&+\frac{\left(m^4 (\xi -1)^2-2 m^2 M^2 \xi  (\xi +1)+M^4 \xi ^2\right) }{4 M^2}H\left(m,m \sqrt{\xi },M \sqrt{\xi }\right)\nn\\
&+\frac{\left(m^4 (\xi -1)^2-2 m^2 M^2 \xi  (\xi +1)+M^4 \xi ^2\right) }{4 M^2}H\left(m \sqrt{\xi },m,M \sqrt{\xi }\right)\nn\\
&+\frac{\left(M^6-4 m^2 M^4 \xi \right) }{4 m^4}H\left(m \sqrt{\xi },m \sqrt{\xi },M\right)+A\left(m \sqrt{\xi }\right)^2 \left(\frac{(d-1) \xi ^2}{d}-\frac{M^4}{4 m^4}+\frac{M^2 \xi }{2 m^2}\right)\nn\\
&+\frac{1}{2} \left(4 d \left(\frac{M^2}{m^2}+\xi -1\right)+\frac{4 \xi }{d}+\frac{M^4}{m^4}-\frac{M^2 (\xi +7)}{m^2}-7 \xi +7\right) A(m) A\left(m \sqrt{\xi }\right)\nn\\
&+\frac{1}{2}\left(d \left(-\frac{4 M^2}{m^2}+4 \xi +4\right)+\frac{4 \xi }{d}-\frac{m^2 (\xi -1)}{M^2}-\frac{M^2 (\xi -7)}{m^2}-6 \xi -7\right)A(M) A\left(m \sqrt{\xi }\right)\nn\\
&+\frac{1}{2} \left(\frac{\xi  (4 d \xi +d-4 \xi )}{d}+\frac{m^2 (\xi -1)}{M^2}\right)A\left(m \sqrt{\xi }\right) A\left(M \sqrt{\xi }\right) \nn\\
&+\frac{1}{2} \left(\frac{4 d M^2}{m^2}-\frac{4}{d}+\frac{m^2 (\xi -1)}{M^2}+\frac{M^2 (\xi -7)}{m^2}-2 \xi +5\right) A(m) A(M)\nn\\
&+\frac{1}{2}  \left(\left(4 d+\frac{4}{d}-7\right) \xi -\frac{m^2 (\xi -1)}{M^2}\right)A(m) A\left(M \sqrt{\xi }\right)\nn\\
&+\left(d \left(3-\frac{2 M^2}{m^2}\right)-\frac{1}{d}-\frac{M^4}{4 m^4}+\frac{7 M^2}{2 m^2}-4\right)A(m)^2
\end{align}

(b) $m_1=m_2=m, m_3=0$:
\begin{align}
&\mathcal{D}_{\mathrm{VVV}}(m, m, 0)\nn\\
=&-\frac{1}{2} m^2 \left((d-3) \xi ^2+3 d+4 \xi -5\right) H\left(m,m \sqrt{\xi },0\right)\nn\\
&+2 (d-1) m^2 (d (\xi -1)-3 \xi +1) H(m,m,0)+\frac{(d-1) \xi ^2}{d} A\left(m \sqrt{\xi }\right)^2\nn\\
&+\left(d (d (\xi -1)-3 \xi +6)-\frac{1}{d}+2 (\xi -3)\right)A(m)^2 \nn\\
&+\frac{1}{2} \left(3 d (\xi -1)+\frac{4 \xi }{d}-5 \xi +5\right) A(m) A\left(m \sqrt{\xi }\right)
\end{align}

\subsection{Feynman diagram---double bubble}

For SS,
we have
\begin{equation}
\begin{aligned}
(\mathrm{SS})&=\frac{1}{8} C_{h h h h}\mathcal{D}_\mathrm{SS}(m_h,m_h)+\frac{1}{8} C_{\chi \chi \chi \chi}\mathcal{D}_\mathrm{SS}(m_{\chi^0},m_{\chi^0}) \\
&+\frac{1}{4} C_{h h \chi \chi} \mathcal{D}_\mathrm{SS}(m_h,m_{\chi^0})+\frac{1}{2} C_{h h \chi^{+} \chi^{-}} \mathcal{D}_\mathrm{SS}(m_h,m_{\chi^\pm}) \\
&+\frac{1}{2} C_{\chi \chi \chi^{+} \chi^{-}}\mathcal{D}_\mathrm{SS}(m_{\chi^0},m_{\chi^\pm})+\frac{1}{2} C_{\chi^{+} \chi^{-} \chi^{+} \chi^{-}}\mathcal{D}_\mathrm{SS}(m_{\chi^\pm},m_{\chi^\pm}),
\end{aligned}
\end{equation}
where
\be
\mathcal{D}_{\mathrm{SS}}(m_1, m_2)=\int_{p,q}\frac{1}{(p^2+m_1^2)(q^2+m_2^2)}=A(m_1) A(m_2)\,.
\ee

For SV,
we have
\be
\begin{aligned}
    (\mathrm{SV})&=\frac{1}{4} C_{Z Z h h} \mathcal{D}_{\mathrm{SV}}(m_h, m_Z)+\frac{1}{4} C_{Z Z \chi \chi} \mathcal{D}_{\mathrm{SV}}(m_{\chi^0}, m_Z)  \\
    & +\frac{1}{2} C_{W^{+} W^{-} h h} \mathcal{D}_{\mathrm{SV}}(m_h, m_W)+\frac{1}{2} C_{W^{+} W^{-} \chi \chi} \mathcal{D}_{\mathrm{SV}}(m_{\chi^0}, m_W) \\
    & +\frac{1}{2} C_{Z Z \chi^{+} \chi^{-}} \mathcal{D}_{\mathrm{SV}}(m_{\chi^\pm}, m_Z)+C_{W^{+} W^{-} \chi^{+} \chi^{-}} \mathcal{D}_{\mathrm{SV}}(m_{\chi^\pm}, m_W),
\end{aligned}
\ee
the characteristic integral function is
\begin{equation}
    \mathcal{D}_{\mathrm{SV}}(m_1, m_2)=\int_{p,q}\frac{\delta_{i j}}{(p^2+m_1^2)(q^2+m_2^2)}\lp \delta_{i j}-(1-\xi)\frac{q_i q_j}{q^2+\xi m_2^2}\rp\;,
\end{equation}
\be
\mathcal{D}_{\mathrm{SV}}(m_1, m_2)=(d-1) A(m_1) A(m_2)+\xi A(m_1) A( m_2 \sqrt{\xi})\;.
\ee

For VV,
we have
\begin{equation}
(\mathrm{VV})=\frac{1}{2} C_{W^{+} W^{-} W^{+} W^{-}} \mathcal{D}_{\mathrm{VV}}(m_W, m_W)+C_{W^{+} W^{-} \mathrm{ZZ}} \mathcal{D}_{\mathrm{VV}}(m_W, m_Z)\;,
\end{equation}
its characteristic integral function is
\be
\begin{aligned}
    \mathcal{D}_{\mathrm{VV}}(m_1, m_2)=&\int_{p,q} \frac{g^{\mu \sigma}g^{\nu \rho}+g^{\mu \nu}g^{\sigma \rho}-2g^{\mu \rho}g^{\nu \sigma}}{(p^2+m_1^2)(q^2+m_2^2)}\lp g_{\mu \nu}-(1-\xi)\frac{p_\mu p_\nu}{p^2+\xi m_1^2}  \rp\,\\
    &\lp g_{\sigma \rho}-(1-\xi)\frac{q_\sigma q_\rho}{q^2+\xi m_2^2}\rp\,,
\end{aligned}
\ee
\be
\begin{aligned}
    \mathcal{D}_{\mathrm{VV}}(m_1,m_2)&=\frac{(d-1)^3}{d}A(m_1) A(m_2)+\xi\frac{(d-1)^2}{d}A( m_1 \sqrt{\xi}) A( m_2)\,\\
    &+\xi\frac{(d-1)^2}{d}A(m_1) A( m_2 \sqrt{\xi})+\xi^2(1-\frac{1}{d})A(m_1 \sqrt{\xi})A(m_2 \sqrt{\xi})\,.
\end{aligned}
\ee

\subsection{Vertex coefficients}
In the above, $C_{\psi_i \psi_j}$ denote vertex coefficients for generic fields $\psi_i$ and are defined as minus the respective coefficient in the Lagrangian, including the combinatorial factors arising from contractions. For momentum-dependent vertices the momentum is absorbed inside the integral definition. An exhaustive list of required vertex coefficients read
\be C_{h h h h}=-(6 \lambda+45 c_6 \phi^2) ,\ee
\be C_{\chi \chi \chi \chi}=-(6 \lambda+9 c_6 \phi^2) ,\ee
\be C_{h h \chi \chi}=C_{h h \chi^+ \chi^-}=-(2 \lambda+9 c_6 \phi^2) ,\ee
\be C_{\chi \chi \chi^+ \chi^-}=-(2\lambda+3 c_6 \phi^2) ,\ee
\be C_{\chi^+ \chi^- \chi^+ \chi^-}=-(4 \lambda+6 c_6 \phi^2) ,\ee
\be C_{Z Z h h}=-\frac{1}{2}(g^2+g'^2),\ee
\be C_{Z Z \chi \chi}=-\frac{1}{2}(g^2+g^{\prime 2}) ,\ee
\be C_{W^+ W^- h h}=-\frac{1}{2}g^2 ,\ee
\be C_{W^+ W^- \chi \chi}=C_{W^+ W^- \chi^+ \chi^-}=-\frac{1}{2}g^2 ,\ee
\be C_{Z Z \chi^+ \chi^-}=-\frac{1}{2}\frac{(g^2-g^{\prime 2})^2}{g^2+g^{\prime 2}}, \ee
\be C_{W^+ W^- W^+ W^-}=-g^2 ,\ee
\be C_{W^+ W^- Z Z}=-\frac{g^4}{g^2+g^{\prime 2}},\ee
\be C_{h h h}=-(6 \lambda \phi+15 c_6 \phi^3) ,\ee
\be C_{h \chi \chi}=C_{h \chi^+ \chi^-}=-(2 \lambda \phi+3 c_6 \phi^3) ,\ee
\be C_{Z h \chi}=-\frac{i}{2}\sqrt{g^2+g^{\prime 2}},\ee
\be C_{Z \chi \chi \chi^-}=-\frac{1}{2}\frac{g^2-g^{\prime 2}}{\sqrt{g^2+g^{\prime 2}}},\ee
\be C_{A \chi^+ \chi^-}=-\frac{g g^{\prime}}{\sqrt{g^2+g^{\prime 2}}} ,\ee
\be C_{W^- h \chi^+}=-C_{W^+ h \chi^-}=\frac{1}{2} g ,\ee
\be C_{W^- \chi \chi^+}=-C_{W^+ \chi \chi^-}=-\frac{i}{2} g ,\ee
\be C_{Z Z h}=-\frac{1}{2}(g^2+g^{\prime 2})\phi ,\ee
\be C_{W^+ W^- h}=-\frac{1}{2}g^2\phi ,\ee
\be C_{W^- Z \chi^+}=C_{W^+ Z \chi^-}=\frac{\phi}{2}\frac{g g^{\prime 2}}{\sqrt{g^2+g^{\prime 2}}},\ee
\be C_{W^- A \chi^+}=C_{W^+ A \chi^-}=-\frac{\phi}{2}\frac{g^2 g^{\prime}}{\sqrt{g^2+g^{\prime 2}}},\ee
\be C_{W^+ W^- Z}=\frac{g^2}{\sqrt{g^2+g^{\prime 2}}},\ee
\be C_{W^+ W^- A}=\frac{g g^{\prime}}{\sqrt{g^2+g^{\prime 2}}},\ee

\be C_{h c_z \bar{c}_Z}=-\frac{g}{2 \cos\theta} m_Z \xi ,\ee
\be C_{h c^+ \bar{c}^-}=C_{h c^- \bar{c}^+}=-\frac{1}{2}g m_W \xi,\ee
\be C_{\chi c^+ \bar{c}^-}=C_{\chi c^- \bar{c}^+}=\frac{i}{2}g m_W \xi ,\ee
\be C_{\chi^+ c^- \bar{c}_Z}=C_{\chi^- c^+ \bar{c}_Z}=\frac{1}{2}g m_Z \xi ,\ee
\be C_{\chi^- c_Z \bar{c}^+}=C_{\chi^+ c_Z \bar{c}^-}=-\frac{g \cos2\theta}{2\cos\theta}m_W \xi ,\ee
\be C_{\chi^- c_A \bar{c}^+}=C_{\chi^+ c_A \bar{c}^-}=-\frac{g g^\prime}{\sqrt{g^2+g^{\prime 2}}}\,, \ee
\be C_{W^+ \bar{c}^- c_Z}=C_{W^- \bar{c}_Z c^+}=-C_{W^+ \bar{c}_Z c^-}=-C_{W^- \bar{c}^+ c_Z}=-\frac{g^2}{\sqrt{g^2+g^{\prime 2}}},\ee
\be C_{W^+ \bar{c}^- c_A}=C_{W^- \bar{c}_A c^+}=-C_{W^+ \bar{c}_A c^-}=-C_{W^- \bar{c}^+ c_A}=-\frac{g g'}{\sqrt{g^2+g^{\prime 2}}},\ee
\be C_{Z \bar{c}^+ c^-}=-C_{Z \bar{c}^- c^+}=-\frac{g^2}{\sqrt{g^2+g^{\prime 2}}}.\ee

\be C_{A \bar{c}^+ c^-}=-C_{A \bar{c}^- c^+}=-\frac{g g^\prime}{\sqrt{g^2+g^{\prime 2}}}.\ee



\subsection{The contributions of $A^a_0$ and $B_0$ at soft level}

\begin{figure}[!htp]
    \centering
    \includegraphics[width=0.8\linewidth]{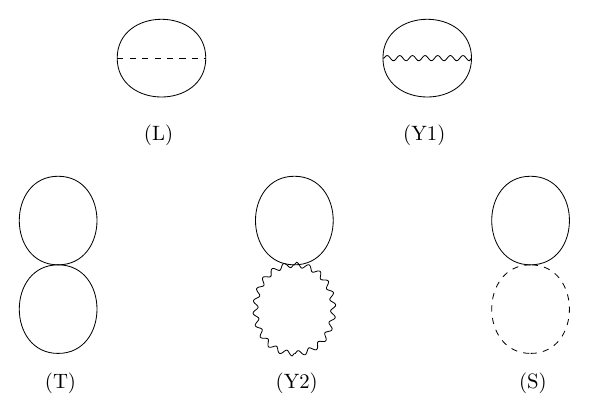}
    \caption{$A^a_0,B_0$ contributing to the potential at the soft scale. Solid line denote $A^a_0,B_0$, dashed line denote scalar, wave line denote gauge boson.}
    \label{fig:V2A0B0}
\end{figure}

At the soft scale,we incorporate the contributions from $A^a_0$ and $B_0$ in the computation of the one-loop effective potential, which is given by:
\begin{equation}
    V^{(A^a_0,B_0)}_{1,3d}=-\frac{1}{12 \pi}\lp 3 m_L^3+m_L^{\prime 3} \rp\;,
\end{equation}
Additionally, we need to calculate the two-loop potential correction requires additionally considering the contributions from $A^a_0,B_0$, see Figure~\ref{fig:V2A0B0}.
The contributions are
\begin{equation}
    V^{(A^a_0,B_0)}_{2,3d}=\lp (T)+(Y1)+(Y2)+(S)+(L)\rp,
\end{equation}
we have
\begin{align}
(T)&=\frac{15}{4}\kappa_1 D_{SS}(m_L,m_L)+\frac{3}{4}\kappa_2 D_{SS}(m'_L,m'_L)+\frac{3}{4}\kappa_3 D_{SS}(m_L,m'_L)\;,\\
(Y1)&=-\frac{3}{2}g^2 D_{SSV}(m_L,m_L,m_T)\;,\\
(Y2)&=3g^2D_{SV}(m_L,m_T)\;,\\
(S)&=\frac{3}{2}h_1 D_{SS}(m_L,m_h)+\frac{6}{2} h_1 D_{SS}(m_L,m_{\chi^{1,2}})+\frac{3}{2} h_1 D_{SS}(m_L,m_{\chi^3})\notag\\
   &+\frac{1}{2} h_2 D_{SS}(m'_L,m_h)+\frac{2}{2} h_2 D_{SS}(m'_L,m_{\chi^{1,2}})+\frac{1}{2} h_2 D_{SS}(m'_L,m_{\chi^3})\\
(L)&=3 (\phi h_1)^2 D_{SSS}(m_L,m_L,m_h)+(\phi h_2)^2 D_{SSS}(m'_L,m'_L,m_h)\notag\;,\\
   &+\frac{1}{2}(\phi h_3)^2 D_{SSS}(m_L,m'_L,m_h)+\frac{2}{2}(\phi h_3)^2 D_{SSS}(m_L,m'_L,m_{\chi^{1,2}})\;,
\end{align}
with related couplings being:
\begin{equation}\label{soft couplings}
\begin{aligned}
&\kappa_1=\frac{5}{3}\frac{g^4}{(4 \pi)^2 T} \;,\quad \kappa_2=-\frac{271}{27}\frac{g^{'4}}{(4 \pi)^2 T} \,,\quad \kappa_3=-\frac{6g^2 g^{'2}}{(4 \pi)^2 T} ,\,\\
  &  h_1=\frac{1}{4}g^2\;,\quad h_2=\frac{1}{4}g^{'2}\;,\quad h_3=\frac{1}{2}g g'\;.
    \end{aligned}
\end{equation}
Related soft scale masses are:
\begin{equation}
\begin{aligned}\label{soft masses}
&m_{\chi^{1,2}}^2=m_{\chi^\pm}^2\,,\quad m_{\chi^3}^2=m_{\chi^0}^2\;,\\
&m_T=m_W=\frac{1}{2}g \phi\,,\quad m_D^2=\frac{11}{6}g^2 T \,,  \\
&m'_T=m_B=\frac{1}{2}g' \phi\,,\quad m_D^{'2}=\frac{11}{6}g^{'2} T \,, \\
&m_L^2=m_D^2+\frac{1}{4} g^2 \phi^2\,,\quad m_L^{'2}=m_D^{'2}+\frac{1}{4} g^{'2} \phi^2\,.
\end{aligned}
\end{equation}

\section{Nielsen identity derivation}
\label{appendix:Nielsen identity derivation}
We present a compact derivation of this identity, following the method of~\cite{Kobes:1990dc}. However, the usual form of the identity is not quite sufficient for our purposes. Instead, what we need is a series of identities, each of which gives the gauge dependence of one of the functions appearing in the derivative Eq.~(\ref{effective Euclidean action}).

We firstly note that $\mathcal{L}$ of Eq.~(\ref{SMEFT}) is invariant under the BRS transformation,
\begin{equation}
	\begin{aligned}\label{BRST}
		\delta A^a_\mu&=\epsilon D^{ab}_\mu c^b\,,\\
		\delta B_\mu&=\epsilon \partial_\mu c^0\,,\\
		\delta c^a&=-\frac{1}{2}\epsilon g f^{abc}c^b c^c\,,\\
		\delta c^0&=0\,,\\
		\delta \bar{c}^a&=-\epsilon \frac{1}{\xi}(\partial_\mu A^{a \mu}+i g \xi(H^\dagger t^a \Phi_0-\Phi_0^\dagger t^a H)) \,,\\
		\delta \bar{c}^0&=-\epsilon \frac{1}{\xi}(\partial_\mu B^\mu+i \frac{g'}{2} \xi(H^\dagger \Phi_0-\Phi_0^\dagger H)) \,,\\
		\delta H&=\epsilon (i g t^a c^a+i g' \frac12 c^0)H\;,
	\end{aligned}
\end{equation}
where $\epsilon$ is an infinitesimal anticommuting c-number
\begin{equation}
	\{\epsilon,c^a\}=0,\qquad \{\epsilon,c^0\}=0\;.
\end{equation}
In order to facilitate the subsequent calculation and derivation of expressions related to the Nielsen identity, we need to rewrite the gauge fixing items and ghost fields as
\begin{equation}
	\mathcal{L}_\mathrm{g.f}=-\frac{1}{2 \xi}(\partial_\mu A^{a \mu}+g v^a_i \varphi_i)^2-\frac{1}{2 \xi}(\partial_\mu B^\mu+g' v_i \varphi_i)^2\,.
\end{equation}
Here, the conventional choice for $v^a_i$ and $v_i$ would be
\begin{equation}
	v^a_i=i \xi t^a_{ij}\bra{0} \varphi_j\ket{0},\qquad v_i=i \xi n'_{ij} \bra{0}\varphi_j\ket{0}\;,
\end{equation}
where $n'_{ij} = n_{ij}/2$, and $\varphi_i(i=1,2,3,4)$ respectively representing $\phi+h$, $\chi^1$, $\chi^2$, and $\chi^3$.
The ghost Lagrangian turns to be
\begin{equation}
	\mathcal{L}_{\mathrm{ghost}}=-\begin{pmatrix}
		\bar{c}^a & \bar{c}^0
	\end{pmatrix} \begin{pmatrix}
		M^{ab} & M^a\\
		M^b & M
	\end{pmatrix}\begin{pmatrix}
		c^b\\
		c^0
	\end{pmatrix}\;,
\end{equation}
where
\begin{equation}
	\begin{aligned}
		M^{ab}&=(\partial^\mu D_\mu^{ab})+i g^2 v^a_i t^b_{ij}\varphi_j\,,\\
		M^a&=igg' v^a_i n'_{ij}\varphi_j\,,\\
		M^b&=igg' v_i t^a_{ij}\varphi_j\,,\\
		M&=\partial^2+ig^{'2}v_i n'_{ij}\varphi_j\;,
	\end{aligned}
\end{equation}
In the formalim, one can consider a system of scalar fields $\phi_i$ that appear in a Lagrangian under a symmetry group G, represented by the transformation
\begin{equation}
	\phi_i \to (1+i g \alpha^a t^a+i g' \frac12 \beta)_{ij}\phi_j\;,
\end{equation}
where
\begin{equation}
	\delta \phi_i=ig\alpha^at^a_{ij}\phi_j+ig'\frac12\beta n_{ij}\phi_j\;.
\end{equation}
Then the group representation matrices $t^a$ can be write as
\begin{equation}
	t^a_{ij}=iT^a_{ij}\,,\quad n_{ij}=iN_{ij}\;.
\end{equation}
They are given by the following matrix:
\begin{equation}
	\begin{aligned}
		[t^1_{ij}]&=\begin{pmatrix}
			0 & 0 & \frac{i}{2} & 0 \\
			0 & 0 & 0 & \frac{i}{2} \\
			-\frac{i}{2} & 0 & 0 & 0 \\
			0 & -\frac{i}{2} & 0 & 0 \\
		\end{pmatrix}\,,\quad
		[t^2_{ij}]=\begin{pmatrix}
			0 & \frac{i}{2} & 0 & 0 \\
			-\frac{i}{2} & 0 & 0 & 0 \\
			0 & 0 & 0 & -\frac{i}{2} \\
			0 & 0 & \frac{i}{2} & 0 \\
		\end{pmatrix}\,,\\
		[t^3_{ij}]&=\begin{pmatrix}
			0 & 0 & 0 & -\frac{i}{2} \\
			0 & 0 & \frac{i}{2} & 0 \\
			0 & -\frac{i}{2} & 0 & 0 \\
			\frac{i}{2} & 0 & 0 & 0 \\
		\end{pmatrix}\,,\quad
		[n_{ij}]=\begin{pmatrix}
			0 & 0 & 0 & i \\
			0 & 0 & i & 0 \\
			0 & -i & 0 & 0 \\
			-i & 0 & 0 & 0 \\
		\end{pmatrix}\;.
	\end{aligned}
\end{equation}
 When taking the derivative of the effective potential with respect to $\xi$, we follow the method of Ref.~\cite{PhysRevD.9.2904}. Where, the explicit $\xi$-dependence of $v_i$ in the gauge fixing and ghost Lagrangian parts would not appear in our following derivation of Nielsen Identity.
 Since $v_i$ does not appear in any of the physical quantities we calculate, so in the following derivation, we treat $v$ as a quantity independent of $\xi$. Therefore, for notational simplicity, Equation~(\ref{BRST}) can be rewritten as
\begin{equation}
	\begin{aligned}\label{BRSt}
		\delta A^a_\mu&=\epsilon D^{ab}_\mu c^b\,,\\
		\delta B_\mu&=\epsilon \partial_\mu c^0\,,\\
		\delta c^a&=-\frac{1}{2}\epsilon g f^{abc}c^b c^c\,,\\
		\delta c^0&=0\,,\\
		\delta \bar{c}^a&=-\epsilon \frac{1}{\xi}(\partial_\mu A^{a \mu}+g v^a_i\varphi_i) \,,\\
		\delta \bar{c}^0&=-\epsilon \frac{1}{\xi}(\partial_\mu B^\mu+g^{\prime} v_i\varphi_i) \,,\\
		\delta \varphi_i&=\epsilon i(g c^a t^a_{ij}\varphi_j+g^{\prime} c^0 n^{\prime}_{ij}\varphi_j)\;,
	\end{aligned}
\end{equation}

Our aim is to extract the explicit dependence of effective potential $V$ on $\xi$, $V$ is obtained from the generating functional of one particle irreducible Green functions, and then expanding which in powers of momenta and consider all external momenta vanish:
\begin{equation}
	\Gamma \lp 0,0,0,\phi ; \xi \rp=-V(\phi,\xi)\int \diff^4 x\;,
\end{equation}
where $\phi$ is x-independent, and $\Gamma$ itself is obtained from the Legendre transformation of the function
$F$:
\begin{equation}
	\Gamma \lp\phi_{a c};\xi \rp=F-\int \diff^4 x J_a \phi_a\;,
\end{equation}
where
\begin{equation}
	\phi_{a c}=\frac{\delta F}{\delta J_a}\;.
\end{equation}
Here, $J_a$ as shorthand for all the sources, and $\phi_a$ for all the fields, and the function F is
\begin{equation}
	F(J_a;\xi)=-i \ln Z\;,
\end{equation}
with the generating functional of Green function being
\begin{equation}\label{eq:Z0}
	Z(J_a;\xi)= \int \mathcal{D}\phi_a e^{i S}\;.
\end{equation}
Here, the action is
\begin{equation}
	S=\int \diff^4 x (\mathcal{L}+J_a \phi_a)\;.
\end{equation}

We need the gauge dependence of the effective action ($\partial \Gamma /\partial \xi$), and we have $\partial \Gamma /\partial \xi=\partial F /\partial \xi$, see also the Abelian Higgs and SU(2) Higgs cases in Ref.~\cite{AITCHISON19841}. The explicit dependence of $\Gamma, F$ and $Z$ on the gauge parameter $\xi$ arise entirely from the gauge-fixing term in $\mathcal{L}$:
\begin{equation}\label{eq:derivation_2}
	\xi \frac{\partial F}{\partial \xi}=\frac{1}{Z}\int \mathcal{D}\phi_a \int\diff^4 x \frac{1}{2 \xi}(\partial_\mu A^{a \mu}+g v^a_i \Phi_i)^2 \exp(i S)\;.
\end{equation}
Nielsen identities can be regarded as a generalization of the Ward-Takahashi identities, whose explicit form can be derived directly through BRS transformation. The BRS transformation takes $\phi_a$ to $\phi_a+\delta \phi_a$, which makes $\mathcal{L}$ and $Z$ invariant. Further, $\delta \phi_a$ can be treated as infinitesimal, we have
\begin{equation}
\begin{aligned}
     Z(J_a; \xi)&=\int \mathcal{D} \phi_a \exp\left[i\int \diff^4 x \mathcal{L}+J_a(\phi_a+\delta \phi_a)\right]\\
     &=Z(J_a;\xi)+\int \mathcal{D} \phi_a \left[ i \int \diff^4 z J_a \delta\phi_a\right]\exp\left[ i\int \diff^4 x(\mathcal{L}+J_a \phi_a)\right]\;,
\end{aligned}
\end{equation}
hence
\begin{equation}\label{eq:derivation_1}
    \int \diff^4 z J_a\left\{\int \mathcal{D} \phi_a \delta\phi_a\exp\left[ i\int \diff^4 x(\mathcal{L}+J_a \phi_a)\right]\right\}=0\;,
\end{equation}
As can be seen from Eq.~(\ref{BRST}), now $\delta \phi_a$ can now be regarded as comprising a linear combination of the fields $\phi_a$ and composite terms $Q_a$ that are functions of $\phi_a$. For the linear part in $\phi_a$, the corresponding source is the original linear source $J_a$. For the composite operator $Q_a$, we introduce a new nonlinear source $K_a$. This allows us to define a modified Lagrangian $\mathcal{L} + K_a Q_a$, where the additional source $K_a$ coupled to $Q_a$ remains invariant under the BST transformation. Eq.~(\ref{eq:derivation_1}) will then take the form
\begin{equation}\label{eq:derivation_4}
    \int \diff^4 z J_a\left\{\int \mathcal{D} \phi_a \delta\phi_a\exp\left[ i\int \diff^4 x(\mathcal{L}+J_a \phi_a+K_a Q_a)\right]\right\}=0\;.
\end{equation}
With the introduction of the nonlinear source $K_a$, Equation (\ref{eq:Z0}) is modified to:
\begin{equation}\label{eq:Z2}
    Z_K=\int \mathcal{D}\phi_a \exp{i \int\diff^4 x \lp \mathcal{L}+J_a \phi_a+K_a Q_a\rp}\;,
\end{equation}
A natural consequence is the relation $F_K = -i \ln Z_K$, where $F_K$ is the generating functional with the $K_a$ source. However, to properly handle the composite operators, we define the corresponding 1PI effective action $\Gamma_K$ via a Legendre transformation with respect to the linear sources $J_a$ only except $K_a$. Finally, we obtain the standard 1PI effective action $\Gamma$ by taking the limit $K_a \to 0$.

We now aim to apply a similar procedure to the generating functional $\xi \partial \Gamma/\partial \xi$. Let us denote the composite operator $\frac{1}{2 \xi}(\partial_\mu A^{a \mu}+g v^a_i\Phi_i)^2$ that appears in $\xi \partial F/\partial \xi$ (Eq.~(\ref{eq:derivation_2}))by $\bar{O}(x)$. We then introduce a corresponding source term $h(x)$ such that
\begin{equation}
    \begin{aligned}\label{eq:derivation_3}
\int \mathcal{D}\phi_a \int \diff^4 x \bar{O}(x)\exp{(iS)}=\frac{\delta}{\delta h(x)}\left[ \int\mathcal{D}\phi_a \int \diff^4 x h \bar{O} \exp{(iS)} \right]\;.
    \end{aligned}
\end{equation}
If we can find an operator $O$ whose BRST variation yields $\delta O = \epsilon \bar{O}$ (where $\epsilon$ is the Grassmann-odd BRST parameter), we can recast this term into the form presented in Eq.~(\ref{eq:derivation_4})
\begin{equation}\label{eq:derivation_5}
    \int \diff^4 z\int \mathcal{D}\phi_a \left\{ (J_a \delta\phi_a+h\bar{O}) \exp{\left[i\int\diff^4 x(\mathcal{L}+J_a\phi_a+K_a Q_a+h O)\right]}\right\}=0\;,
\end{equation}
and we find the corresponding operator $O$ is
\begin{equation}\label{operatorO}
	O=-\frac{1}{2}\bar{c}^a(\partial_\mu A^{a\mu}+g v^a_i \Phi_i)\;,
\end{equation}
since $\delta O$ is not exactly equal to $\epsilon \bar{O}$, we denote their relation as $\delta O = \epsilon \hat{O}$, While one term in $\delta O$ indeed corresponds to the operator $\bar{O}$ appearing in $\xi (\partial \Gamma/\partial \xi)$, the other term can be combined into $-\frac{1}{2}\bar{c}^a \eta^a$, where $\eta^a$ denotes the ghost current. This combination is realized upon using the ghost equation of motion and subsequently setting the ghost source to zero.

With the introduction of the new source $h(x)$, we define, in analogy with Eq.~(\ref{eq:Z2}), a new generating function $\widetilde{Z}_K$,
\begin{equation}
     \tilde{Z}_K=\int \mathcal{D}\phi_a \exp{i \int\diff^4 x \lp \mathcal{L}+J_a \phi_a+K_a Q_a+h O\rp}\;.
\end{equation}
Introducing the connected generating function $\tilde{F}_K$ via $\tilde{F}_K=-i \ln{\tilde{Z}_K}$. When performing the Legendre transformation, we again treat the sources $h$ and $K_a$ as spectators—they are not transformed. Consequently, the resulting effective action $\widetilde{\Gamma}_K$ retains explicit dependence on both $K_a$ and $h$, analogous to how the original expression $\xi \partial \Gamma/\partial \xi$ depends on them. Our primary focus then shifts to examining the dependence of $\widetilde{\Gamma}$ on these sources, $K_a$ and $h$. For these explicit depedences on sources which are not Legendre transformed, we have
\begin{equation}
    \frac{\delta \tilde{\Gamma}_K}{\delta K_a}=\frac{\delta \tilde{F}_K}{\delta K_a}, \quad \frac{\delta \tilde{\Gamma}_K}{\delta h}=\frac{\delta \tilde{F}_K}{\delta h}, \quad \frac{\partial \tilde{\Gamma}_K}{\partial \xi}=\frac{\partial \tilde{F}_K}{\partial \xi}.
\end{equation}

Finally, by differentiating $\tilde{\Gamma}_K$ with respect to $h$ (and invoking the relation from Eq.~(\ref{eq:derivation_5}), and subsequently setting the sources $h$ and $K_a$ to zero and $A^a_{\mu c}=c^a_c=\bar{c}^a_c=0$, we can obtain the Nielsen identity, see the following detailed derivations.

Introducing the corresponding nonlinear source term for the composite operator in Equation~(\ref{BRSt}), the generating function $\tilde{Z}_K$ is
\begin{equation}
    \tilde{Z}_K(J_a,K_a,h;\xi)=\int\mathcal{D}\phi_a \exp{i \tilde{S}_K}\;,
\end{equation}
where
\begin{equation}
    \begin{aligned}\label{SKtilde}
        \tilde{S}_K=\int \diff^4 x\Big(\mathcal{L}&+K_{1i}\lp i(g c^a t^a_{ij}\varphi_j+g^{\prime} c^0 n^{\prime}_{ij}\varphi_j) \rp +K^{a \mu}_2 D^{ab}_\mu c^b+ K_3^a(-\frac{1}{2}gf^{abc}c^bc^c)\\
        &+ J^{a \mu} A^a_\mu+J^{\prime \mu} B_\mu +\bar{\eta}^a c^a+\bar{\eta}^0 c^0+\bar{c}^a \eta^a+\bar{c}^0 \eta^0+f_i \varphi_i +h O\Big)\;,
    \end{aligned}
\end{equation}
then Eq.~(\ref{eq:derivation_5}) becomes
\begin{equation}
    \begin{aligned}\label{eq:shizi6}
&\int \diff^4 x \mathcal{D}\phi_a \left[J^{a \mu}(D^{ab}_\mu c^b)+J^{\prime \mu}\partial_\mu c^0+\bar{\eta}^a( -\frac{1}{2}gf^{abc}c^bc^c)+\eta^a \lp-\frac{1}{\xi}(\partial_\mu A^{a \mu}+g v^a_i \varphi_i)\rp\right.\\
&+\left. \eta \lp-\frac{1}{\xi}(\partial_\mu B^ \mu+g^\prime v_i \varphi_i)\rp+f_i\lp i g c^a t^a_{ij}\varphi_j+i g^{\prime} c^0 n^{\prime}_{ij}\varphi_j+h(x)\hat{O}(x)\rp\right]\exp{i \tilde{S}_K}=0\;.
    \end{aligned}
\end{equation}
Introducing the connected generating function $\tilde{F}_K$, Eq.~(\ref{eq:shizi6}) becomes
\begin{equation}
    \begin{aligned}\label{eq:shizi7}
\int \diff^4 x &\left[ J^{a\mu} \frac{\delta }{\delta K^{a \mu}_2}+J^{\prime \mu}\partial_\mu\frac{\delta }{\delta \bar{\eta}^0} +\bar{\eta}^a  \frac{\delta }{\delta K^a_3}-\eta^a\frac{1}{\xi}\lp\partial_\mu \frac{\delta}{\delta J^a_\mu}+g v^a_i\frac{\delta}{\delta f_i} \rp \right.\\
&\left.-\eta\frac{1}{\xi}\lp\partial_\mu \frac{\delta}{\delta J^\prime_\mu}+g^\prime v_i\frac{\delta}{\delta f_i} \rp+f_i \frac{\delta}{\delta K_{1i}}\right]\tilde{F}_K\\
&=-\frac{1}{\tilde{Z}_k}\int \diff^4x \mathcal{D}\phi_a h(x) \hat{O}(x)\exp{(i\tilde{S}_K)}\;.
    \end{aligned}
\end{equation}
Finally, we introduce the effective action $\tilde{\Gamma}_K$ via a Legendre transformation on the liner sources $J^{a\mu},J^{\prime \mu},\bar{\eta}^a,\bar{\eta}^0,\eta^a,\eta^0$ and $f_i$.
\begin{equation}
    \tilde{\Gamma}_K(\phi_{ac},K_a,h;\xi)=\tilde{F}_K(J_a,K_a,h;\xi)-\int \diff^4x\lp J_a \phi_a\rp\;,
\end{equation}
where the c-subscripted quantities are defined as
\begin{equation}
\begin{aligned}\label{leg_field}
    &A^a_{\mu c}=\frac{\delta \tilde{F}_K}{\delta J^{a \mu}}\;,\quad B_{\mu c}=\frac{\delta \tilde{F}_K}{\delta J^{\prime \mu}}\;,\quad c^a=\frac{\delta \tilde{F}_K}{\delta \bar{\eta}^a}\;,\quad c^0=\frac{\delta \tilde{F}_K}{\delta \bar{\eta}^0}\;,\\
    &\bar{c}^a=-\frac{\delta \tilde{F}_K}{\delta \eta^a}\;,\quad \bar{c}^0=-\frac{\delta \tilde{F}_K}{\delta \eta^0}\;,\quad \varphi_i=\frac{\delta \tilde{F}_K}{\delta f_i}\;.
\end{aligned}
\end{equation}
Conversely, we have
\begin{equation}
    \begin{aligned}\label{leg_source}
&J^{a\mu}=-\frac{\delta \tilde{\Gamma}_K}{\delta A^a_{\mu c}}\;,\quad J^{\prime\mu}=-\frac{\delta \tilde{\Gamma}_K}{\delta B_{\mu c}}\;,\quad \bar{\eta}^a=\frac{\delta \tilde{\Gamma}_K}{\delta c^a}\;,\quad \bar{\eta}^0=\frac{\delta \tilde{\Gamma}_K}{\delta c^0}\;,\quad \\
&\eta^a=-\frac{\delta \tilde{\Gamma}_K}{\delta \bar{c}^a}\;,\quad \eta^0=-\frac{\delta \tilde{\Gamma}_K}{\delta \bar{c}^0}\;,\quad f_i=-\frac{\delta \tilde{\Gamma}_K}{\delta \varphi_i}\;.
    \end{aligned}
\end{equation}

Since no Legendre transformation is performed with respect to the sources $K_a$ and $h$, we obtain the following relations for these explicit dependence items,
\begin{equation}
    \frac{\delta \tilde{\Gamma}_K}{\delta K_{1i}}=\frac{\delta \tilde{F}_K}{\delta K_{1i}}\;,\quad \frac{\delta \tilde{\Gamma}_K}{\delta K^{a\mu}_2}=\frac{\delta \tilde{F}_K}{\delta K^{a\mu}_2}\;,\quad  \frac{\delta \tilde{\Gamma}_K}{\delta K^a_3}=\frac{\delta \tilde{F}_K}{\delta K^a_3}\;,\quad  \frac{\delta \tilde{\Gamma}_K}{\delta h}=\frac{\delta \tilde{F}_K}{\delta h}\;,\quad  \frac{\delta \tilde{\Gamma}_K}{\delta \xi}=\frac{\delta \tilde{F}_K}{\delta \xi}\;.
\end{equation}
Use Eqs.~(\ref{leg_field}) and (\ref{leg_source}), Eq.~(\ref{eq:shizi7}) becomes
\begin{equation}
    \begin{aligned}\label{eq:shizi8}
&\int \diff^4 x\left[-\frac{\delta \tilde{\Gamma}_K}{\delta A^a_{\mu c}}\frac{\delta \tilde{\Gamma}_K}{\delta K^{a\mu}_2}-\frac{\delta \tilde{\Gamma}_K}{\delta B_{\mu c}}\partial_\mu c^0_c+\frac{\delta \tilde{\Gamma}_K}{\delta c^a_c}\frac{\delta \tilde{\Gamma}_K}{\delta K^a_3}+\frac{\delta \tilde{\Gamma}_K}{\delta \bar{c}^a_c}\frac{1}{\xi}(\partial_\mu A^{a\mu}_c+gv^a_i \varphi_{ic})\right. \\
&\left.+\frac{\delta \tilde{\Gamma}_K}{\delta \bar{c}^0}\frac{1}{\xi}(\partial_\mu B^\mu_c+g^\prime v_i \varphi_{ic})-\frac{\delta \tilde{\Gamma}_K}{\delta \varphi_{ic}}\frac{\delta\tilde{\Gamma}_K}{\delta K_{1i}} \right]=-\frac{1}{\tilde{Z}_K}\int\diff^4x\mathcal{D}\phi_a h(x)\hat{O}(x)\exp{i\tilde{S}_K}\;,
    \end{aligned}
\end{equation}
Subsequently, we will differentiate these expressions with respect to $h$, we have
\begin{equation}
    \frac{\delta \tilde{F}_K}{\delta h(x)}=\frac{1}{\tilde{Z}_K}\int \mathcal{D}\phi_a O(x) \exp{i\tilde{S}_K}\;.
\end{equation}
Since $h$ serves as an external source and does not participate in the Legendre transformation, the computation of $\frac{\delta \tilde{\Gamma}_K}{\delta h}$ retains the explicit dependence on $O(x)$, We denote this quantity as $\tilde{\Gamma}_K(O)$. Then differentiating Eq.~(\ref{eq:shizi8}) with respect to the explicit dependence on $h(x)$ yields
\begin{equation}
    \begin{aligned}\label{eq:shizi9}
&\int \diff^4 x\left[\frac{\delta \tilde{\Gamma}_K(O)}{\delta A^a_{\mu c}}\frac{\delta \tilde{\Gamma}_K}{\delta K^{a\mu}_2}+\frac{\delta \tilde{\Gamma}_K}{\delta A^a_{\mu c}}\frac{\delta \tilde{\Gamma}_K(O)}{\delta K^{a\mu}_2}+\frac{\delta \tilde{\Gamma}_K(O)}{\delta K^{a\mu}_2}\partial_\mu c^0_c-\frac{\delta \tilde{\Gamma}_K(O)}{\delta c^a_c}\frac{\delta \tilde{\Gamma}_K}{\delta K^a_3}-\frac{\delta \tilde{\Gamma}_K}{\delta c^a_c}\frac{\delta \tilde{\Gamma}_K(O)}{\delta K^a_3}\right.\\
&\left. -\frac{\delta \tilde{\Gamma}_K(O)}{\delta \bar{c}^a_c}\frac{1}{\xi}(\partial_\mu A^{a\mu}_c+gv^a_i \varphi_{ic})-\frac{\delta \tilde{\Gamma}_K(O)}{\delta \bar{c}^0}\frac{1}{\xi}(\partial_\mu B^\mu_c+g^\prime v_i \varphi_{ic})+\frac{\delta \tilde{\Gamma}_K(O)}{\delta \varphi_{ic}}\frac{\tilde{\Gamma}_K}{\delta K_{1i}}+\frac{\delta \tilde{\Gamma}_K}{\delta \varphi_{ic}}\frac{\tilde{\Gamma}_K(O)}{\delta K_{1i}}\right]\\
&=\frac{1}{\tilde{Z}_K}\int\diff^4x \mathcal{D}\phi_a \hat{O}(x) \exp{i \tilde{S}_K}\;.
    \end{aligned}
\end{equation}

Now, We consider a specific operator $O$,
\begin{equation}\label{operatorO}
    O=-\frac{1}{2}\bar{c}^a(\partial_\mu A^{a \mu}+g v^a_i \varphi_i)-\frac{1}{2}\bar{c}^0(\partial_\mu B^\mu g^\prime v_i \varphi_i)\;,
\end{equation}
then we proceed to derive the specific form of $\delta O$
\begin{equation}
    \begin{aligned}\label{eq:deltaO}
\delta O&=\epsilon\left\{\frac{1}{2 \xi}(\partial_\mu A^{a\mu}+gv^a_i \varphi_i)^2+\frac{1}{2 \xi}(\partial_\mu B^\mu +g^\prime v_i \varphi_i)^2\right.\\
&-\frac{1}{2}\bar{c}^a\left[\partial^\mu(\partial_\mu-gf^{abc}c^bA^c_\mu)+gv^a_i(i gc^a t^a_{ij}\varphi_j+ig^\prime c^0n^\prime_{ij}\varphi_j) \right]\\
&\left.-\frac{1}{2}\bar{c}^0\left[\partial^\mu \partial_\mu c^0+g^\prime v_i(i gc^a t^a_{ij}\varphi_j+ig^\prime c^0n^\prime_{ij}\varphi_j) \right]\right\}\;,
    \end{aligned}
\end{equation}
this $\epsilon \hat{O}$ is precisely the one mentioned above, the last two lines in Eq.~(\ref{eq:deltaO}) correspond to the ghost Lagrangian, we identify them as $-\frac{1}{2}\bar{c}^a \eta^a$ and $-\frac{1}{2}\bar{c}^0 \eta^0$ respectively, where $\eta^a$ and $\eta^0$ are the ghost currents. Therefore, Equation~(\ref{eq:deltaO}) can be simplified to
\begin{equation}
\delta O=\epsilon \lp\bar{O}-\frac{1}{2}\bar{c}^a \eta^a-\frac{1}{2}\bar{c}^0 \eta^0\rp=\epsilon \hat{O}\;,
\end{equation}
substituting the $\hat{O}$ into the right-hand side of Eq.~(\ref{eq:shizi9}) with $h \to 0$, we obtain
\begin{equation}\label{eq:zuo1}
    \frac{1}{Z_K}\int\diff^4x \mathcal{D}\phi_a \lp \bar{O}-\frac{1}{2}\bar{c}^a \eta^a-\frac{1}{2}\bar{c}^0 \eta^0 \rp\exp{i S_K}=\xi \frac{\partial F_K}{\partial \xi}-\frac{1}{2}\int \diff^4x \lp \eta^a \frac{\delta F_K}{\delta \eta^a} +\eta^0 \frac{\delta F_K}{\delta \eta^0} \rp\;,
\end{equation}
the right-hand side of Eq.~(\ref{eq:zuo1}) becomes,
\begin{equation}\label{eq:shizi10}
    \xi\frac{\partial \Gamma_K}{\partial \xi}-\frac{1}{2}\int \diff^4x \lp \frac{\delta \Gamma_K}{\delta \bar{c}^a}\bar{c}^a_c+\frac{\delta \Gamma_K}{\delta \bar{c}^0}\bar{c}^0_c \rp\;,
\end{equation}
after Legendre transformation and set $h \to 0$.
Combining Equation~(\ref{eq:shizi9}) and Equation~(\ref{eq:shizi10}), and setting $h$ and $K_a$ to zero, we obtain
\begin{equation}
    \begin{aligned}\label{eq:shizi11}
\xi\frac{\partial \Gamma}{\partial \xi}&-\frac{1}{2}\int \diff^4x \lp \frac{\delta \Gamma}{\delta \bar{c}^a}\bar{c}^a_c+\frac{\delta \Gamma}{\delta \bar{c}^0}\bar{c}^0_c \rp \\
&=\int \diff^4 x \int \diff^4 z \left[ \frac{\delta \Gamma(O(x))}{\delta A^a_{\mu c}(z)}\frac{\delta \Gamma}{\delta K^{a\mu}_2(z)}+\frac{\delta \Gamma}{\delta A^a_{\mu c}(z)}\frac{\delta \Gamma(O(x))}{\delta K^{a\mu}_2(z)}+\frac{\delta \Gamma(O(x))}{\delta K^{a\mu}_2(z)}\partial_\mu c^0_c(z)\right.\\
&-\frac{\delta \Gamma(O(x))}{\delta c^a_c(z)}\frac{\delta \Gamma}{\delta K^a_3(z)}-\frac{\delta \Gamma}{\delta c^a_c(z)}\frac{\delta \Gamma(O(x))}{\delta K^a_3(z)}-\frac{\delta \Gamma(O(x))}{\delta \bar{c}^a_c(z)}\frac{1}{\xi}\lp \partial_\mu A^{a\mu}_c(z)+gv^a_i \varphi_{ic}(z)\rp\\
&\left. -\frac{\delta \Gamma(O(x))}{\delta \bar{c}^0(z)}\frac{1}{\xi}(\partial_\mu B^\mu_c(z)+g^\prime v_i \varphi_{ic}(z))+\frac{\delta \Gamma(O(x))}{\delta \varphi_{ic}(z)}\frac{\delta \Gamma}{\delta K_{1i}(z)}+\frac{\delta \Gamma}{\delta \varphi_{ic}(z)}\frac{ \delta \Gamma(O(x))}{\delta K_{1i}(z)}\right]\;,\\
    \end{aligned}
\end{equation}
then set $\varphi_{1c}=\phi$, $\varphi_{ic}=0$ (for $i=2,3,4$), and $A^a_{\mu c}=B_{\mu c}=c^a_c=c^0_c=\bar{c}^a_c=\bar{c}^0_c=0$, so that many terms will immediately disappear from Eq.~(\ref{eq:shizi11}), we obtain
\begin{equation}
    \begin{aligned}
        \xi\frac{\partial \Gamma}{\partial \xi}=\int \diff^4x \diff^4z \left[\frac{\delta \Gamma}{\delta \varphi_{ic}(z)}\frac{ \delta \Gamma(O(x))}{\delta K_{1i}(z)}\right]\;.
    \end{aligned}
\end{equation}
Refer to Eq.~(\ref{Nielsenkey}), we have
\begin{equation}
    K_i[\phi]=-\int \diff^4x \frac{\delta \Gamma(O(x))}{\delta K_{1i}}\,,
\end{equation}
the quantity $\delta \Gamma(O(x))/\delta K_{1i}$ is obtained via a Legendre transformation from $\delta^2 \tilde{F}_K/\delta h(x) \delta K_{1i}$, where the function dervatives are understood to operate on $h$ and $K_{1i}$ dependence in $\tilde{S}_K$, and $h$ and $K_{1i}$ are set to zero finally.

From Eqs.~(\ref{SKtilde}) and~(\ref{operatorO}), we have
\begin{equation}
    \begin{aligned}
        \lim_{h \to 0,K_{1i} \to 0}\frac{\delta^2 \tilde{Z}_K}{\delta h \delta K_{1i}}&=\lp\frac{i}{\hbar} \rp^2 \int \mathcal{D} \phi_a \left[ \lp-\frac{1}{2}\bar{c}^a(\partial_\mu A^{a \mu}+g v^a_i \varphi_i)-\frac{1}{2}\bar{c}^0(\partial_\mu B^\mu g^\prime v_i \varphi_i)\rp\right. \\
        &\left.\lp-\frac{1}{2}\bar{c}^a(\partial_\mu A^{a \mu}+g v^a_i \varphi_i)-\frac{1}{2}\bar{c}^0(\partial_\mu B^\mu g^\prime v_i \varphi_i)\rp\exp{\frac{iS}{\hbar}}
        \right]\,.
    \end{aligned}
\end{equation}
Following the method of Ref.~\cite{Jackiw:1974cv}, we can derive the loop expansion for the $K[\phi,\xi]$ that appears in Eq.~(\ref{Nielsenkey}), the effective action
\begin{equation}
	S_\mathrm{eff}(\Phi,\Phi_c)=S_{\mathscr{L}}(\Phi+\Phi_c)-S_{\mathscr{L}}(\Phi_c)-\int \diff^4 x \Phi(x)\left.\frac{\delta S_{\mathscr{L}}}{\delta \Phi(x)}\right|_{\Phi=\Phi_c}\;,
\end{equation}
or, equivalently, the effective Lagrangian
\begin{equation}
	\mathcal{L}_\mathrm{eff}(\Phi,\Phi_c)=\mathcal{L}(\Phi+\Phi_c)-\mathcal{L}(\Phi_c)-\Phi(x)\left.\frac{\delta \mathcal{L}}{\delta \Phi(x)}\right|_{\Phi=\Phi_c}\;.
\end{equation}
Then we get
\begin{equation}
	\begin{aligned}\label{nielsenK}
		K_j[\phi]=&-\int \diff^4 x i \hbar \bra{0}T \lp\frac{i}{\hbar}\rp^2\left[ \frac{1}{2} \bar{c}^a(x)\lp\partial_\mu A^{a \mu}+g v^a_i \varphi_i\rp igc^b(0) t^b_{jk}\varphi_k(0) \exp \frac{i}{\hbar}S_\mathrm{eff}\right]\ket{0}\,\\
		&-\int \diff^4 x i \hbar \bra{0}T \lp\frac{i}{\hbar}\rp^2\left[ \frac{1}{2} \bar{c}^0(x)\lp\partial_\mu B^\mu+g' v_i \varphi_i\rp ig' c^0(0) n'_{jk}\varphi_k(0) \exp \frac{i}{\hbar}S_\mathrm{eff}\right]\ket{0}\,,
	\end{aligned}
\end{equation}
The parts of right-hand side in Eq.~(\ref{nielsenK}) come from the $O(x)$ as given by Ee.~(\ref{operatorO}) and the transformation part of the scalar field corresponding to the source $K_a$.
\subsection{The Calculation of D in Nielsen identity}\label{appendix:D}
\begin{figure}[!htp]
	\centering
	\includegraphics[width=0.8\linewidth]{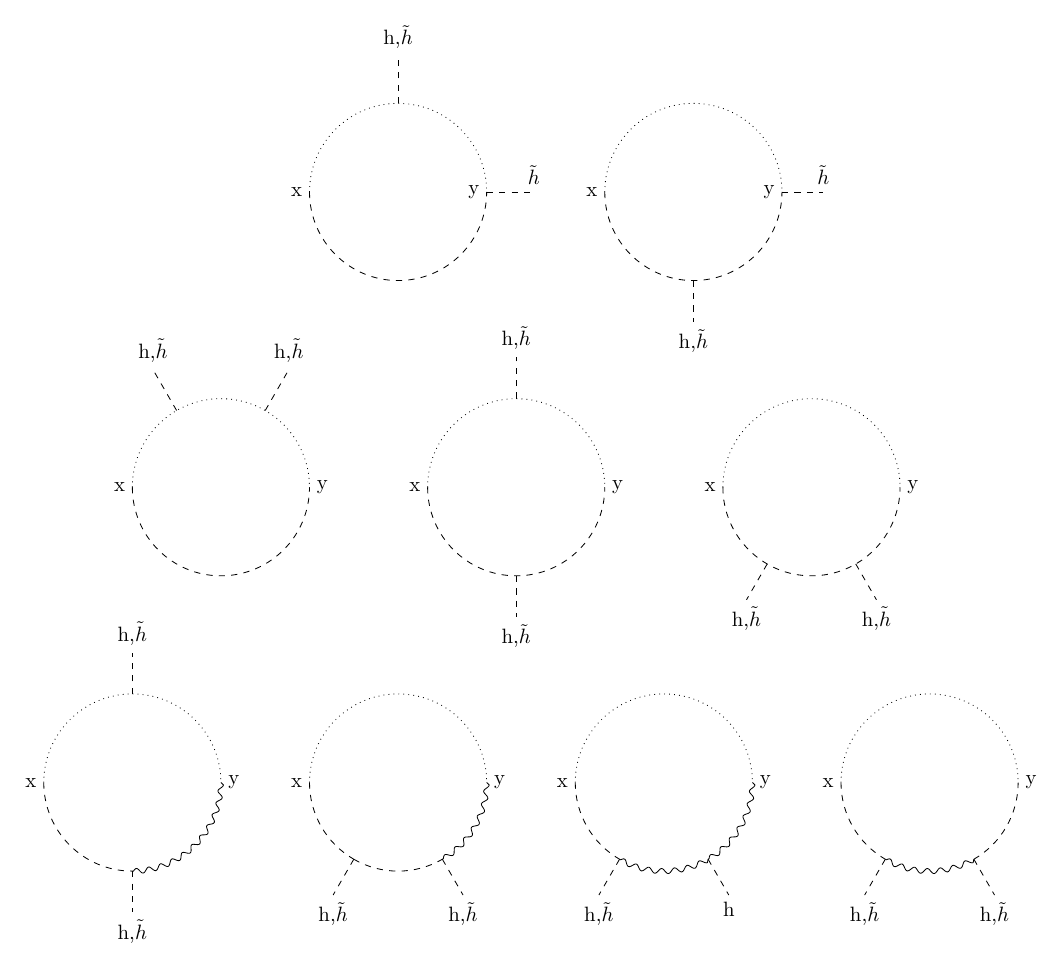}
	\caption{The diagrams contribution to the function D.}
	\label{fig:D factor}
\end{figure}

When working at the leading order of the potential, which is of order $g^2$, power counting dictates that we must compute the factors $D$ and $\tilde{D}$, the corresponding Feynman diagrams about $D$ factor are shown in Figure~\ref{fig:D factor}. We now list the characteristic integrals relevant to the calculation.

\begin{equation}
	\begin{aligned}
		F^{(D)}_1(M, m)&=\int_p \frac{1}{(p^2+M^2)(p^2+m^2)^2[(p+k)^2+m^2]}\,\\
		&=\frac{3 m+M}{32 \pi  m^3 (m+M)^3}+k^2 \lp \frac{25 m^3-29 m^2 M-15 m M^2-3 M^3}{384 \pi  m^5 (m+M)^5}\rp+\mathcal{O}(k^4)\;,
	\end{aligned}
\end{equation}

\begin{equation}
	\begin{aligned}
		F^{(D)}_2(M, m)&=\int_p \frac{1}{(p^2+M^2)(p^2+m^2)[(p+k)^2+M^2][(p+k)^2+m^2]}\,\\
		&=\frac{1}{8 \pi m M (m+M)^3}+k^2 \lp -\frac{m^4+5 m^3 M+12 m^2 M^2+5 m M^3+M^4}{96 \pi  m^3 M^3 (m+M)^5}\rp+\mathcal{O}(k^4)\,,
	\end{aligned}
\end{equation}

\begin{equation}
	\begin{aligned}
		F^{(D)}_3(M, m)&=\int_p \frac{1}{(p^2+m^2)(p^2+M^2)^2[(p+k)^2+M^2]}=F^{(D)}_1(m, M)\,\\
		&=\frac{3 M+m}{32 \pi  M^3 (m+M)^3}+k^2 \lp \frac{25 M^3-29 M^2 m-15 M m^2-3 m^3}{384 \pi  M^5 (m+M)^5}\rp+\mathcal{O}(k^4)\,,
	\end{aligned}
\end{equation}

here, $m$ is scaler mass, $M$ is ghost mass.

\begin{equation}
	\begin{aligned}
		F^{(D)}_4(M, m_1, m_2)&=\int_p \frac{p^\mu(2k+p)^\nu\left[g^{\mu\nu}-(1-\xi)\frac{p^\mu p^\nu}{p^2+\xi m_2^2}\right]}{(p^2+M^2)[(p+k)^2+M^2](p^2+m_2^2)[(p+k)^2+m_1^2]}\,\\
		&=\frac{\xi  (M+3 m_1)}{32 \pi  M (M+m_1)^3}+k^2\lp -\frac{ \xi  \left(13 M^3+33 M^2 m_1+15 M m_1^2+3 m_1^3\right)}{128 \pi  M^3 (M+m_1)^5}\rp+\mathcal{O}(k^4)\,,
	\end{aligned}
\end{equation}

\begin{equation}
	F^{(D)}_5(M, m_1, m_2)=\int_{p} \frac{(p-k)^\mu \left[g^{\mu\nu}-(1-\xi)\frac{(p+k)^\mu (p+k)^\nu}{(p+k)^2+\xi m_2^2}\right] g^{\nu \rho}\left[g^{\rho\sigma}-(1-\xi)\frac{p^\rho p^\sigma}{p^2+\xi m_2^2}\right] p^\sigma}{(p^2+M^2)(p^2+m_1^2)[(p+k)^2+m_2^2](p^2+m_2^2)}\,,
\end{equation}
Here, $M$ is ghost mass, $m_1$is scalar mass, $m_2$ is boson mass. For the convenience of the subsequent calculation, we have $M^2=\xi m_2^2$,because the full calculation formulas are rather lengthy, we only present the terms of $k^2$ ,
\begin{align}
F^{(D)}_5(M, m_1, m_2)&=\frac{k^2}{192 \pi  M m_2 (M-m_1) (M+m_1)^5 (M+m_2)^2 (m_1+m_2)}\times \bigg(M^6 (13-19 \xi )\nn\\
&+M^5 (m_1 (52-50 \xi )+26 m_2 \xi )+M^4 m_1 (24 m_1 (\xi +2)+m_2 \xi  (117-19 \xi ))\nn\\
&+4 M^3 m_1^2 (5 m_1 (7 \xi -1)+m_2 \xi  (37-3 \xi ))+M^2 m_1^3 (3 m_1 (33 \xi -7)\nn\\
&+8 m_2 \xi  (6 \xi +1))+2 M m_1^4 \xi  (11 m_1+m_2 (22 \xi -31))+m_1^5 m_2 \xi  (11 \xi -21)\bigg)\nn\\
&+\cdots\,,
	\end{align}
\begin{equation}
	\begin{aligned}
F^{(D)}_6(M, m_1, m_2)&=\int_p \frac{p^\mu(2k+p)^\nu\left[g^{\mu\nu}-(1-\xi)\frac{p^\mu p^\nu}{p^2+\xi m_2^2}\right]}{(p^2+M^2)(p^2+m_1^2)[(p+k)^2+m_1^2](p^2+m_2^2)}\\
&=\frac{\xi }{8 \pi  (M+m_1)^3}+k^2\lp -\frac{ \xi  (11 M+19 m_1)}{96 \pi  m_1 (M+m_1)^5}\rp+\mathcal{O}(k^4)\,,
	\end{aligned}
\end{equation}

\begin{equation}
	\begin{aligned}
F^{(D)}_7(M, m_1, m_2)=\frac{(p-k)_\mu (p-k)_\nu \left[g^{\mu\nu}-(1-\xi)\frac{(p+k)^\mu (p+k)^\nu}{(p+k)^2+\xi m_2^2}\right]}{(p^2+M^2)(p^2+m_1^2)^2[(p+k)^2+m_2^2]}\,,\\
	\end{aligned}
\end{equation}
we only show the terms of $k^2$,
\begin{equation}
	\begin{aligned}
F^{(D)}_7(M, m_1, m_2)&=k^2\Bigg(\frac{8 M^2+12 M m_2+5 m_2^2}{12 \pi  \left(M^2-m_1^2\right)^2 (M+m_2)^3}+\frac{-2 M^2-6 m_1^2-12 m_1 m_2-5 m_2^2}{12 \pi  (M-m_1)^2 (M+m_1)^2 (m_1+m_2)^3}\\
&-\frac{(1-\xi)M^2 m_1^2}{6 \pi  (M-m_1)^3 (M+m_1)^3 (M+m_2)^3}+\frac{(1-\xi)M^2 m_1^2}{6 \pi  (M-m_1)^3 (M+m_1)^3 (m_1+m_2)^3}\\
&-\frac{(1-\xi)\lp M^2+m_1^2\rp}{2 \pi  (M-m_1)^2 (M+m_1)^3 (M+m_2) (m_1+m_2)}\Bigg)+\cdots\,,
	\end{aligned}
\end{equation}

\begin{equation}
	\begin{aligned}
		F^{(D)}_8(M, m)&=\int_p \frac{1}{(p^2+M^2)[(p+k)^2+M^2](p^2+m^2)}\,\\
		&=\frac{1}{8 \pi M(M+m)^2}+k^2\lp -\frac{7 M^2+4M m+m^2}{96 \pi M^3(M+m)^4} \rp+\mathcal{O}(k^4)\;,
	\end{aligned}
\end{equation}

\begin{equation}
	\begin{aligned}
		F^{(D)}_9(M, m_1, m_2)&=\int_p \frac{p^\mu p^\nu\left[g^{\mu\nu}-(1-\xi)\frac{p^\mu p^\nu}{p^2+\xi m_2^2}\right]}{(p^2+M^2)[(p+k)^2+M^2](p^2+m_2^2)[(p+k)^2+m_1^2]}\,\\
		&=\frac{\xi  (M+3 m_1)}{32 \pi  M (M+m_1)^3}+k^2 \lp \frac{\xi  (M-m_1) \left(17 M^2+6 M m_1+m_1^2\right)}{384 \pi  M^3 (M+m_1)^5}\rp+\mathcal{O}(k^4)\;,
	\end{aligned}
\end{equation}

\begin{equation}
	\begin{aligned}
		F^{(D)}_{10}(M, m_1, m_2)&=\int_p \frac{p^\mu p^\nu\left[g^{\mu\nu}-(1-\xi)\frac{p^\mu p^\nu}{p^2+\xi m_2^2}\right]}{(p^2+M^2)(p^2+m_1^2)[(p+k)^2+m_1^2](p^2+m_2^2)}\\
		&=\frac{\xi }{8 \pi  (M+m_1)^3}+k^2 \lp \frac{\xi  (M-7 m_1)}{96 \pi  m_1 (M+m_1)^5}\rp+\mathcal{O}(k^4)\,,
	\end{aligned}
\end{equation}

\begin{equation}
	\begin{aligned}
		F^{(D)}_{11}(M, m_1, m_2)=\int_{p} \frac{(p+k)^\mu \left[g^{\mu\nu}-(1-\xi)\frac{(p+k)^\mu (p+k)^\nu}{(p+k)^2+\xi m_2^2}\right] g^{\nu \rho}\left[g^{\rho\sigma}-(1-\xi)\frac{p^\rho p^\sigma}{p^2+\xi m_2^2}\right] p^\sigma}{(p^2+M^2)(p^2+m_1^2)[(p+k)^2+m_2^2](p^2+m_2^2)}\,,
	\end{aligned}
\end{equation}

\begin{equation}
	\begin{aligned}
		F^{(D)}_{11}(M, m_1, m_2)&=k^2 \Bigg( \frac{M (\xi -1) \left(11 M^2-m_1^2\right) \left(M^2 (\xi -2)+m_2^2\right)}{192 \pi  (M-m_1)^2 (M+m_1)^2 (M-m_2)^3 (M+m_2)^3}\\
		&+\frac{M^2 (\xi -1) \left(M^2+3 M m_1+m_1^2\right) \left(M^2-m_1^2 \xi \right)}{12 \pi  (M-m_1)^2 (M+m_1)^5 (M-m_2) (M+m_2) (m_1-m_2) (m_1+m_2)}\Bigg)+\cdots\,,
	\end{aligned}
\end{equation}

\begin{equation}
	\begin{aligned}
		F^{(D)}_{12}(M, m_1, m_2)&=\frac{(p+k)_\mu (p-k)_\nu \left[g^{\mu\nu}-(1-\xi)\frac{(p+k)^\mu (p+k)^\nu}{(p+k)^2+\xi m_2^2}\right]}{(p^2+M^2)(p^2+m_1^2)^2[(p+k)^2+m_2^2]}\\
		&=k^2\lp\frac{M (\xi -1) \left(5 M^2+20 M m_1+11 m_1^2\right)}{96 \pi  (M-m_1) (M+m_1)^5 (M-m_2) (M+m_2)}\rp+\cdots\,,
	\end{aligned}
\end{equation}

\begin{equation}
	\begin{aligned}
		F^{(D)}_{13}(M, m_1, m_2)&=\frac{(p+k)_\mu (p+k)_\nu \left[g^{\mu\nu}-(1-\xi)\frac{(p+k)^\mu (p+k)^\nu}{(p+k)^2+\xi m_2^2}\right]}{(p^2+M^2)(p^2+m_1^2)^2[(p+k)^2+m_2^2]}\\
		&=k^2\lp -\frac{M (\xi -1) \left(7 M^2+4 M m_1+m_1^2\right)}{96 \pi  (M-m_1) (M+m_1)^5 (M-m_2) (M+m_2)}\rp+\cdots\,.
	\end{aligned}
\end{equation}

\subsection{the result of D}
By incorporating the characteristic integrals listed above, we calculate $D$, which is divided into three parts: $\Pi^D_{h,h}$, $\Pi^D_{h,\tilde{h}}$, and $\Pi^D_{\tilde{h},\tilde{h}}$. Their expressions are as follows:

\begin{equation}
	\begin{aligned}
-\Pi^D_{h,h}&=2 X_W C_{h c^+ c^-}^2 F^{(D)}_1(m_{c_W}, m_{\chi^\pm})+X_Z C_{h c_Z c_Z}^2 F^{(D)}_1(m_{c_Z}, m_{\chi^0})\\
&+2 X_W C_{h c^+ c^-} C_{h \chi^+ \chi^-} F^{(D)}_2(m_{c_W}, m_{\chi^\pm})+X_Z C_{h c_Z c_Z} C_{H \chi \chi} F^{(D)}_2(m_{c_Z}, m_{\chi^0})\\
&+2 X_W C_{h \chi^+ \chi^-}^2 F^{(D)}_3(m_{c_W}, m_{\chi^\pm})+X_Z C_{h \chi \chi}^2 F^{(D)}_3(m_{c_Z}, m_{\chi^0})\\
&+2 Q_W C_{h c^+ c^-} C_{h W^\pm G^\pm} F^{(D)}_4(m_{c_W}, m_{\chi^\pm}, m_W)+Q_Z C_{h c_Z c_Z} C_{H Z \chi} F^{(D)}_4(m_{c_Z}, m_{\chi^0}, m_Z)\\
&+2 Q_W C_{h W^\pm \chi^\pm} C_{h W^+ W^-} F^{(D)}_5(m_{c_W}, m_{\chi^\pm}, m_W)+Q_Z C_{h Z \chi} C_{h Z Z} F^{(D)}_5(m_{c_Z}, m_{\chi^0}, m_Z)\\
&+2 Q_W C_{h \chi^+ \chi^-} C_{h W^\pm \chi^\pm} F^{(D)}_6(m_{c_W}, m_{\chi^\pm}, m_W)+Q_Z C_{h \chi \chi} C_{h Z \chi} F^{(D)}_6(m_{c_Z}, m_{\chi^0}, m_Z)\\
&+2 X_W C_{h W^\pm \chi^\pm}^2 F^{(D)}_7(m_{c_W}, m_{\chi^\pm}, m_W)+X_Z C_{h Z \chi}^2 F^{(D)}_7(m_{c_Z}, m_{\chi^0}, m_Z)\;,
	\end{aligned}
\end{equation}

	\begin{align}
		-\Pi^D_{h,\tilde{h}}&=2 Q_W C_{h c^+ c^-} C_{\tilde{h} \chi^\pm \bar{c}^\pm} F^{(D)}_8(m_{c_W}, m_{\chi^\pm})+ Q_Z C_{h c_Z c_Z} C_{\tilde{h} \chi \bar{c}_Z} F^{(D)}_8(m_{c_Z}, m_{\chi^0})\nn\\
		&+2 Q_W C_{h \chi^+ \chi^-} C_{\tilde{h} \chi^\pm \bar{c}^\pm} F^{(D)}_8(m_{\chi^\pm},m_{c^\pm})+ Q_Z C_{h \chi \chi} C_{\tilde{h} \chi \bar{c}_Z} F^{(D)}_8(m_{\chi^0},m_{c_Z})\nn\\
		&+4 X_W C_{h c^+ c^-} C_{\tilde{h} c^+ c^-} F^{(D)}_1(m_{c_W}, m_{\chi^\pm})+2 X_Z C_{h c_Z c_Z} C_{\tilde{h} c_Z c_Z} F^{(D)}_1(m_{c_Z}, m_{\chi^0})\nn\\
		&+4 X_W C_{h c^+ c^-} C_{\tilde{h} \chi^+ \chi^-} F^{(D)}_2(m_{c_W}, m_{\chi^\pm})+2 X_Z C_{h c_Z c_Z} C_{\tilde{h} \chi \chi} F^{(D)}_2(m_{c_Z}, m_{\chi^0})\nn\\
		&+4 X_W C_{h \chi^+ \chi^-} C_{\tilde{h} \chi^+ \chi^-} F^{(D)}_3(m_{c_W}, m_{\chi^\pm})+2 X_Z C_{h \chi \chi} C_{\tilde{h} \chi \chi} F^{(D)}_3(m_{c_Z}, m_{\chi^0})\nn\\
		&+2 Q_W C_{\tilde{h} c^+ c^-} C_{h W^\pm \chi^\pm} F^{(D)}_4(m_{c_W}, m_{\chi^\pm}, m_W)+Q_Z C_{\tilde{h} c_Z c_Z} C_{h Z \chi} F^{(D)}_4(m_{c_Z}, m_{\chi^0}, m_Z)\nn\\
		&+2 Q_W C_{h c^+ c^-} C_{\tilde{h} W^\pm \chi^\pm} F^{(D)}_9(m_{c_W}, m_{\chi^\pm}, m_W)+Q_Z C_{h c_Z c_Z} C_{\tilde{h} Z \chi} F^{(D)}_9(m_{c_Z}, m_{\chi^0}, m_Z)\nn\\
		&+2 Q_W C_{\tilde{h} \chi^+ \chi^-} C_{h W^\pm \chi^\pm} F^{(D)}_6(m_{c_W}, m_{\chi^\pm}, m_W)+Q_Z C_{\tilde{h} \chi \chi} C_{h Z \chi} F^{(D)}_6(m_{c_Z}, m_{\chi^0}, m_Z)\nn\\
		&+2 Q_W C_{h \chi^+ \chi^-} C_{\tilde{h} W^\pm \chi^\pm} F^{(D)}_{10}(m_{c_W}, m_{\chi^\pm}, m_W)+Q_Z C_{h \chi \chi} C_{\tilde{h} Z \chi} F^{(D)}_{10}(m_{c_Z}, m_{\chi^0}, m_Z)\nn\\
		&+2 Q_W C_{\tilde{h} W^\pm \chi^\pm} C_{h W^+ W^-} F^{(D)}_{11}(m_{c_W}, m_{\chi^\pm}, m_W)+Q_Z C_{\tilde{h} Z \chi} C_{h Z Z} F^{(D)}_{11}(m_{c_Z}, m_{\chi^0}, m_Z)\nn\\
		&+4 X_W C_{h W^\pm \chi^\pm} C_{\tilde{h} W^\pm \chi^\pm} F^{(D)}_{12}(m_{c_W}, m_{\chi^\pm}, m_W)+2 X_Z C_{h Z \chi} C_{\tilde{h} Z \chi} F^{(D)}_{12}(m_{c_Z}, m_{\chi^0}, m_Z)\;,
	\end{align}

\begin{equation}
	\begin{aligned}
		-\Pi^D_{\tilde{h},\tilde{h}}&=2 Q_W C_{\tilde{h} c^+ c^-} C_{\tilde{h} \chi^\pm \bar{c}^\pm} F^{(D)}_8(m_{c_W}, m_{\chi^\pm})+Q_Z C_{\tilde{h} c_Z c_Z} C_{\tilde{h} \chi \bar{c}_Z} F^{(D)}_8(m_{c_Z}, m_{\chi^0})\\
		&+2 Q_W C_{\tilde{h} \chi^+ \chi^-} C_{\tilde{h} \chi^\pm \bar{c}^\pm} F^{(D)}_8(m_{\chi^\pm},m_{c_W})+Q_Z C_{\tilde{h} \chi \chi} C_{\tilde{h} \chi \bar{c}_Z} F^{(D)}_8(m_{\chi^0},m_{c_Z})\\
		&+2 X_W C_{\tilde{h} c^+ c^-}^2 F^{(D)}_1(m_{c_W}, m_{\chi^\pm})+X_Z C_{\tilde{h} c_Z c_Z}^2 F^{(D)}_1(m_{c_Z}, m_{\chi^0})\\
		&+2 X_W C_{\tilde{h} c^+ c^-} C_{\tilde{h} \chi^+ \chi^-} F^{(D)}_2(m_{c_W}, m_{\chi^\pm})+X_Z C_{\tilde{h} c_Z c_Z} C_{\tilde{h} \chi \chi} F^{(D)}_2(m_{c_Z}, m_{\chi^0})\\
		&+2 X_W C_{\tilde{h} \chi^+ \chi^-}^2 F^{(D)}_3(m_{c_W}, m_{\chi^\pm})+X_Z C_{\tilde{h} \chi \chi}^2 F^{(D)}_3(m_{c_Z}, m_{\chi^0})\\
		&+2 Q_W C_{\tilde{h} c^+ c^-} C_{\tilde{h} W^\pm \chi^\pm} F^{(D)}_9(m_{c_W}, m_{\chi^\pm}, m_W)+Q_Z C_{\tilde{h} c_Z c_Z} C_{\tilde{h} Z \chi} F^{(D)}_9(m_{c_Z}, m_{\chi^0}, m_Z)\\
		&+2 Q_W C_{\tilde{h} \chi^+ \chi^-} C_{\tilde{h} W^\pm \chi^\pm} F^{(D)}_{10}(m_{c_W}, m_{\chi^\pm}, m_W)+Q_Z C_{\tilde{h} \chi \chi} C_{\tilde{h} Z \chi} F^{(D)}_{10}(m_{c_Z}, m_{\chi^0}, m_Z)\\
		&+2 X_W C_{\tilde{h} W^\pm \chi^\pm}^2 F^{(D)}_{13}(m_{c_W}, m_{\chi^\pm}, m_W)+X_Z C_{\tilde{h} Z \chi}^2 F^{(D)}_{13}(m_{c_Z}, m_{\chi^0}, m_Z)\;.
	\end{aligned}
\end{equation}
Correspondingly, we perform an expansion in the external momentum $k$ and extract the contributions proportional to $k^2$. This yields
\begin{equation}
    D=\frac{\partial }{\partial k^2}\lp \Pi^D_{h,h}+\Pi^D_{h,\tilde{h}}+\Pi^D_{\tilde{h},\tilde{h}}\rp\;.
\end{equation}

\subsection{The Calculation of $\tilde{D}$}
\begin{figure}[h!]
	\centering
	\includegraphics[width=0.7\linewidth]{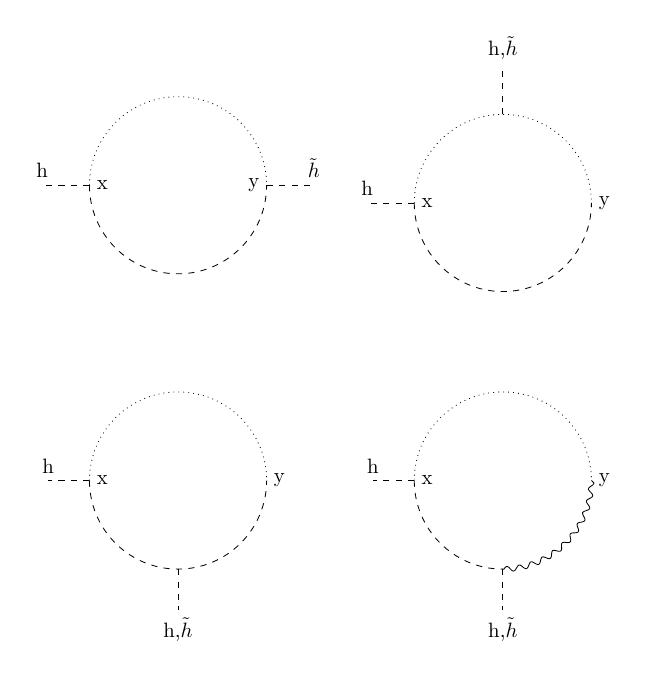}
	\caption{The diagrams contribution to the function $\tilde{D}$.}
	\label{fig:Dtilde factor}
\end{figure}

The corresponding Feynman diagrams about $\tilde{D}$ factor are shown in Figure~\ref{fig:Dtilde factor}.We begin by enumerating the characteristic integrals required for the computation.

\begin{equation}
	J^{(\tilde{D})}_1(m_1, m_2)=\int_{p}\frac{1}{(p^2+m_1^2)(p^2+m_2^2)[(p+k)^2+m_2^2]}\;,
\end{equation}

we have
\begin{equation}
	J^{(\tilde{D})}_1(m_1, m_2)=\frac{1}{8 \pi m_2(m_1+m_2)^2}+k^2\lp -\frac{m_1^2+4 m_1 m_2+7 m_2^2}{96 \pi  m_2^3 (m_1+m_2)^4}\rp+\mathcal{O}(k^4)\;,
\end{equation}

\begin{equation}
	J^{(\tilde{D})}_2(M, m_1, m_2)=\int_{p}\frac{p_i (2k+p)_j[\delta_{ij}-(1-\xi)\frac{p_i p_j}{p^2+\xi m_2^2}]}{(p^2+M^2)[(p+k)^2+m_1^2](p^2+m_2^2)}\;,
\end{equation}
It can be simplified as
\begin{equation}
	J^{(\tilde{D})}_2(M, m_1, m_2)=\int_p\frac{\xi(2 k+p)\cdot p}{(p^2+M^2)[(p+k)^2+m_1^2](p^2+\xi m_2^2)}\;,
\end{equation}
Considering specific quality relationships, there is $\xi m_2^2=M^2$

we have
\begin{equation}
	J^{(\tilde{D})}_2(M, m_1, m_2)=\frac{\xi  (M+2 m_1)}{8 \pi  (M+m_1)^2}+k^2\lp-\frac{\xi (M+2 m_1)}{8 \pi  (M+m_1)^4} \rp+\mathcal{O}(k^4)\;,
\end{equation}

\begin{equation}
	\begin{aligned}
		J^{(\tilde{D})}_3(M, m)&=\int_{p}\frac{1}{(p^2+M^2)[(p+k)^2+m^2]}\\
		&=\frac{1}{4 \pi(M+m)}+k^2\lp \frac{1}{12 \pi (M+m)^3} \rp+\mathcal{O}(k^4)\;,
	\end{aligned}
\end{equation}

\begin{equation}
	\begin{aligned}
		J^{(\tilde{D})}_4(M, m_1, m_2)&=\int_{p}\frac{p_i p_j[\delta_{ij}-(1-\xi)\frac{p_i p_j}{p^2+\xi m_2^2}]}{(p^2+M^2)[(p+k)^2+m_1^2](p^2+m_2^2)}\\
		&=\frac{\xi(M+2 m_1)}{8 \pi(M+m_1)^2}+k^2 \lp \frac{\xi (M-2 m_1)}{24 \pi(M+m_1)^4} \rp+\mathcal{O}(k^4)\;.
	\end{aligned}
\end{equation}

\subsection{The result of $\tilde{D}$}
\label{appendix:Dtilde}
The contributions are

\begin{equation}
	\begin{aligned}
		-\Pi^{\tilde{D}}_{h}&=2 X_W C_{h c^+ c^-}J^{(\tilde{D})}_1(m_{\chi^\pm}, m_{c_W})+X_Z C_{h c_Z c_Z}J^{(\tilde{D})}_1(m_{\chi^0}, m_{c_Z})\,\\
		&+2 X_W C_{h \chi^+ \chi^-}J^{(\tilde{D})}_1( m_{c_W},m_{\chi^\pm})+X_Z C_{h \chi \chi}J^{(\tilde{D})}_1(m_{c_Z}, m_{\chi^0})\,\\
		&+2 Q_W C_{h W^\pm \chi^\pm}J^{(\tilde{D})}_2(m_{c_W}, m_{\chi^\pm}, m_W)+Q_Z C_{h Z \chi}J^{(\tilde{D})}_2(m_{c_Z}, m_{\chi^0}, m_Z)\;,
	\end{aligned}
\end{equation}

\begin{equation}
	\begin{aligned}
		-\Pi^{\tilde{D}}_{\tilde{h}}&=2 Q_W C_{\tilde{h} \chi^\pm \bar{c}^\pm} J^{(\tilde{D})}_3(m_{\chi^\pm}, m_{c_W})+Q_Z C_{\tilde{h} \chi \bar{c}_Z} J^{(\tilde{D})}_3(m_{\chi^0}, m_{c_Z})\\
		&+2 X_W C_{\tilde{h} c^+ c^-}J^{(\tilde{D})}_1(m_{\chi^\pm}, m_{c_W})+X_Z C_{\tilde{h} c_Z c_Z}J^{(\tilde{D})}_1(m_{\chi^0}, m_{c_Z})\\
		&+2 X_W C_{\tilde{h} \chi^+ \chi^-}J^{(\tilde{D})}_1( m_{c_W},m_{\chi^\pm})+X_Z C_{\tilde{h} \chi \chi}J^{(\tilde{D})}_1(m_{c_Z}, m_{\chi^0})\,\\
		&+2 Q_W C_{\tilde{h} W^\pm \chi^\pm}J^{(\tilde{D})}_4(m_{c_W}, m_{\chi^\pm}, m_W)+Q_Z C_{\tilde{h} Z \chi}J^{(\tilde{D})}_4(m_{c_Z}, m_{\chi^0}, m_Z)\;.
	\end{aligned}
\end{equation}
Similarly, by extracting the terms proportional to $k^2$, we obtain
\begin{equation}
    \tilde{D}=\frac{\partial }{\partial k^2}\lp \Pi^{\tilde{D}}_{h}+\Pi^{\tilde{D}}_{\tilde{h}}\rp\;.
\end{equation}
The relevant additional vertices reads
\be
C_{h c^+ c^-}=-\frac{1}{2}g m_W \xi\,,
\ee
\be
C_{h c_Z c_Z}=-\frac{g}{2 \cos\theta} m_Z \xi\,,
\ee
\be
C_{h \chi^+ \chi^-}=-(2 \lambda \phi+3 c_6 \phi^3)\phi\,,
\ee
\be
C_{h \chi \chi}=-(2 \lambda \phi+3 c_6 \phi^3) \phi\,,
\ee
\be
C_{h W^\pm \chi^\pm}=-\frac{g}{2}
\ee
\be
C_{h Z \chi}=-\frac{1}{2}\sqrt{g^2+g^{'2}}\,,
\ee

\be
C_{\tilde{h} \chi^\pm \bar{c}^\pm}=\frac{1}{2} g \xi \,,
\ee
\be
C_{\tilde{h} \chi \bar{c}_Z}=\frac{1}{2}\sqrt{g^2+g^{'2}} \xi\,,
\ee
\be
C_{\tilde{h} c^+ c^-}=-\frac{1}{4} g^2 \xi \phi\,,
\ee
\be
C_{\tilde{h} c_Z c_Z}=-\frac{1}{4}\lp g^2+g^{'2} \rp \xi \phi\,,
\ee
\be
C_{\tilde{h} \chi^+ \chi^-}=-\frac{1}{2}g^2 \xi \tilde{\phi}\,,
\ee
\be
C_{\tilde{h} \chi \chi}=-\frac{1}{2}\lp g^2+g^{'2} \rp \xi \tilde{\phi}\,,
\ee
\be
C_{\tilde{h} W^\pm \chi^\pm}=\frac{g}{2}\,,
\ee
\be
C_{\tilde{h} Z \chi}=\frac{1}{2}\sqrt{g^2+g^{'2}}\,.
\ee

\subsection{C factor of Nielsen Identity at two-loop level}
\label{appendix:C factor two loop}
	
\begin{figure}[h!]
	\centering
	\includegraphics[width=0.8\linewidth]{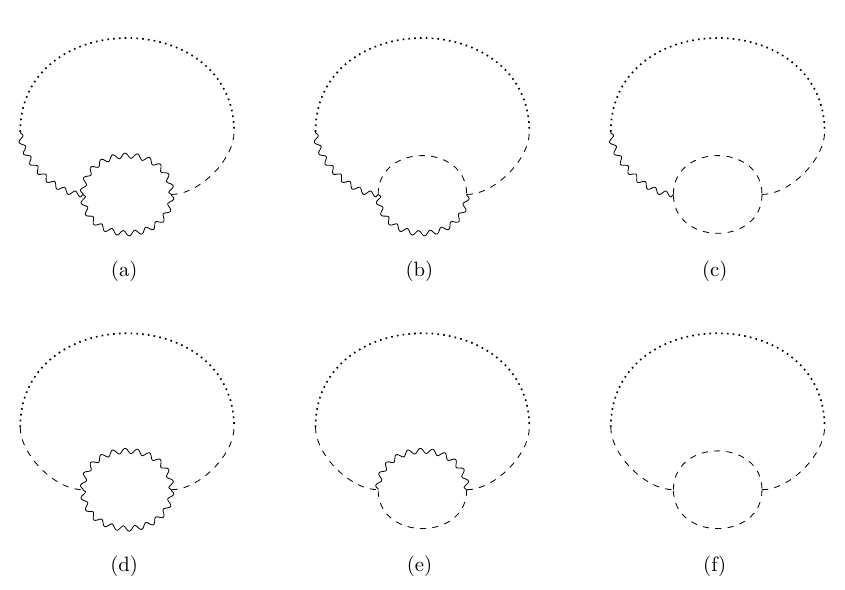}
	\caption{Diagrams for $C(\phi, \xi)$ at two loop level.}
	\label{fig:C:two loop}
\end{figure}

We expand $\exp((i/\hbar)S_\mathrm{eff})$, Calculate the C factor at two loop level. The relevant Feynman diagram is shown in Figure \ref{fig:C:two loop}.

The characteristic integral is derived from the schematic in Figure~\ref{fig:C:two loop}.(a):
\begin{equation}
	\begin{aligned}
    \label{eq:D_etavvvs}
		\mathcal{D}_{\eta VV VS}=&\int_{p,q}\frac{p^\mu}{p^2+m_1^2}\frac{g_{\mu\nu}-(1-\xi)\frac{p_\mu p_\nu}{p^2+\xi m_4^2}}{p^2+m_4^2}\frac{g_{\sigma\rho}-(1-\xi)\frac{q_\sigma q_\rho}{q^2+\xi m_2^2}}{q^2+m_2^2}\,\\
		&\times \frac{g_{r s}-(1-\xi)\frac{(p+q)_r (p+q)_s}{(p+q)^2+\xi m_3^2}}{(p+q)^2+m_3^2}\frac{g^{\rho s}}{p^2+m_5^2} \,\\
		&\times \lp g^{\sigma\nu}(p-q)^r-g^{\nu r}(2p+q)^\sigma+g^{\sigma r}(2q+p)^\nu \rp\;,
	\end{aligned}
\end{equation}
to perform the two-loop integral reduction, we first address the denominators. We have
\begin{equation}
	\label{eq:split fenmu1}
	\frac{1}{(p^2+x)(p^2+y)(p^2+z)}=\frac{\frac{1}{(x-y)(x-z)}}{p^2+x}+\frac{\frac{1}{(y-x)(y-z)}}{p^2+y}+\frac{\frac{1}{(z-x)(z-y)}}{p^2+z}\;.
\end{equation}
We begin by addressing the numerator of Eq.~\ref{eq:D_etavvvs}:
\begin{equation}
	\begin{aligned}
		\text{numerator}&=(D-1) \lp p^2+2 p\cdot q \rp\,\\
		&-(1-\xi)\lp\frac{p^2 q^2- (p \cdot q)^2}{q^2+\xi m_2^2}+\frac{(p \cdot q)^2-p^2 q^2}{(p+q)^2+\xi m_3^2}+\frac{(D-1) p^2 \left(2( p\cdot q)+p^2 \right)}{p^2+\xi m_4^2}\rp\,\\
		&+(1-\xi)^2\lp \frac{p^2 \left(p^2 q^2-(p\cdot q)^2\right)}{(q^2+\xi m_2^2) (p^2+\xi m_4^2)}+\frac{p^2 \left((p \cdot q)^2-p^2 q^2\right)}{[(p+q)^2+\xi m_3^2] (p^2+\xi m_4^2)}\rp\;,
	\end{aligned}
\end{equation}
according to Eq.~(\ref{eq:split fenmu1}), we insert the numerator and reduce it to more fundamental characteristic integrals, we set $m_2^2\to y, m_3^2\to z, m_4^2\to s$ and have
\begin{equation}
	\mathcal{Z}(x,y,z,s)=\int_{p,q}\frac{\text{numerator}}{(p^2+x)(q^2+y)[(p+q)^2+z]}\;.
\end{equation}
More exactly,
\begin{align}
	\mathcal{Z}(x,y,z,s)&=-\frac{\xi  (s-x) (y-z) \left(-2 z ((3-2 d) y+x)+(x-y)^2+z^2\right)}{4 y z (x-\xi  s)}H(x,y,z)\nn\\
	&-\frac{(\xi -1) \xi  s (y-z) \left((4 d-6) y z-2 \xi  s y+(z-\xi  s)^2+y^2\right) }{4 y z (x-\xi  s)}H(\xi s,y,z)\nn\\
	&+\frac{\xi  (s-x) \left(x^2-2 x (y+\xi  z)+(y-\xi  z)^2\right) }{4 z (x-\xi  s)}H(x,y,\xi z)\nn\\
	&-\frac{\xi  (s-x) \left(x^2-2 x (\xi  y+z)+(z-\xi  y)^2\right) }{4 y (x-\xi  s)}H(x,\xi y,z)\nn\\
	&+\frac{(\xi -1) \xi  s \left(-2 \xi  y (s+z)+\xi ^2 (s-z)^2+y^2\right) }{4 z (x-\xi  s)}H(\xi s,y,\xi z)\nn\\
	&-\frac{(\xi -1) \xi  s \left(\xi ^2 (s-y)^2-2 \xi  z (s+y)+z^2\right) }{4 y (x-\xi  s)}H(\xi s,\xi y,z)\nn\\
	&+\frac{\xi  (s-x) (-y (4 d+\xi -6)+x-z)}{4 y (x-\xi  s)}A(x) A(y)-\frac{\xi  (s-x) (x+\xi  y-z)}{4 y (x-\xi  s)}A(x)A(\xi y)\nn\\
	&-\frac{\xi  (s-x) (-z (4 d+\xi -6)+x-y)}{4 z (x-\xi  s)}A(x)A(z)+\frac{\xi  (s-x) (x-y+\xi  z)}{4 z (x-\xi  s)} A(x)A(\xi z)\nn\\
    &+\frac{\xi  (y-z) (-\xi  s+s-x+y+z)}{4 y z}A(y)A(z)-\frac{\xi  (-\xi  s+s-x+y+\xi  z)}{4 z}A(y)A(\xi z)\nn\\
	&-\frac{(\xi -1) \xi  s (y (4 d+\xi -6)-\xi  s+z)}{4 y (x-\xi  s)}A(y)A(\xi s)+\frac{\xi  (-\xi  s+s-x+\xi  y+z)}{4 y}A(z)A(\xi y)\nn\\
	&+\frac{(\xi -1) \xi  s (z (4 d+\xi -6)-\xi  s+y)}{4 z (x-\xi  s)}A(z)A(\xi s)+\frac{(\xi -1) \xi  s (z-\xi  (s+y))}{4 y (x-\xi  s)}A(\xi s)A(\xi y)\nn\\
	&+\frac{(\xi -1) \xi  s (\xi  (s+z)-y)}{4 z (x-\xi  s)}A(\xi s)A(\xi z)\;.
\end{align}

For the case where $x=\xi s$ arises, the above equation requires additional treatment, with the variable x retained
\begin{equation}
    \mathcal{Z}(x,y,z)=\mathcal{Z}(x=\xi s,y,z,s)\;,
\end{equation}

we have
\begin{align}
	\mathcal{Z}(x,y,z)&=\lp \frac{(d-1) \left(x^2 (d (\xi -1)-2 \xi +3)-x (d (\xi -1)-\xi +3) (y+z)+\xi  (y-z)^2\right)}{x^2-2 x (y+z)+(y-z)^2}\right.\nn\\
	&\left.+ \frac{x^2 (d (\xi -1)+1)-x (d (\xi -1)+\xi +1) (y+z)+\xi  (y-z)^2}{4yz} \rp (y-z)H(x,y,z)\nn\\
	&+\frac{x^2 (d (\xi -1)+1)-x (d (\xi -1)+\xi +1) (\xi  y+z)+\xi  (z-\xi  y)^2 }{4 y}H(x,\xi  y,z)\nn\\
	&+\frac{x^2 (d (-\xi )+d-1)+x (d (\xi -1)+\xi +1) (y+\xi  z)-\xi  (y-\xi  z)^2 }{4 z}H(x,y,\xi z)\nn\\
	&+\frac{1}{4 y \left(x^2-2 x (y+z)+(y-z)^2\right)}\biggl( x^3 (d (1-\xi)-1)+\xi  (y-z)^2 \Big(y (4 d+\xi -6)\nn\\
    &+z\Big)+x^2 \left(y \left(2 d^2 (\xi -1)+4 d+\xi ^2-2 \xi -2\right)+z (2 d (\xi -1)+\xi +2)\right) \nn\\
	&-x \Bigl(2 y z \left(3 d^2 (\xi -1)+d (10-6 \xi )+\xi ^2+\xi -7\right)+z^2 (d (\xi -1)+2 \xi +1)\nn\\
	&+y^2 \left(2 d^2 (\xi -1)+d (3 \xi +5)+2 \xi ^2-8 \xi -3\right)\Bigr)\biggr) A(x)A(y)\nn\\
	&+\frac{1}{4 z \left(x^2-2 x (y+z)+(y-z)^2\right)}\biggl( x^3 (d (-\xi )+d-1)-\xi(y-z)^2 \Big(z (4 d+\xi -6)\nn\\
    &+y\Big)-x^2 \left(z \left(2 d^2 (\xi -1)+4 d+\xi ^2-2 \xi -2\right)+y (2 d (\xi -1)+\xi +2)\right) \nn\\
	&+x \Bigl(2 y z \left(3 d^2 (\xi -1)+d (10-6 \xi )+\xi ^2+\xi -7\right)+y^2 (d (\xi -1)+2 \xi +1)\nn\\
	&+z^2 \left(2 d^2 (\xi -1)+d (3 \xi +5)+2 \xi ^2-8 \xi -3\right)\Bigr)\biggr) A(x)A(z)\nn\\
	&+\frac{d (\xi -1) x+x+\xi  (\xi  y-z)}{4 y} A(x)A(\xi y)-\frac{d (\xi -1) x+x+\xi  (\xi  z-y)}{4z}A(x)A(\xi z)\nn\\
	&+\lp\frac{(y-z)^2 (x (d (1-\xi )-1)+\xi  (y+z))}{4 y z}-\frac{(d^2-3d+2) (\xi -1) x(y-z)}{x^2-2 x (y+z)+(y-z)^2} \rp A(y)A(z)\nn\\
	&+\frac{d (\xi -1) x+x-\xi  (y+\xi  z)}{4 z}A(y)A(\xi z)-\frac{d (\xi -1) x+x-\xi  (\xi  y+z)}{4 y}A(\xi y)A(z)\;,
\end{align}

finally,
\begin{equation}
	\mathcal{D}_{\eta VV VS}=\frac{\mathcal{Z}(m_1^2,m_2^2,m_3^2)}{(m_1^2-m_4^2)(m_1^2-m_5^2)}+\frac{\mathcal{Z}(m_4^2,m_2^2,m_3^2,m_4^2)}{(m_4^2-m_1^2)(m_4^2-m_5^2)}+\frac{\mathcal{Z}(m_5^2,m_2^2,m_3^2,m_4^2)}{(m_5^2-m_1^2)(m_5^2-m_4^2)}\;.
\end{equation}

In the subsequent process, when we encounter calculations where the mass of photon is 0, it will cause the above calculations to diverge, which is the main source
Caused by $m_3 = 0(z = 0)$. We are now considering this situation.

\begin{align}
		\mathcal{Z}(x,y,0,s)&=\left(\frac{(\xi -1) \xi  s \left((5-4 d) y^2+\xi ^2 s^2-2 \xi  s y\right)}{4 y (x-\xi  s)}-\frac{(\xi -1)^2 \xi  s \left(\xi ^2 s^2-2 \xi  s y+y^2\right)}{4 (x-\xi  s)}\right)H(\xi  s,y,0)\nn\\
		&+\left(\frac{\xi  (s-x) \left(-(4 d-5) y^2+x^2-2 x y\right)}{4 y (x-\xi  s)}+\frac{(\xi -1) \xi  (x-s) \left(x^2-2 x y+y^2\right)}{4 (x-\xi  s)}\right)H(x,y,0)\nn\\
		&+\frac{\xi  (x-s) \left(x^2-2 \xi  x y+\xi ^2 y^2\right) }{4 y (x-\xi  s)}H(x,\xi  y,0)-\frac{(\xi -1) \xi ^3 s (s-y)^2 }{4 y (x-\xi  s)}H(\xi  s,\xi  y,0)\,\\
		&-\frac{(\xi -1) \xi  s  (y (4 d+(\xi -1) \xi  s+(\xi -1) y-5)-\xi  s)}{4 y (x-\xi  s)}A(y) A(\xi s)\nn\\
		&+ \frac{\xi (x-s) (y (4 d+(\xi -1) y-5)+x ((\xi -1) y-1))}{4 y (x-\xi  s)}A(x)A(y)\nn\\
		&+\frac{\xi (x-s)  (x+\xi  y)}{4 y (x-\xi  s)}A(x)A(\xi  y)-\frac{(\xi -1) \xi ^2 s (s+y) }{4 y (x-\xi s)}A(\xi  s) A(\xi  y)\;,
	\end{align}

\begin{equation}
	\begin{aligned}
		\mathcal{Z}(x,y,0)=&\frac{\left(x^2 (d (\xi -1)+1)-\xi  x y (d (\xi -1)+\xi +1)+\xi ^3 y^2\right) }{4 y}H(x,\xi  y,0)\,\\
		&+ \Biggl(\frac{(d-1) (\xi -1) y \left((d-2) x^2-(d-1) x y+y^2\right)}{x^2-2 x y+y^2}+(d-1) y\,\\
		&+\frac{1}{4} (\xi -1) \left(x^2 (d (\xi -1)+1)-x y (d (\xi -1)+\xi +1)+\xi  y^2\right)\,\\
		&+\frac{(\xi -1) \left(d x (y-x)+x y-y^2\right)}{4 y}-\frac{x^2-2 x y+y^2}{4 y}\Biggr)H(x,y,0)\,\\
		&+\biggl(\frac{(d-1) (\xi -1) (d x (x-y)-2 y (x-y))}{2(x-y)^2}+d-1-\frac{x+y}{4y}\,\\
		&+\frac{(\xi -1)^2 (d x+y)}{4}-\frac{(\xi -1) (d x+y)}{4y}+\frac{(\xi -1) (x+y)}{4}\biggr)A(x)A(y)\,\\
		&+\frac{\left(d (\xi -1) x+x+\xi ^2 y\right)}{4 y}A(x)A(\xi y) \;,
	\end{aligned}
\end{equation}
for $\mathcal{D}_{\eta VVVS}[m_3=0,m_1^2=\xi m_4^2]$
\begin{equation}
	\mathcal{D}_{\eta VV VS}=\frac{\mathcal{Z}(m_1^2,m_2^2,0)}{(m_1^2-m_4^2)(m_1^2-m_5^2)}+\frac{\mathcal{Z}(m_4^2,m_2^2,0,m_4^2)}{(m_4^2-m_1^2)(m_4^2-m_5^2)}+\frac{\mathcal{Z}(m_5^2,m_2^2,0,m_4^2)}{(m_5^2-m_1^2)(m_5^2-m_4^2)}\;,
\end{equation}

For (b)
\begin{equation}
	\mathcal{D}_{\eta SVVS}=\int_{p,q}\frac{p_\mu \lp g^{\mu \nu}-(1-\xi)\frac{p^\mu p^\nu}{p^2+\xi m_4^2}\rp g_{\nu \rho} \lp g^{\rho \sigma}-(1-\xi)\frac{q^\rho q^\sigma}{q^2+\xi m_3^2}\rp (2p+q)_\sigma }{(p^2+m_1^2)(p^2+m_4^2)(q^2+m_3^2)[(p+q)^2+m_2^2](p^2+m_5^2)}\;,
\end{equation}

Correspondingly, we can handle more basic integrals. Considering the form of the integral, we still retain the four-variable form.

\begin{equation}
	\mathcal{K}(x,y,z,s)=\int_{p,q}\frac{p_\mu \lp g^{\mu \nu}-(1-\xi)\frac{p^\mu p^\nu}{p^2+\xi s}\rp g_{\nu \rho} \lp g^{\rho \sigma}-(1-\xi)\frac{q^\rho q^\sigma}{q^2+\xi y}\rp (2p+q)_\sigma}{(p^2+x)(q^2+y)[(p+q)^2+z]}\;,
\end{equation}

we have
\begin{equation}
	\begin{aligned}
		\mathcal{K}(x,y,z,s)&=-\frac{\xi  (s-x) \left(x^2-2 x (y+z)+(y-z)^2\right)}{2 y (x-\xi  s)} H(x,y,z)\,\\
		&-\frac{(\xi -1) \xi  s \left(-2 y (\xi  s+z)+(z-\xi  s)^2+y^2\right) }{2 y (x-\xi  s)}H(\xi  s,y,z)\,\\
		&+\frac{\xi  (s-x) (x-z) (x+\xi  y-z) }{2 y (x-\xi  s)}H(x,\xi  y,z)\,\\
		&+\frac{(\xi -1) \xi  s (\xi  s-z) (\xi  (s+y)-z) }{2 y (x-\xi  s)}H(\xi  s,\xi  y,z)\,\\
		&-\frac{\xi  (s-x) (x+y-z)}{2 y (x-\xi  s)}A(x) A(y) +\frac{\xi  (s-x) (x-z) }{2 y (x-\xi  s)}A(x)A(\xi  y)  \,\\
		&+\frac{\xi (-\xi  s+s-x+y+z)}{2 y}A(y) A(z)-\frac{(\xi -1) \xi  s  (\xi  s+y-z)}{2 y (x-\xi  s)}A(y)A(\xi  s)\,\\
		&+\frac{(\xi -1) \xi  s }{2 (x-\xi  s)}A(\xi  s)A(z)+\frac{\xi  ((\xi -1) s+x+2 \xi  y-z)}{2 y}A(\xi y)A(z) \,\\
		&+\frac{(\xi -1) \xi  s  (\xi  s-z)}{2 y (x-\xi  s)}A(\xi  s) A(\xi  y)+\frac{\xi  (x-s)}{2 \xi  s-2 x}A(x)A(z) \;,
	\end{aligned}
\end{equation}

For the case where $x=\xi s$ arises, the above equation requires additional treatment, with the variable x retained

\begin{equation}
	\mathcal{K}(x,y,z)=\mathcal{K}(x=\xi s,y,z,s)\;,
\end{equation}

We have
\begin{align}
		\mathcal{K}(x,y,z)&=\frac{x^2 (d (\xi -1)+1)-x (d (\xi -1)+\xi +1) (y+z)+\xi  (y-z)^2}{2 y}H(x,y,z)\nn\\
		&+\frac{1}{2 y \left(x^2-2 x (\xi  y+z)+(z-\xi  y)^2\right)}\Bigg(x^3 (z (3 d (\xi -1)+\xi +3)+\xi  (4 \xi -3) y)\nn\\
		&-x^2 \left(\xi ^2 y^2 (d (-\xi )+d+2 \xi -3)+3 z^2 (d (\xi -1)+\xi +1)+\xi  (2-3 \xi ) y z\right)\nn\\
		&+x (z-\xi  y) \left(\xi  y z (d (\xi -1)-7 \xi +6)+z^2 (d (\xi -1)+3 \xi +1)+\xi ^2 (2 \xi -1) y^2\right)\nn\\
		&+x^4 (d (-\xi )+d-1)-\xi  z (z-\xi  y)^3\Bigg)H(x,\xi  y,z)\nn\\
		&-\frac{\xi  \left(x^2 (2 d (\xi -1)-4 \xi +5)-2 x (z (d (\xi -1)-2 \xi +3)+\xi y)+(z-\xi y)^2\right)}{2 \left(x^2-2 x (\xi  y+z)+(z-\xi  y)^2\right)}A(x)A(z) \nn\\
        &+\frac{ (x (d (-\xi )+d-1)+\xi  (y+z))}{2 y}A(y) A(z)+\frac{ (d (\xi -1) x+x+\xi  (y-z))}{2 y}A(x)A(y)\nn\\
		&+\frac{1}{2 y \left(x^2-2 x (\xi y+z)+(z-\xi  y)^2\right)} \Bigg(x^3 (d (-\xi )+d-1)+\xi  z (z-\xi  y)^2\nn\\
		&-x \left((d-4) (\xi -1) \xi  y z+z^2 (d (\xi -1)+2 \xi +1)+\xi ^2 (2 \xi -1) y^2\right)\nn\\
		&+x^2 (\xi  y (d (\xi -1)+2 \xi )+z (2 d (\xi -1)+\xi +2))\Bigg)A(x)A(\xi y)\nn\\
		&+\frac{1}{2 y \left(x^2-2 x (\xi  y+z)+(z-\xi  y)^2\right)}\Bigg(d (\xi -1) x^3+(d-4) (\xi -1) \xi  x^2 y \nn\\
		&-x^2 z (2 d (\xi -1)+\xi +2)+\xi  x y z (d (-\xi )+d-4 \xi )+x z^2 (d (\xi -1)+2 \xi +1)\nn\\
		&+x^3-\xi ^2 (2 \xi +1) x y^2-\xi  (z-2 \xi  y) (z-\xi  y)^2\Bigg)A(z) A(\xi  y)\;,
\end{align}

finally,
\begin{equation}
	\mathcal{D}_{\eta SV VS}=\frac{\mathcal{K}(m_1^2,m_3^2,m_2^2)}{(m_1^2-m_4^2)(m_1^2-m_5^2)}+\frac{\mathcal{K}(m_4^2,m_3^2,m_2^2,m_4^2)}{(m_4^2-m_1^2)(m_4^2-m_5^2)}+\frac{\mathcal{K}(m_5^2,m_3^2,m_2^2,m_4^2)}{(m_5^2-m_1^2)(m_5^2-m_4^2)}\;,
\end{equation}

when we deal with the case where $m_3=0$ such that $y=0$, we have

\begin{equation}
	\begin{aligned}
		\mathcal{K}(x,0,z,s)&=-\frac{\xi  (s-x) (d (\xi -1) x+d (\xi -1) z-4 \xi  x+x-2 \xi  z+z)}{2 (x-\xi  s)}H(x,0,z)\,\\
		&-\frac{(\xi -1) \xi  s (\xi  s (d (\xi -1)-4 \xi +1)+z (d (\xi -1)-2 \xi +1))}{2 (x-\xi  s)}H(\xi s,0,z)\,\\
		&-\frac{\xi  (d (\xi -1)-2 \xi +1) (s-x)}{2 (x-\xi  s)}A(x)A(z)\,\\
		&-\frac{(\xi -1) \xi  s (d (\xi -1)-2 \xi +1)}{2 (x-\xi s)}A(\xi s)A(z)\;,
	\end{aligned}
\end{equation}

	\begin{align}
		\mathcal{K}(x,0,z)&=\frac{1}{2 (x-z)}\Bigg(x^2 \left(d^2 (\xi -1)^2+d \left(-5 \xi ^2+8 \xi -3\right)+4 \xi ^2-9 \xi +2\right)\nn\\
		&+x z \left(d^2 (\xi -1)^2+d \left(-6 \xi ^2+11 \xi -5\right)+2 \left(6 \xi ^2-7 \xi +2\right)\right)\nn\\
		&+\xi  z^2 (d (-\xi )+d+2 \xi -1)\Bigg)H(x,0,z)\nn\\
		&+\frac{x \left(d^2 (\xi -1)^2+d \left(-4 \xi ^2+7 \xi -3\right)+4 \xi ^2-7 \xi +2\right)+\xi  z (d (-\xi )+d+2 \xi -1)}{2 (x-z)}A(x)A(z)\;,
	\end{align}

for $\mathcal{D}_{\eta SVVS}[m_3=0,m_1^2=\xi m_4^2]$
\begin{equation}
	\mathcal{D}_{\eta SV VS}=\frac{\mathcal{K}(m_1^2,0,m_2^2)}{(m_1^2-m_4^2)(m_1^2-m_5^2)}+\frac{\mathcal{K}(m_4^2,0,m_2^2,m_4^2)}{(m_4^2-m_1^2)(m_4^2-m_5^2)}+\frac{\mathcal{K}(m_5^2,0,m_2^2,m_4^2)}{(m_5^2-m_1^2)(m_5^2-m_4^2)}\;,
\end{equation}

For (c)
\begin{equation}
	\mathcal{D}_{\eta SSVS}=\int_{p,q}\frac{p^\mu (2q+p)^\nu \lp g^{\mu \nu}-(1-\xi)\frac{p^\mu p^\nu}{p^2+\xi m_4^2}\rp}{(p^2+m_1^2)(q^2+m_2^2)[(p+q)^2+m_3^2](p^2+m_4^2)(p^2+m_5^2)}\;,
\end{equation}
following the same approach as before, due to specific quality relations,according to Eq.~(\ref{eq:split fenmu1}), we divide it into two types of basic integrals,

\begin{equation}
	\mathcal{R}(x,y,z,s)=\int_{p,q}\frac{p^\mu (2q+p)^\nu \lp g^{\mu \nu}-(1-\xi)\frac{p^\mu p^\nu}{p^2+\xi s}\rp}{(p^2+x)(q^2+y)[(p+q)^2+z]}\;,
\end{equation}
we have
\begin{equation}
	\begin{aligned}
		\mathcal{R}(x,y,z,s)=&\frac{\xi  (x-s) (y-z) }{x-\xi  s}H(x,y,z)+\frac{(\xi -1) \xi  s (y-z) }{\xi  s-x}H(\xi  s,y,z)\,\\
		&+\frac{(\xi -1) \xi  s }{\xi s-x}A(y) A(\xi s)+\frac{\xi(s-x)}{\xi s-x}A(x)A(y)\,\\
		&+\frac{\xi  (x-s)}{\xi s-x}A(x) A(z)-\frac{(\xi -1) \xi  s }{\xi  s-x}A(z) A(\xi  s)\;.
	\end{aligned}
\end{equation}
For the case where $x=\xi s$ arises, the above equation requires additional treatment, with the variable x retained
\begin{equation}
	\mathcal{R}(x,y,z)=\mathcal{R}(x=\xi s,y,z,s)\;,
\end{equation}
we have
\begin{equation}
	\begin{aligned}
		\mathcal{R}(x,y,z)=&\left(\frac{(\xi -1) (y-z) \left((d-2) x^2-(d-1) x (y+z)+(y-z)^2\right)}{x^2-2 x (y+z)+(y-z)^2}+y-z\right)H(x,y,z)\,\\
		&+\left(\frac{(\xi -1) (d x (x-y-3 z)-2 (y-z) (x-y+z))}{2 \left(x^2-2 x (y+z)+(y-z)^2\right)}+1\right)A(x)A(y)\,\\
		&+\left(\frac{(\xi -1) (d x (-x+3 y+z)-2 (y-z) (x+y-z))}{2 \left(x^2-2 x (y+z)+(y-z)^2\right)}-1\right)A(x)A(z)\,\\
		&+\frac{(2-d) (\xi -1) x  (y-z)}{x^2-2 x (y+z)+(y-z)^2}A(y) A(z)\;.
	\end{aligned}
\end{equation}

Finally, all contributions are
\begin{equation}
	\mathcal{D}_{\eta SS VS}=\frac{\mathcal{R}(m_1^2,m_2^2,m_3^2)}{(m_1^2-m_4^2)(m_1^2-m_5^2)}+\frac{\mathcal{R}(m_4^2,m_2^2,m_3^2,m_4^2)}{(m_4^2-m_1^2)(m_4^2-m_5^2)}+\frac{\mathcal{R}(m_5^2,m_2^2,m_3^2,m_4^2)}{(m_5^2-m_1^2)(m_5^2-m_4^2)}\;.
\end{equation}

In the subsequent calculations, we will have the values of $m_4$ and $m_5$ being the same. We need to reprocess the denominator. We have
\begin{equation}
	\label{eq:split fenmu2}
	\frac{1}{(p^2+x)(p^2+y)^2}=\frac{1}{(x-y)^2}\lp\frac{1}{p^2+x}-\frac{1}{p^2+y}\rp+\frac{1}{x-y}\frac{1}{(p^2+y)^2}\;,
\end{equation}
for $(d)$
\begin{equation}
	\mathcal{D}_{\eta VVSS}=\int_{p,q}\frac{\lp g^{\mu \nu}-(1-\xi)\frac{q^\mu q^\nu}{q^2+\xi m_2^2}\rp \lp g_{\mu \nu}-(1-\xi)\frac{(p+q)_\mu (p+q)_\nu}{(p+q)^2+\xi m_3^2}\rp }{(p^2+m_1^2)(q^2+m_2^2)[(p+q)^2+m_3^2](p^2+m_4^2)(p^2+m_5^2)}\;.
\end{equation}
For $m_4=m_5$, according to Eq.~(\ref{eq:split fenmu2}), we deal with more fundamental integrals
\begin{equation}
	\mathcal{X}(x,y,z)=\int_{p,q}\frac{\lp g^{\mu \nu}-(1-\xi)\frac{q^\mu q^\nu}{q^2+\xi y}\rp \lp g_{\mu \nu}-(1-\xi)\frac{(p+q)_\mu (p+q)_\nu}{(p+q)^2+\xi z}\rp }{(p^2+x)(q^2+y)[(p+q)^2+z]}\,,
\end{equation}
we obtain the final expression,
\begin{equation}
	\begin{aligned}
		\mathcal{X}(x,y,z)=&\left(d-2+\frac{(-x+y+z)^2}{4 y z}\right) H(x,y,z)+\frac{(x-\xi  (y+z))^2 }{4 y z}H(x,\xi  y,\xi  z)\,\\
		&+\left(\xi -\frac{(-x+y+\xi  z)^2}{4 y z}\right) H(x,y,\xi  z)+\left(\xi -\frac{(-x+\xi  y+z)^2}{4 y z}\right) H(x,\xi  y,z)\,\\
		&+\frac{\xi -1}{4y}A(x)A(y)-\frac{\xi-1}{4y}A(x)A(\xi y)+\frac{\xi-1}{4z}A(x)A(z)-\frac{\xi-1}{4z}A(x)A(\xi z)\,\\
		&+\frac{-x+y+z}{4 yz}A(y)A(z)-\frac{-x+y+\xi z}{4yz}A(y)A(\xi z)\,\\
		&+\frac{ (\xi  (y+z)-x)}{4 y z}A(\xi y) A(\xi z)-\frac{-x+\xi y+z}{4 y z}A(z) A(\xi y)\;,
	\end{aligned}
\end{equation}
and,
\begin{equation}
	\mathscr{X}(x,y,z)=\int_{p,q}\frac{\lp g^{\mu \nu}-(1-\xi)\frac{q^\mu q^\nu}{q^2+\xi y}\rp \lp g_{\mu \nu}-(1-\xi)\frac{(p+q)_\mu (p+q)_\nu}{(p+q)^2+\xi z}\rp }{(p^2+x)^2(q^2+y)[(p+q)^2+z]}\;,
\end{equation}
with the exact expression reads
	\begin{align}
	\mathscr{X}(x,y,z)=&\frac{(x-\xi  (y+z))[(d-1)(x^2-2x(y+z)\xi+(y^2+z^2)\xi^2)+2(d-5)yz\xi^2]}{4 y z \left(x^2-2 \xi  x (y+z)+(\xi  y-\xi  z)^2\right)}H(x,\xi  y,\xi  z)\nn\\
	&+\frac{(d-1) (-x+y+z) \left(-2 z ((7-2 d) y+x)+(x-y)^2+z^2\right) }{4 y z \left(x^2-2 x (y+z)+(y-z)^2\right)}H(x,y,z)\nn\\
	&-\frac{(d-1) (-x+y+\xi  z)}{4 y z}H(x,y,\xi  z)-\frac{(d-1) (-x+\xi  y+z) }{4 y z}H(x,\xi  y,z)\nn\\
	&-\frac{(d-2)\xi ^2 (x+\xi  (z-y))}{2 x \left(x^2-2 \xi  x (y+z)+\xi ^2 (y-z)^2\right)}A(x) A(\xi  y)-\frac{d-1}{4yz}A(y)A(\xi z)\nn\\
	&-\frac{(d-2)\xi ^2 (x+\xi  (y-z))}{2 x \left(x^2-2 \xi  x (y+z)+\xi ^2 (y-z)^2\right)}A(x) A(\xi z)-\frac{d-1}{4 y z} A(z) A(\xi  y)\nn\\
	&-\frac{(d-2) (d-1)  (x-y+z)}{2 x \left(x^2-2 x (y+z)+(y-z)^2\right)}A(x)A(y)\nn\\
	&-\frac{(d-2) (d-1)  (x+y-z)}{2 x \left(x^2-2 x (y+z)+(y-z)^2\right)}A(x)A(z)\nn\\
	&+ \left(\frac{(d-2) \xi ^2}{x^2-2 \xi  x (y+z)+\xi ^2 (y-z)^2}+\frac{d-1}{4 y z}\right)A(\xi y) A(\xi z)\nn\\
	&+\lp\frac{(d-2)(d-1)}{x^2-2 x (y+z)+(y-z)^2}+\frac{d-1}{4yz} \rp A(y)A(z)\;.
	\end{align}
Finally, the summation reads

\begin{equation}
	\mathcal{D}_{\eta VVSS}=\frac{\mathcal{X}(m_1^2,m_2^2,m_3^2)}{(m_1^2-m_4^2)^2}-\frac{\mathcal{X}(m_4^2,m_2^2,m_3^2)}{(m_1^2-m_4^2)^2}+\frac{\mathscr{X}(m_4^2,m_2^2,m_3^2)}{m_1^2-m_4^2}\,.
\end{equation}

When we deal with the case where $m_3 = 0$ such that $z = 0$, we have
\begin{equation}
	\begin{aligned}
		\mathcal{X}(x,y,0)=&\frac{(d-1) ((\xi -1) x+(\xi +3) y)}{4 y}H(x,y,0)\,\\
		&+\lp \xi-\frac{(\xi -1) ((d-1) x+(d-5) \xi  y)}{4 y}\rp H(x,\xi  y,0)\,\\
		&+\frac{(d-1) (\xi -1) }{4 y}A(x)A(y)-\frac{(d-1) (\xi -1) }{4 y}A(x)A(\xi y)\;,
	\end{aligned}
\end{equation}

\begin{equation}
	\begin{aligned}
		\mathscr{X}(x,y,0)=&-\frac{(d-1) ((d-2) (\xi -1) x+d (\xi +3) y-4 (\xi +2) y)}{4 y (x-y)}H(x,y,0) \,\\
		&\frac{\xi  y \left(-\left(d^2 (\xi -1)\right)+d (9 \xi -5)-16 \xi +4\right)-\left(d^2-3 d+2\right) (\xi -1) x}{4 y (\xi  y-x)}H(x,\xi  y,0)\,\\
		&+\frac{(d-2) \left((d-1) (\xi -1) x-2 \xi ^2 y\right)}{4 x y (x-\xi  y)}A(x) A(\xi y)\,\\
		& -\frac{(d-2) (d-1) ((\xi -1) x+2 y)}{4 x y (x-y)}A(x)A(y)\;.
	\end{aligned}
\end{equation}
So, the expression of $\mathcal{D}_{\eta VVSS}[m_3=0,m_4=m_5]$ is,

\begin{equation}
	\begin{aligned}
		\mathcal{D}_{\eta VVSS}[m_3=0,m_4=m_5]=\frac{\mathcal{X}(m_1^2,m_2^2,0)}{(m_1^2-m_4^2)^2}-\frac{\mathcal{X}(m_4^2,m_2^2,0)}{(m_1^2-m_4^2)^2}+\frac{\mathscr{X}(m_4^2,m_2^2,0)}{m_1^2-m_4^2}\;.
	\end{aligned}
\end{equation}
The (e) contribution reads,
\begin{equation}
	\mathcal{D}_{\eta VS SS}=\int_{p,q}\frac{(2p+q)^\mu (2p+q)^\nu \lp g_{\mu \nu}-(1-\xi)\frac{q_\mu q_\nu}{q^2+\xi m_2^2}\rp}{(p^2+m_1^2)(q^2+m_2^2)[(p+q)^2+m_3^2](p^2+m_4^2)(p^2+m_5^2)}\,,
\end{equation}
For $m_4=m_5$, according to Eq.~(\ref{eq:split fenmu2}), we deal with more fundamental integrals
\begin{equation}
	\mathcal{Y}(x,y,z)=\int_{p,q}\frac{(2p+q)^\mu (2p+q)^\nu \lp g_{\mu \nu}-(1-\xi)\frac{q_\mu q_\nu}{q^2+\xi y}\rp}{(p^2+x)(q^2+y)[(p+q)^2+z]}\,,
\end{equation}
we have
\begin{equation}
	\begin{aligned}
		\mathcal{Y}(x,y,z)=&-\frac{(x-z)^2}{y}H(x,\xi y,z)+\left(\frac{(x-z)^2}{y}-2 x+y-2 z\right) H(x,y,z)\,\\
		&+\frac{-x+\xi y+z}{y}A(x)A(\xi y)+\frac{x+y-z}{y}A(x)A(y)-A(x)A(z)\,\\
		&+\left(\xi +\frac{x-z}{y}\right)A(z) A(\xi  y)+\frac{-x+y+z}{y}A(y) A(z)\;.
	\end{aligned}
\end{equation}

And,
\begin{equation}
	\mathscr{Y}(x,y,z)=\int_{p,q}\frac{(2p+q)^\mu (2p+q)^\nu \lp g_{\mu \nu}-(1-\xi)\frac{q_\mu q_\nu}{q^2+\xi y}\rp}{(p^2+x)^2(q^2+y)[(p+q)^2+z]}\,,
\end{equation}
which related with other integrals as
\begin{equation}
	\begin{aligned}
		\mathscr{Y}(x,y,z)=&\frac{(x-z) \left(-\xi  y (d x-d z+x+7 z)+(d-1) (x-z)^2+2 \xi ^2 y^2\right) }{y \left(x^2-2 x (\xi  y+z)+(z-\xi  y)^2\right)}H(x,\xi  y,z)\,\\
		&+\frac{(d-1) (-x+y+z) }{y}H(x,y,z)+\frac{1-d}{y}A(x)A(y)+\frac{(d-1) A(y) A(z)}{y}\,\\
		&-\frac{(d-1) (x-z)^2-2 \xi  y (x+z)+\xi ^2 y^2}{y \left(x^2-2 x (\xi  y+z)+(z-\xi  y)^2\right)}A(z) A(\xi  y)\,\\
		&+\frac{(d-2)\xi (x-z) (3 x-\xi  y+z)}{2 x \left(x^2-2 x (\xi  y+z)+(z-\xi  y)^2\right)}A(x)A(z)\,\\
		&+\left(\frac{(d-2) \xi  (x (\xi  y+8 z)+\xi  y (z-\xi  y))}{2 x \left(x^2-2 x (\xi  y+z)+(z-\xi  y)^2\right)}+\frac{d-1}{y}\right)A(x)A(\xi y)\;,
	\end{aligned}
\end{equation}
Then, we have
\begin{equation}
	\mathcal{D}_{\eta VS SS}=\frac{\mathcal{Y}(m_1^2,m_2^2,m_3^2)}{(m_1^2-m_4^2)^2}-\frac{\mathcal{Y}(m_4^2,m_2^2,m_3^2)}{(m_1^2-m_4^2)^2}+\frac{\mathscr{Y}(m_4^2,m_2^2,m_3^2)}{m_1^2-m_4^2}\,.
\end{equation}

When we deal with the case where $m_2 = 0$ such that $y = 0$, we have
\begin{equation}
	\begin{aligned}
		\mathcal{Y}(x,0,z)=&(d-3) (\xi -1) (x+z) H(x,0,z)-2 (x+z) H(x,0,z)\,\\
		&+ (d (\xi -1)-2 \xi +1)A(x) A(z)\;,
	\end{aligned}
\end{equation}
and
\begin{equation}
	\begin{aligned}
		\mathscr{Y}(x,0,z)=&-\frac{(d (\xi -1)-3 \xi +1) ((d-2) x+(d-4) z)}{x-z}H(x,0,z)\,\\
		&-\frac{(d-2)(x (2 d (\xi -1)-5 \xi +2)-\xi  z)}{2 x (x-z)}A(x) A(z)\;.
	\end{aligned}
\end{equation}
Thus, for $\mathcal{D}_{\eta VSSS}[m_2=0,m_4=m_5]$, we have
\begin{equation}
	\mathcal{D}_{\eta VSSS}[m_2=0,m_4=m_5]=\frac{\mathcal{Y}(m_1^2,0,m_3^2)}{(m_1^2-m_4^2)^2}-\frac{\mathcal{Y}(m_4^2,0,m_3^2)}{(m_1^2-m_4^2)^2}+\frac{\mathscr{Y}(m_4^2,0,m_3^2)}{m_1^2-m_4^2}\,,
\end{equation}

The (f) contribution is,
\begin{equation}
	\mathcal{D}_{\eta SS SS}=\int_{p,q}\frac{1}{(p^2+m_1^2)(q^2+m_2^2)[(p+q)^2+m_3^2](p^2+m_4^2)(p^2+m_5^2)}\;.
\end{equation}
When $m_4=m_5$, according to Eq.~(\ref{eq:split fenmu2}), we deal with more fundamental integrals
\begin{equation}
	\mathcal{H}(x,y,z)=\int_{p,q}\frac{1}{(p^2+x)(q^2+y)[(p+q)^2+z]}=H(x,y,z)\;,
\end{equation}
and
\begin{equation}
	\mathscr{H}(x,y,z)=\int_{p,q}\frac{1}{(p^2+x)^2(q^2+y)[(p+q)^2+z]}\;,
\end{equation}
after calculation, we then have
\begin{equation}
	\begin{aligned}
		\mathscr{H}(x,y,z)=&-\frac{(d-3) (x-y-z) }{x^2-2 x (y+z)+(y-z)^2}H(x,y,z)\,\\
		&-\frac{(d-2)  (x-y+z)}{2 x \left(x^2-2 x (y+z)+(y-z)^2\right)}A(x) A(y)\,\\
		&-\frac{(d-2)  (x+y-z)}{2 x \left(x^2-2 x (y+z)+(y-z)^2\right)}A(x) A(z)\,\\
		&+\frac{(d-2) }{x^2-2 x (y+z)+(y-z)^2}A(y) A(z)\;.
	\end{aligned}
\end{equation}
And, therefore,
\begin{equation}
	\mathcal{D}_{\eta SS SS}=\frac{\mathcal{H}(m_1^2,m_2^2,m_3^2)}{(m_1^2-m_4^2)^2}-\frac{\mathcal{H}(m_4^2,m_2^2,m_3^2)}{(m_1^2-m_4^2)^2}+\frac{\mathscr{H}(m_4^2,m_2^2,m_3^2)}{m_1^2-m_4^2}\,.
\end{equation}
So there is,
\begin{equation}
	\begin{aligned}
		C^{(2),1}&=2 Q_W C_{W^+ W^- Z} C_{W^- Z \chi^+} \mathcal{D}_{\eta VVVS}(m_{cW},m_W,m_Z,m_W,m_{\chi^\pm})\,\\
		&+2 Q_W C_{W^+ W^- A} C_{W^- A \chi^+} \mathcal{D}_{\eta VVVS}(m_{cW},m_W,m_A,m_W,m_{\chi^\pm})\,\\
		&+2 Q_W C_{W^+ \chi^- Z} C_{\chi^- Z \chi^+} \mathcal{D}_{\eta SVVS}(m_{cW},m_{\chi^\pm},m_Z,m_W,m_{\chi^\pm})\,\\
		&+2 Q_W C_{W^+ \chi^- A} C_{\chi^- A \chi^+} \mathcal{D}_{\eta SVVS}(m_{cW},m_{\chi^\pm},m_A,m_W,m_{\chi^\pm})\,\\
		&+Q_Z C_{h Z Z} C_{h \chi Z} \mathcal{D}_{\eta SVVS}(m_{cZ},m_h,m_Z,m_Z,m_{\chi^0})\,\\
		&+2 Q_W C_{W^+ \chi^- h} C_{h \chi^- \chi^+} \mathcal{D}_{\eta SSVS}(m_{cW},m_{\chi^\pm},m_h,m_W,m_{\chi^\pm})\,\\
		&+Q_Z C_{h \chi Z} C_{h \chi \chi} \mathcal{D}_{\eta SSVS}(m_{cZ},m_{\chi^0},m_h,m_Z,m_{\chi^0})\,,\\
	\end{aligned}
\end{equation}
and
\begin{equation}
	\begin{aligned}
		C^{(2),2}&=2 X_W C_{W^+ \chi^- Z}^2 \mathcal{D}_{\eta VVSS}(m_{cW},m_W,m_Z,m_{\chi^\pm},m_{\chi^\pm})\,\\
		&+2 X_W C_{W^+ \chi^- A}^2 \mathcal{D}_{\eta VVSS}(m_{cW},m_W,m_A,m_{\chi^\pm},m_{\chi^\pm})\,\\
		&+2 X_W C_{W^+ \chi^- \chi}^2 \mathcal{D}_{\eta VSSS}(m_{cW},m_W,m_{\chi^0},m_{\chi^\pm},m_{\chi^\pm})\,\\
		&+2 X_W C_{Z \chi^- \chi^+}^2 \mathcal{D}_{\eta VSSS}(m_{cW},m_Z,m_{\chi^0},m_{\chi^\pm},m_{\chi^\pm})\,\\
		&+2 X_W C_{A \chi^- \chi^+}^2 \mathcal{D}_{\eta VSSS}(m_{cW},m_A,m_{\chi^0},m_{\chi^\pm},m_{\chi^\pm})\,\\
		&+2 X_Z C_{W^+ \chi^- \chi}^2 \mathcal{D}_{\eta VSSS}(m_{cZ},m_W,m_{\chi^\pm},m_{\chi^0},m_{\chi^0})\,\\
		&+X_Z C_{h \chi Z}^2 \mathcal{D}_{\eta VSSS}(m_{cZ},m_Z,m_h,m_{\chi^0},m_{\chi^0})\,\\
		&+2 X_W C_{h \chi^- \chi^+}^2 \mathcal{D}_{\eta SSSS}(m_{cW},m_{\chi^\pm},m_h,m_{\chi^\pm},m_{\chi^\pm})\,\\
		&+X_Z C_{h \chi \chi}^2 \mathcal{D}_{\eta SSSS}(m_{cZ},m_{\chi^0},m_h,m_{\chi^0},m_{\chi^0})\,,\\
	\end{aligned}
\end{equation}

where $Q_i, X_i(i=W,Z)$ are the additional coefficient for calculating the factor integral of C,
\begin{equation}
	\begin{aligned}
		Q_W=\frac{1}{4} g,& \,\quad &Q_Z&=\frac{1}{4} \sqrt{g^2+g^{'2}},\,\\
		X_W=\frac{1}{4} g \xi m_W,& \,\quad &X_Z&=\frac{1}{4} \sqrt{g^2+g^{'2}}\xi m_Z.\,\\
	\end{aligned}
\end{equation}

Finally,
\begin{equation}
	C^{(2)}=C^{(2),1}+C^{(2),2}\;.
\end{equation}

\section{Nielsen Identities in 4d}
\label{appendix:4D Nilsen C}
for $\xi_W=\xi_B=\xi$,with the $C(\phi,\xi)$ functions given by

\begin{align}
\label{CW}
C_{W}(\phi,\xi)&=\frac{1}{2} g \int \frac{ \xi m_W}{\lp k^2-m^2_{\chi^+}\rp \lp k^2-m^2_{cW} \rp}\ , \\
\label{CZ}
C_{Z}(\phi,\xi)&=\frac{1}{2} g' \int \frac{\frac{1}{2}\xi m_B}{\lp k^2-m^2_{\chi^0}\rp \lp k^2-m^2_{cZ} \rp} + \frac{1}{2} g \int \frac{ \frac{1}{2}\xi m_W}{\lp k^2-m^2_{\chi^0}\rp \lp k^2-m^2_{c^Z} \rp} \, ,
\end{align}
with
\be
C(\phi,\xi)=C_{W}(\phi,\xi)+C_{Z}(\phi,\xi)\,.
\ee
With the previous expressions it is  straightforward to check that the one-loop Nielsen identities
\be
\xi \frac{\partial V_1(\phi)}{\partial \xi} + C(\phi,\xi) \frac{\partial V_0(\phi)}{\partial \phi}=0\ ,
\ee
are indeed fulfilled. For the numerical analyses in the following sections discuss the dependence of different quantities with $\xi$ at the EW scale (at $\bmu=M_h$).

The main loop integral is
\be
I(m_1,m_2)=i \int\frac{1}{(k^2-m^2_1)(k^2-m^2_2)}\;,
\ee
slove it:
\be
I(m_1,m_2)=\frac{m_2^2 \left(\Delta_\epsilon+2 \log (\mu )-2 \log \left(m_2\right)+1\right)-m_1^2 \left(\Delta_\epsilon+2 \log (\mu )-2 \log \left(m_1\right)+1\right)}{16 \pi ^2 \left(m_1^2-m_2^2\right)}\;,
\ee
where we introduced the modified minimal subtraction term $(\MSbar)$,

\be
\Delta_\epsilon=\frac{1}{\epsilon}-\gammaE+\log(4 \pi)\;,
\ee
hence, after the renormalization procedure, the function in the $\MSbar$-scheme reads :
\be
I(m_1,m_2)=\frac{m^2_1\lp\ln\frac{m^2_1}{\mu^2}-1\rp-m^2_2\lp\ln\frac{m^2_2}{\mu^2}-1\rp}{16 \pi^2(m^2_1-m^2_2)}\;,
\ee

One has
\be
\xi \frac{\partial V_T}{\partial \xi}+C_T( \phi,T,\xi) V_T'=0 \,,
\label{dVdxiT}
\ee
the expressions for $C_T$ in the BSM (for $R_\xi$ gauge), generalize the $T=0$ ones given in Eqs.~(\ref{CZ},\ref{CW}).
Going to momentum space and using the imaginary time formalism as before we arrive at the thermally corrected expressions (at one loop),
\begin{align}
\label{CWT}
C_{W}(\phi,T,\xi)&=\frac{1}{2} g \Tint{k} \frac{\xi m_W}{\lp k^2-m^2_{\chi^+}\rp \lp k^2-m^2_{cW} \rp}
\ , \\
\label{CZT}
C_{Z}(\phi,T,\xi)&=\frac{1}{2} g' \Tint{k} \frac{\frac{1}{2}\xi m_B}{\lp k^2-m^2_{\chi^0}\rp \lp k^2-m^2_{cZ} \rp} + \frac{1}{2} g \Tint{k} \frac{\frac{1}{2}\xi m_W}{\lp k^2-m^2_{\chi^0}\rp \lp k^2-m^2_{c^Z} \rp}\,,
\end{align}
with
\be
C_T(\phi,T,\xi)=C_{W}(\phi,T,\xi)+C_{Z}(\phi,T,\xi)\;,
\ee
the main loop integral is
\be
I_b(m_1,m_2)=i \Tint{k}\frac{1}{(k^2-m^2_1)(k^2-m^2_2)}\;,
\ee
\be
I_b(m_1,m_2)=I'_b(m_1,m_2)+I_3(m_1,m_2)\;,
\ee
where $I'_b(m_1,m_2)$ and $I_3(m_1,m_2)$ integral is
\be
\begin{aligned}
I'_b(m_1,m_2)&=i \Tint{k}'\frac{1}{(k^2-m^2_1)(k^2-m^2_2)}
\; ,\\
I_3(m_1,m_2)&=T\int\frac{d^d k}{(2 \pi)^d}\frac{1}{(k^2+m^2_1)(k^2+m^2_2)}\;,
\end{aligned}
\ee
slove it:
\be
I'_b(m_1,m_2)=\frac{1}{16 \pi ^2 \epsilon _b}+\frac{m_1^4 \zeta (5)}{1024 \pi ^6 T^4}+\frac{m_2^2 m_1^2 \zeta (5)}{1024 \pi ^6 T^4}+\frac{m_2^4 \zeta (5)}{1024 \pi ^6 T^4}-\frac{m_1^2 \zeta (3)}{128 \pi ^4 T^2}-\frac{m_2^2 \zeta (3)}{128 \pi ^4 T^2}\;,
\ee
\be
I_3(m_1,m_2)=\frac{T}{4 \pi  m_1+4 \pi  m_2}\;,
\ee
where we introduced the modified minimal subtraction term $(\MSbar)$,

\be
\frac{1}{\epsilon_b}=\frac{1}{\epsilon}+\ln{\frac{\mu^2}{T^2}}-(\ln{(4 \pi)}-\gammaE)\;,
\ee
with $\zeta(n)$ the Riemann zeta function and employing the shorthand notation
\begin{align}
	L_b& \equiv \ln \lp \frac{\bar{\mu}^2}{T^2}\rp-2\lp \ln(4\pi)-\gammaE \rp,\,\\
	L_f& \equiv L_b+4 \ln 2\;.\end{align}

Hence, after the renormalization procedure, the function in the $\MSbar$ scheme reads :
\begin{equation}
	I_b(m_1,m_2)=\frac{L_b}{16 \pi^2}+\frac{T}{4 \pi (m_1+m_2)}\;.
\end{equation}

In 4D framework,
\begin{equation}
    C_{\mathrm{LO}}^{4D}=\frac{\xi \phi}{4} g^2(\frac{L_b}{16 \pi^2}+\frac{T}{m_{c_W}+m_{\chi^\pm}})+\frac{\xi \phi}{4} \lp g^2+g^{'2}\rp (\frac{L_b}{16 \pi^2}+\frac{T}{m_{c_Z}+m_{\chi^0}})\,,
\end{equation}
at leading order in our power counting (one can set $m_{\chi^\pm} \to m_{c_W}, m_{\chi^0} \to m_{c_Z}$)
\begin{equation}
    C_{\mathrm{LO}}^{4D}=\frac{\xi \phi}{4} g^2(\frac{L_b}{16 \pi^2}+\frac{T}{\sqrt{\xi} g \phi})+\frac{\xi \phi}{4} \lp g^2+g^{'2}\rp (\frac{L_b}{16 \pi^2}+\frac{T}{\sqrt{\xi}\sqrt{g^2+g^{'2}}\phi})\;.
\end{equation}
Returning to the $Z$-factor computed earlier within the 4D framework (see Equation~(\ref{4D Z}), and retaining only the dominant contributions, we obtain
\begin{equation}
    Z_{\mathrm{NLO}}^{4D}=\frac{1}{(4 \pi)^2}\lp \frac{L_b}{4}(3(3-\xi)g^2+(3-\xi)g^{'2}-3 L_f y_t^2\rp\;,
\end{equation}
we have
 \begin{equation}
     \xi \frac{\partial Z_{\mathrm{NLO}}^{4D}}{\partial \xi}=-2\frac{\partial C_{\mathrm{LO}}^{4D}}{\partial \phi}\;.
 \end{equation}

\bibliographystyle{JHEP}
\bibliography{reference.bib}
\end{document}